\documentclass[manuscript]{aastex61}
\usepackage{etoolbox,color,float,amsmath, threeparttable, longtable,booktabs,rotating,gensymb}
\usepackage{blindtext}
\usepackage[]{graphicx}
\usepackage{scrextend}
\usepackage{ifpdf}
\usepackage{multirow}
\usepackage{enumerate} 

\newtoggle{figs_eps.color}
\toggletrue{figs_eps.color}
\togglefalse{figs_eps.color}

\newcommand{\enig}{\texttt{\detokenize{enigma_1189}}\xspace}

\newcommand{\maket}{\texttt{\detokenize{makeTracklets}}\xspace}

\newcommand{\au}{\,\mathrm{au}}
\newcommand{\linkt}{\texttt{\detokenize{linkTracklets}}\xspace}
\newcommand{\vtree}{\texttt{\detokenize{vtree_thresh}}\xspace}
\newcommand{\pred}{\texttt{\detokenize{pred_thresh}}\xspace}

\newcommand{\platew}{\texttt{\detokenize{plate_width}}\xspace}

\submitjournal{AJ}

\shorttitle{NEO orbit linking with LSST}
\shortauthors{Vere\v{s} and Chesley}

\begin{document}

\title{Near-Earth Object Orbit Linking\\ with the Large Synoptic Survey Telescope}

\correspondingauthor{Peter Vere\v{s}}
\email{peter.veres@cfa.harvard.edu}

\author{Peter Vere\v{s}}
\altaffiliation{Present address: {\em Minor Planet Center, Harvard-Smithsonian Center for Astrophysics, 60 Garden Street, Cambridge, MA 02138}}
\affil{Jet Propulsion Laboratory, California Institute of Technology, 4800 Oak Grove Drive, Pasadena, CA, 91109}

\author{Steven R. Chesley}
\affiliation{Jet Propulsion Laboratory, California Institute of Technology, 4800 Oak Grove Drive, Pasadena, CA, 91109}




 




\begin{abstract}

We have conducted a detailed simulation of LSST's ability to link near-Earth and main belt asteroid detections into orbits. The key elements of the study were a high-fidelity detection model and the presence of false detections in the form of both statistical noise and difference image artifacts. We employed the Moving Object Processing System (MOPS) to generate tracklets, tracks and orbits with a realistic detection density for one month of the LSST survey. The main goals of the study were to understand whether a) the linking of Near-Earth Objects (NEOs) into orbits can succeed in a realistic survey, b) the number of false tracks and orbits will be manageable, and c) the accuracy of linked orbits would be sufficient for automated processing of discoveries and attributions. We found that the overall density of asteroids was more than 5000 per LSST field near opposition on the ecliptic, plus up to 3000 false detections per field in good seeing. We achieved 93.6\% NEO linking efficiency for $H<22$ on tracks composed of tracklets from at least three distinct nights within a 12-day interval. The derived NEO catalog was comprised of 96\% correct linkages. Less than 0.1\% of orbits included false detections, and the remainder of false linkages stemmed from main belt confusion, which was an artifact of the short time span of the simulation. The MOPS linking efficiency can be improved by refined attribution of detections to known objects and by improved tuning of the internal kd-tree linking algorithms.
\end{abstract}


\section{Introduction} \label{sec:intro}

The origin of asteroid orbit determination is closely tied to the discovery of Ceres by Giuseppe Piazzi on January 1, 1801. The new object, believed to be a comet, was followed briefly and lost after going through solar conjunction. To resolve the problem, Carl Friedrich Gauss developed a new method of orbit determination later the same year. With his prediction, the orbit of Ceres was determined and it was recovered at the end of 1801. 

Nowadays, more than 700,000 asteroid orbits are known and managed by The International Astronomical Union's Minor Planet Center (MPC). With the current CCD surveys, dedicated search telescopes, sub-arcsecond astrometry, and improved orbit determination techniques, the derivation of an initial orbit for a single object observed multiple times within a few days is relatively simple, with some known caveats. Subsequent follow-up and archival matching usually extends the observed arc to months or even years. However, new telescopes and deeper search, mostly for Near-Earth Objects (NEOs), make contributions more challenging for the follow-up community. 

NEOs have been a focus of attention for at least two decades, mostly motivated by the goal of reducing the Earth impact hazard by cataloging the population, but lately also for sample-return missions, proposed asteroid mining and crewed missions. Therefore, the MPC's NEO Confirmation Page is being saturated, often leaving objects either unconfirmed or with very short arcs, necessitating archival linking or searches. In 2022, the Large Synoptic Survey Telescope \citep[LSST,][]{Ivezic2014} will start to operate and provide unprecedented numbers of asteroid and comet detections at magnitudes beyond the reach of the current follow-up telescopes. LSST will operate in almost real-time in an automated mode, including identification of known moving objects and linking detections of newly discovered asteroids. The complexity of the problem lies in the fact that instead of careful treatment of a single object, LSST will have to treat millions of detections, including the spurious and false ones, and successfully link them into orbits, while rejecting false linkages. The latter part has yet to be demonstrated on a real asteroid survey comparable to LSST.

The idea of automated asteroid detection, linkage and identification was implemented for Pan-STARRS \citep{2002SPIE.4836..154K,2004SPIE.5489..667H} in its Moving Object Processing System \citep[MOPS,][]{2013PASP..125..357D}. MOPS was developed as an end-to-end collection of algorithms and programs, able to process individual detections, link detections into single-night tracklets, combine tracklets into tracks and then derive multi-night orbits. Its features and capabilities include propagation and simulation of synthetic orbits, efficiency studies, identification of detections with a catalog of known or derived orbits, providing alerts and an interactive web-interface. Its track-finding algorithm is based on scalable and variable kd-tree algorithms \citep{2007ASPC..376..395K}. The MOPS linking performance was tested in simulations, including all types of Solar System objects (150,000 orbits by \citet{2007ASPC..376..395K}) and even with expected false detection and asteroid density rates (15 million orbits by \citet{2013PASP..125..357D}). The resulting linking efficiency was close to 100\% and the high accuracy suggested that MOPS will work for an advanced asteroid survey. However, the Pan-STARRS project was never completed into its original design of 4-telescopes, its 3-year mission with a single telescope was not Solar System optimized and was different from the proposed and tested survey cadence, its limiting magnitude was below the predicted value, and the rates of spurious detections were orders of magnitudes larger than predicted. It was the false detection rates in particular, that did not allow MOPS to derive orbits due to the dramatic increase in the number of false tracks that overwhelmingly outnumbered the real ones. Notably, this experience might be a source of skepticism regarding LSST 's ability to manage the large load of real and false detections into a working linking algorithm, which has not been demonstrated yet. Still, some components of MOPS are being successfully used by many (Pan-STARRS1, Pan-STARRS2, NEOWISE, DECAM) and MOPS is planned to be used with its full capabilities, including linking, for LSST. However, the effectiveness of MOPS is crucial. Without successful linking, the expected numbers of LSST discovered Solar System objects could be significantly decreased.

Our goal was to test MOPS for LSST with a realistic density of moving objects and false detections and to understand whether MOPS can handle the expected large number of false tracks. We emphasized the realistic approach by employing the baseline survey cadence, the exact shape of the field of view, several observational constraints and parameters, as well as the most recent population models of the Solar System objects. 

\section{Large Synoptic Survey Telescope}

LSST is a next generation all-sky survey telescope currently being constructed atop Cerro  Pach\'{o}n in Chile. Its first light is scheduled for 2020 and its 10-year nominal survey will start 2 years later. This 8.4-meter, wide-field telescope, with a 3.2 Gigapixel camera and real-time image processing, data acquisition and transport, is mainly motivated by the study of Dark Matter and Dark Energy. However, its nightly coverage of $6,000$ square degrees, mostly done in two visits to a given field in the same night, provides an excellent opportunity for a deep survey of the small bodies of the Solar System. Because of its limiting magnitude, reaching to 24.5 in r-band in 30-seconds, and large load of detected asteroids and comets, the discovery and characterization of Solar System objects must be done automatically, by identifying with known objects and correct linking of new objects.

\begin{figure}[tbh]
  \centering
    \includegraphics[width=0.7\textwidth]{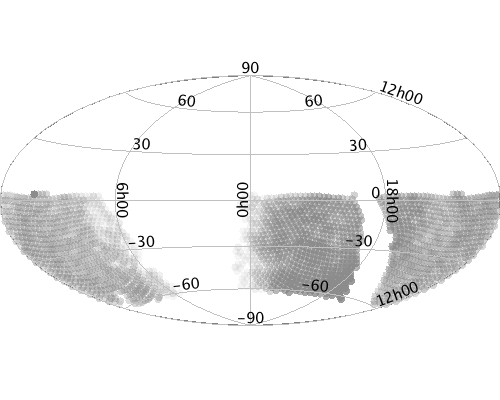}
     \caption{LSST coverage of the sky in OC 28 in equatorial coordinates.}
    \label{fig.focal_plane}
\end{figure}

For our simulations, we selected one month of the 10-year \enig baseline survey (see \citet{2017Veres_1} for a description of \enig) created by the Operations Simulator \citep[OpSim,][]{2014SPIE.9150E..15D}. OpSim provides a list of fields with information on their positions, epochs, limiting magnitudes, filters, seeing, etc. Fields also avoid the Moon and filters are sensitive to the phase of the Moon and its presence above the horizon. The selected dates covered the 28th observing cycle (OC 28) of \enig. An observing cycle is a MOPS-defined interval of time, from a full moon to a full moon. OC 28 spanned  through the months of May when the ecliptic has the largest altitude above the horizon around midnight and also the nights are the longest in the summer at the LSST site (Figure~\ref{fig.focal_plane}). Thus, the density of NEOs and Main Belt Asteroids (MBAs) is at its greatest. Some nights were removed by OpSim to simulate weather, resulting in 27 clear nights.  A small fraction of fields were observed only once per night; these singletons were removed from our simulation. The mean and maximum limiting magnitude of the selected observing cycle as well as the time spent in individual filters are denoted in Table~\ref{tab.surveys_mag}. The survey spends most its time in the r, i and z-band, and only 3\% of time in the u-band.

\begin{table}[tbh]
\small
\begin{center}
\caption{SNR=5 limiting magnitudes ($m_5$) of one month of the \enig survey and fraction of time spent in individual filters.}
\begin{tabular}{c|ccc}
\tableline\tableline
Filter & Average $m_5$ & Max $m_5$ & Time spent (\%)\\
\hline
u&23.54$\pm$0.32&24.16&3\\
g&24.69$\pm$0.31&25.36&9\\
r&24.32$\pm$0.30&24.93&25\\
i&23.65$\pm$0.36&24.34&26\\
z&22.30$\pm$0.39&23.57&20\\
y&21.49$\pm$0.23&21.94&17\\
\tableline
\end{tabular}
\label{tab.surveys_mag}
\end{center}
\end{table}

The LSST camera consists of 21 platforms called rafts, each consisting of a $3\times3$ array of 9 CCD chips, yielding a total of 189 CCDs. Each chip comprises $4096\times4096$ 10-micron pixels, and so the total number of active pixels is 3,170,893,824. Because there are gaps between chips within the $3\times3$ rafts and also between the rafts, some fraction of the focal plane is not useable. The total active area is equal to $9.50\,\mathrm{deg}^2$, whereas the total raft area yields to $10.45\,\mathrm{deg}^2$, resulting in a fill factor of 0.9089. Gaps can be simulated by an exact mask or by a statistical treatment of detections. The pixel mask approach is computationally more expensive, because it requires building and matching up the fields with the 3.2 billion pixels. This work used the probabilistic approach, where fill factor represents the probability of a potential detection to be found in a single frame. To simulate the field, we employed a square layout of 25 rafts with the area of $12.445\mathrm{deg}^2$ and then applied a mask for the four corner rafts to obtain the above mentioned $10.45\,\mathrm{deg}^2$. Finally, 90.89\% of detections were randomly selected to form the detection list.

LSST utilizes an altitude-azimuthal mount and the camera is able to rotate, and thus the fields are not generally aligned with the local RA-DEC frame. In fact, due to desired dithering, each exposure is observed in a randomized field orientation. The field rotation affects the probability of the detection to be visible in multiple visits, because some of the detection can hit the masked area in the second visit. This aspect of the survey is fully modeled in our simulations. 

\section{Field density}

\subsection{Asteroid detections}

We generated synthetic detections for NEO and MBA population models  by propagation of the orbits to the epochs of the OpSim fields. The propagation used JPL's small body codes with the DE405 planetary ephemerides, where all planets, plus Pluto and the Moon were perturbing bodies. We did not use any asteroids as perturbers. Only detections inside of the rotated field were analyzed and filtered based on the field limiting magnitude and other selected parameters of the detection model. Some details of the detection model are described in \citet{2017Veres_1}. 

\begin{figure}[tbh]
  \centering
    \epsscale{0.6}
    \plotone{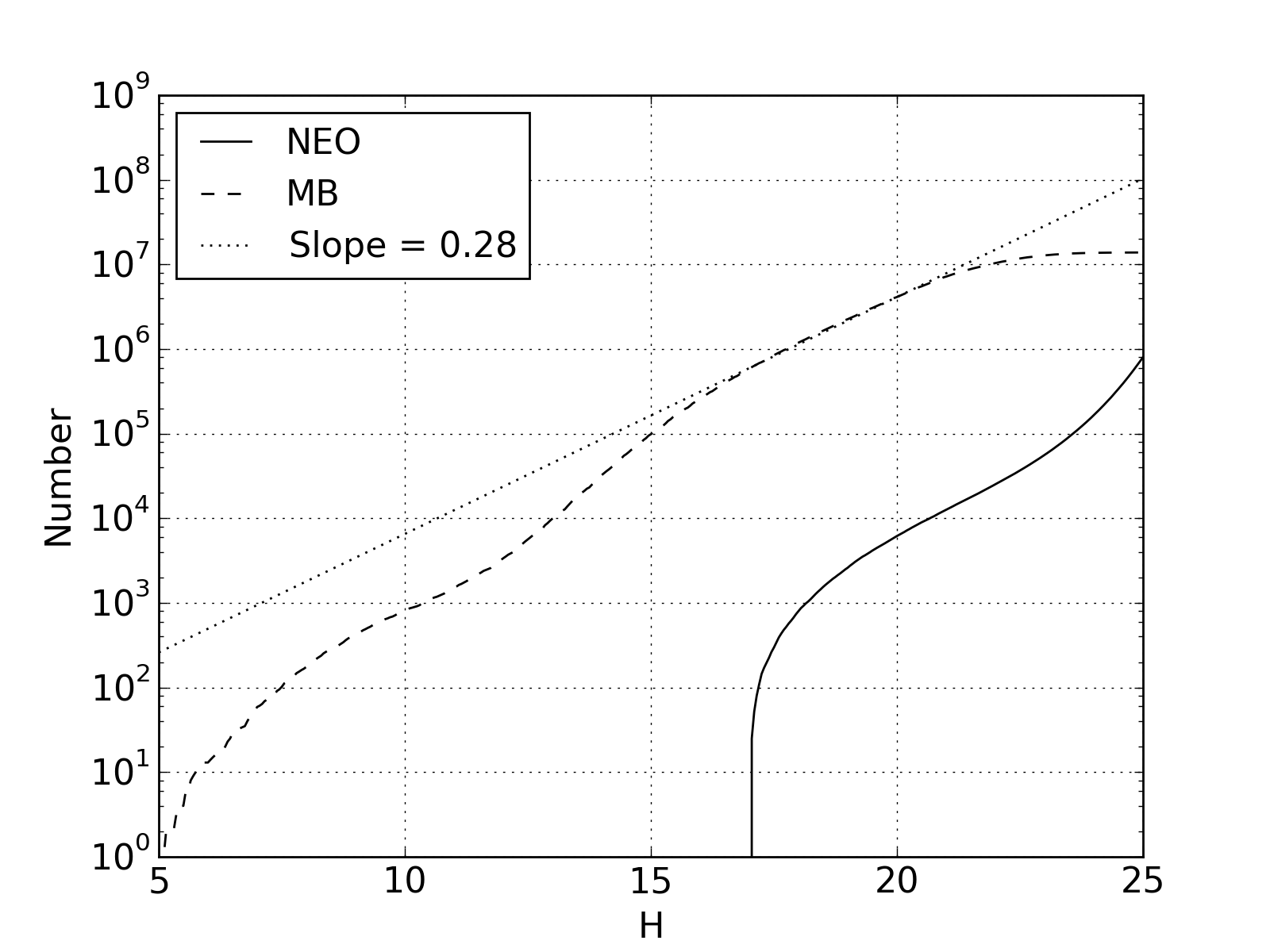}
     \caption{Comparison of MBA \citep{2011PASP..123..423G} and NEO \citep{2016Natur.530..303G} size-frequency distributions, where the model MBA slope change at $H\sim21$ is an artifact of designing a population of Pan-STARRS4 accessible MBAs.}
    \label{MB_model}
\end{figure}

We utilized a \citet{2016Natur.530..303G} NEO population containing 801,959 Keplerian orbits with absolute magnitude down to $H<25$. The distribution of its orbital elements is roughly similar to earlier work by \citet{2002Icar..156..399B}, however, the \citet{2016Natur.530..303G} population is size-dependent and its size-frequency distribution covers the $H>22$ space better than the previous work which underestimated the count. (See Figure \ref{MB_model}.) The orbital and size-frequency distribution properties of ``Granvik's" NEO population were derived from analysis of NEO observations by the Catalina Sky Survey and Mt. Lemmon telescopes. Our NEO population is artificially deficient of large NEO with $H<17$; however, these are believed to be essentially all discovered and there are only about 500 of them, thus they are a negligible fraction of other detections in an LSST field of view. Initially, we were also using the earlier NEO model by \citet{2002Icar..156..399B}, which we denote as ``Bottke's'', that also contains objects down to $H<25$; however, its total number is only $268,896$ and is thus deficient in small objects, particularly for $H>22$.

MBAs will dominate the number density of moving objects in the LSST field of view, and they represent a source of background noise and possible confusion for NEO identification. In our LSST simulations, we used the \citet{2011PASP..123..423G} model of the main-belt population (see Figure \ref{MB_model}). This population contains 13,883,361 orbits and is the most robust population model available to date.

In the Grav MBA model, the cumulative distribution slope is equal to $\alpha=0.28\pm0.01$ for H between 16 and 20. However, the population was created for a Pan-STARRS4-like survey with a limiting magnitude of $m_{V}=24.5$, and so it is truncated to remove MBAs that are fainter than $m_{V}=24.5$ when at perihelion and at opposition. This truncation results in an artificial break, seen in Figure~\ref{MB_model}, in the Grav population size-frequency distribution at $H\simeq21$. 

To investigate how this break affects the areal density of MBAs in the LSST survey simulation, we compared the simulated MBA density in LSST fields to the predicted number density by \citet{2009Icar..202..104G} who had observed MBAs with the 3.8-meter Mayall telescope at Kitt Peak National Observatory in 2001 within the so-called SKADS survey. SKADS detected asteroids in a fixed 8.4 deg$^2$ patch of sky around the opposition point in the Johnson-Cousins R-band down to limiting magnitude of 23.0--23.5 on six nights spanning an 11-night baseline. Based on \citet{2009Icar..202..104G}, the debiased cumulative  number of MBAs follows the equation 
\begin{equation}
N(>H)\propto10^{\alpha H}
\end{equation}
where $\alpha=0.30\pm0.02$. This slope was derived for H in the range 15--18, with assumed validity to at least H=20. \citet{2009Icar..202..104G} derived the areal density of MBAs as
\begin{equation}
\label{eq.gladman}
N(<m_{R})=210*10^{0.27*(m_{R}-23)}
\end{equation}
where $N(<m_{R})$ is the cumulative number of asteroids brighter than $m_{R}$ per square degree. The derived detection efficiency was $98\%$ at $m_{R}$=17.

To compare with our modeled number density of MBAs, we selected LSST fields with  solar elongation greater than $178\deg$ and within one degree from the ecliptic from OC 28, yielding 27 fields. This simulation was run with fill factor of $\epsilon_{0}=90.89\%$, fading and color transformation assuming all asteroids are of a spectroscopic S-type (see \citet{2017Veres_1}). There was a slight difference in the definition of detection efficiency.  Our modeled detections are subject to so-called fading that reduces detection efficiency as
\begin{equation} 
  \label{eq.fading1}
	\epsilon(m) 
	   = \frac{\epsilon_{0}}{1+e^{\frac{m-m_5}{w}}}
\end{equation}
where $\epsilon(m)$ is the detection efficiency, $m$ the apparent magnitude, $m_5$ the limiting magnitude defined for $\mathrm{SNR}=5$ and $w=0.1$ the width of the fading function. SKADS defined its detection efficiency by a similar relation
\begin{equation} 
  \label{eq.fading2}
	\epsilon(m) 
	   = \frac{\eta_{0}-c(m-17)^{2}}{1+e^{\frac{m-m_5}{w}}}
\end{equation}
where, based on observations, $\eta\approx0.98$ and $c\approx0.005$. Here $c$ measures the strength of the quadratic drop and the remaining parameters are the same as in the previous equation.

Additionally, there are a number of sources of uncertainties that must to be considered in the estimate of the MBA density:

\begin{enumerate}[a)]
\item A different slope of the population. $\alpha=0.28$ and $0.30$ for Grav and Gladman, respectively.
\item  The transformation from the LSST bands and the SKADS R-band to V-band. The term $V-R$ in SKADS was $0.37\pm0.15$ mag, leading to relative uncertainty of about 9\% in areal density when transforming to V-band.
\item  The scaling of the detection efficiency. This work used a different model than SKADS for fading.
\end{enumerate}

\begin{figure}[tbh]
  \centering
  \epsscale{0.6}
    \plotone{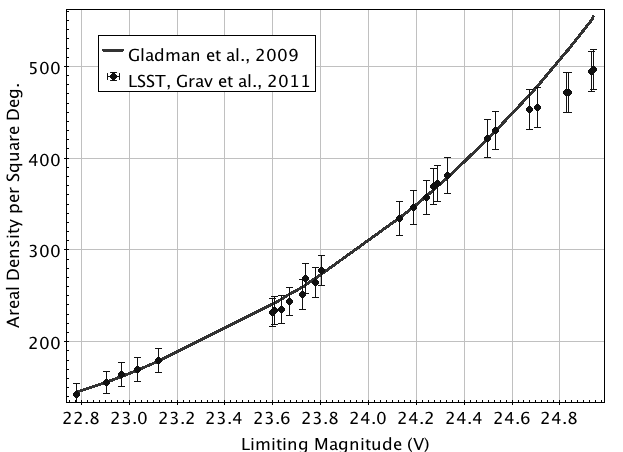}
     \caption{Number of MBAs per square degree at opposition on the ecliptic based on \citet{2009Icar..202..104G} and the \citet{2011PASP..123..423G} population used in this work.}
    \label{MB_comp}
\end{figure}

\begin{figure}[tbh]
  \centering
  \epsscale{0.6}
   \plotone{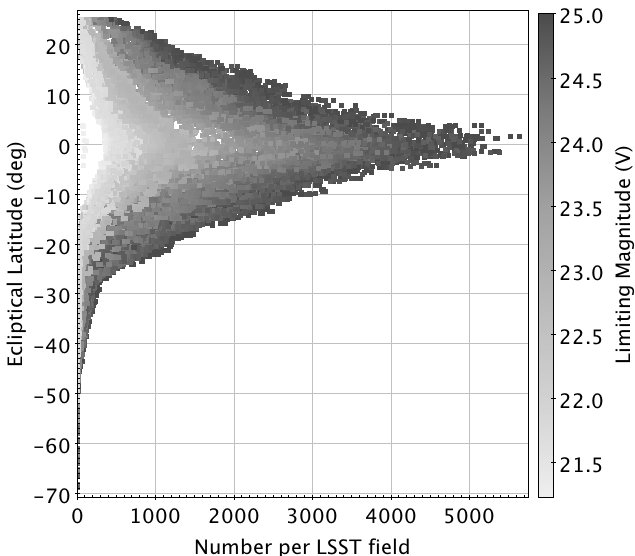}
     \caption{Number of detected MBAs per LSST field as a function of limiting magnitude (V) and ecliptic latitude.}
    \label{MB_density}
\end{figure}

Figure~\ref{MB_comp} shows the number density of MBAs near opposition as a function of limiting magnitude of the field in V-band based on the SKADS survey and the simulated LSST survey with the synthetic Grav MBA population. Note that at $m_5>24.5$ the simulated MBA density drops because of the artificially truncated Grav's population. In \enig, 14\% of the fields have a limiting magnitude fainter than 24.5 in V-band. Depending on the limiting magnitude and the elongation from ecliptic and opposition, the MBA density in our simulation was underestimated by up to 12\% in those fields. However, few of the 14\% fields fainter than 24.5 mag were taken at opposition near the ecliptic, and so the effect of the truncation in Grav's MBA population is presumed negligible. The density of MBAs decreases significantly as a function of ecliptic latitude (Figure~\ref{MB_density}).

\subsection{Measurement errors}

Each ephemeris-based position in the field was altered by adding realistic astrometric and photometric errors based on the computed signal-to-noise ratio (SNR). The limiting magnitude of the field $m_5$ is defined for SNR=5. The SNR of a detection \citep{2009AAS...21346003I} is computed from the difference between the computed magnitude $m$ and $m_5$ as
\begin{equation}
\label{eq.snr}
\mathrm{SNR} = \frac{1}{\sqrt{(0.04-\gamma).\chi+\gamma \chi^2}}
\end{equation}
where $\gamma=0.038$ and $\chi=10^{0.5(m-m_5)}$. Then, photometric uncertainty is derived as
\begin{equation}
\sigma_m=2.5\log_{10}{\left(1+\frac{1}{\mathrm{SNR}}\right)} .\end{equation}
and the computed $m$ is combined with an error drawn from a normal distribution with a mean of zero and variance $\sigma_{m}^2$.

We have assumed that LSST astrometry is measured relative to a post-Gaia star catalog and so absolute systematic errors are negligible while relative errors are expected at a floor level of 10\,mas. The astrometric error $\sigma_{\mathrm{astr}}$ for any detection is therefore computed as quadrature combination of $10\,\mathrm{mas}$ and the ratio of the seeing $\Theta$ and SNR
\begin{equation}
\sigma_{\mathrm{astr}}^2=(10\,\mathrm{mas})^{2}+\left({\frac{\Theta}{\mathrm{SNR}}}\right)^2.
\label{eq.snr}
\end{equation}
Asteroids are moving targets and so, depending on the rate of motion, their shape deviates from a stellar PSF and is in fact a convolution of the motion and the PSF. The faster the object moves, the larger the astrometric error. Therefore, if the trail length $L>\Theta$, the seeing term $\Theta$ in Eq.~\ref{eq.snr} is replaced by the geometric mean of seeing and trail length as $\Theta'=\sqrt{\Theta L}$.

\begin{figure}[tbh]
  \centering
    \includegraphics[width=0.49\textwidth]{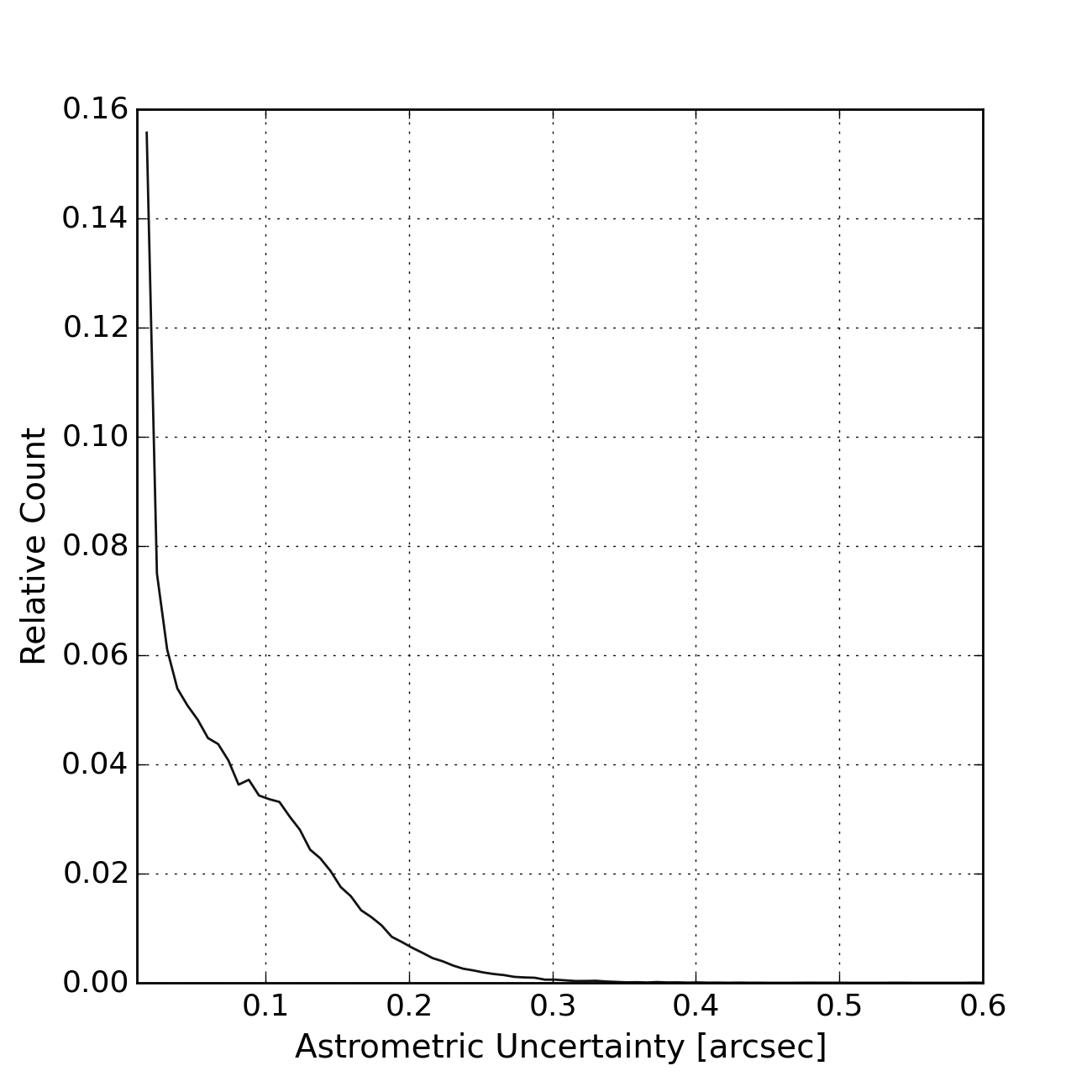}
        \includegraphics[width=0.49\textwidth]{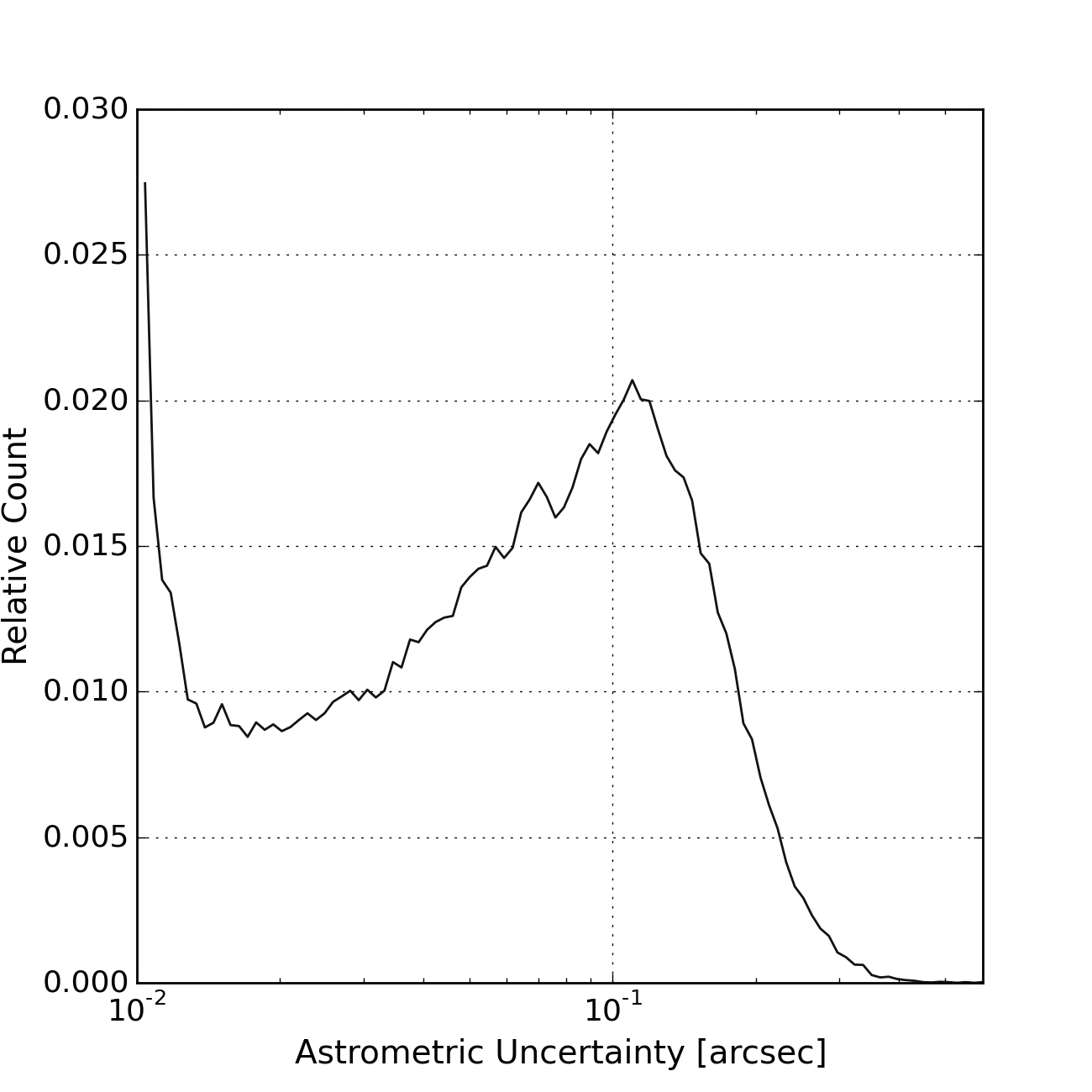}
     \caption{Distribution of astrometric uncertainties of NEOs - normal scale (left) and logarithmic scale (right).}
    \label{fig.unc}
\end{figure}

To obtain realistic astrometry, we combine the computed position with an astrometric error term drawn from a normal distribution with a zero mean and variance of $\sigma_{\mathrm{astr}}^2$. Figure~\ref{fig.unc} shows histograms of astrometric uncertainties, in both linear and log-scale. The latter shows that there are two populations of NEA detections, those with high SNR and therefore low uncertainty, around $10\,\mathrm{mas}$, and another centered around $100\,\mathrm{mas}$ from low SNR detections, which presumably also includes most of the objects with relatively fast rates of motion. The median astrometric error obtained for NEOs is $47\,\mathrm{mas}$. 

To simulate observational constraints and limitations of the LSST processing pipeline and CCD effects, we employed a set of filters that determined whether a detection that fulfilled the limiting magnitude was still visible. We included vignetting, which reduces sensitivity to detections that are far from the optical axis of the field. The LSST optical design minimizes vignetting, with only 7\% of the collecting area having a penalty above 0.1 mag. In CCD surveys the limiting magnitude does not behave like a step function that strictly determines the visibility. In fact, the detection limit follows a fading function, e.g., Eq. (\ref{eq.fading1}) that defines the limiting magnitude as a 50\% probability of detection. In our work, this value is taken at SNR=5 and denoted as $m_5$. The fading function is multiplied by a fill factor, simulating the focal plane gaps. Because of the sidereal tracking rate, all asteroids will move, and particularly fast moving NEOs will look trailed. The detected trails are described by a convolution of a point-spread-function with the motion vector. The longer the trail, the fainter the peak signal and the SNR decreases. This loss effectively decreases the magnitude of asteroids as a function of their on-sky rate of motion. We assumed that all NEOs and MBAs are of S-types for the purpose of the ephemeris computed V-band magnitude transformed to the LSST filter system. The details of this detection model are discussed by \citet{2017Veres_1}.

\subsection{False detections}

The LSST transient detection data stream will include many detections that are not associated with solar system objects, and the objective of linking only real LSST detections of moving objects to form tracks and orbits represents a significant challenge. There are three broad categories of non-solar system transients that are expected from LSST. The first category of LSST transient detections arise from real astrophysical phenomena (e.g., variable stars, supernovae, etc.) that appear in the same location in multiple instances. Such astrophysical transients will be filtered out of the MOPS input stream by virtue of their stationary appearance and thus will not affect the asteroid linking problem. 

\begin{figure}[tbh]
  \centering
    \includegraphics[width=0.6\textwidth]{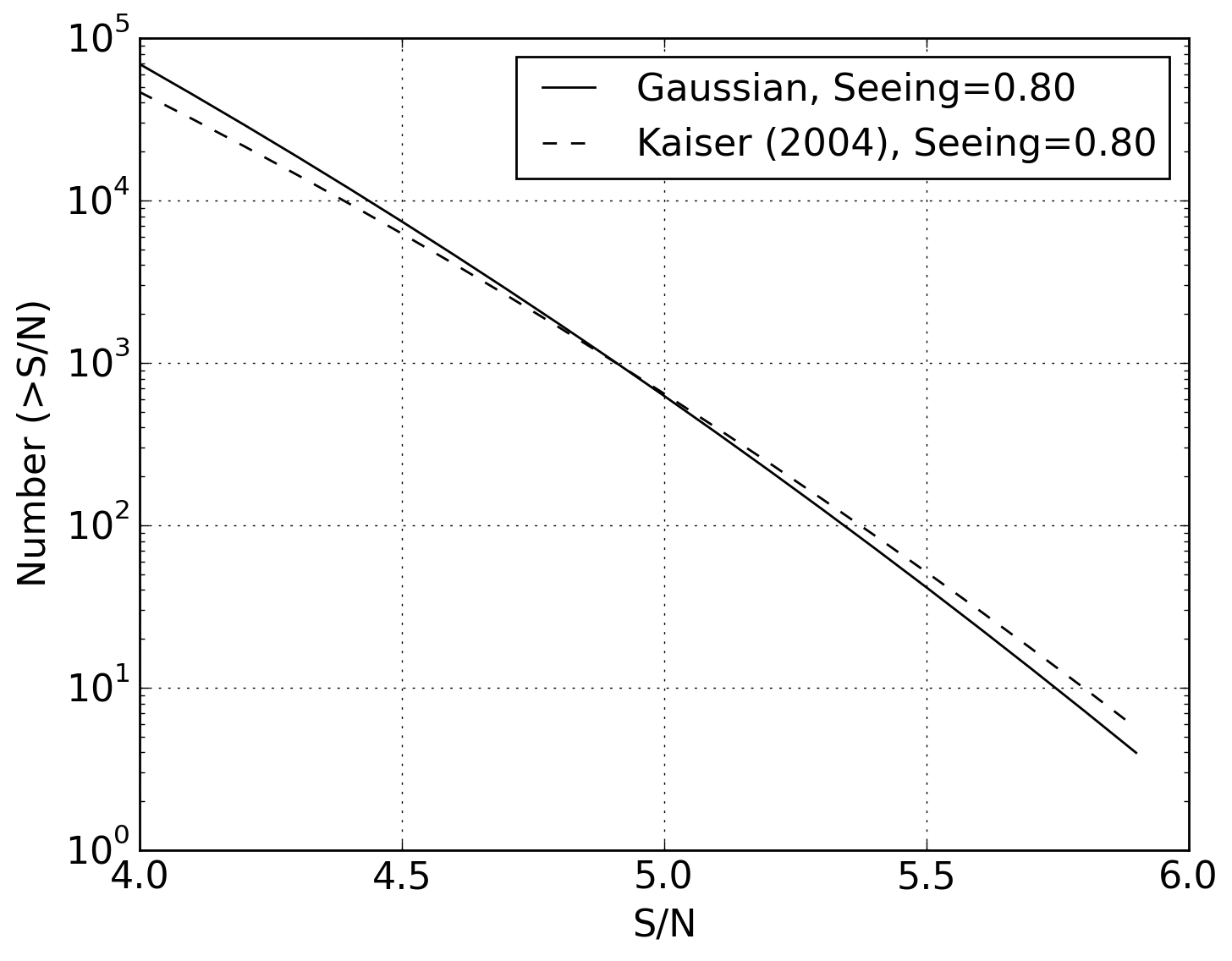}
    \includegraphics[width=0.6\textwidth]{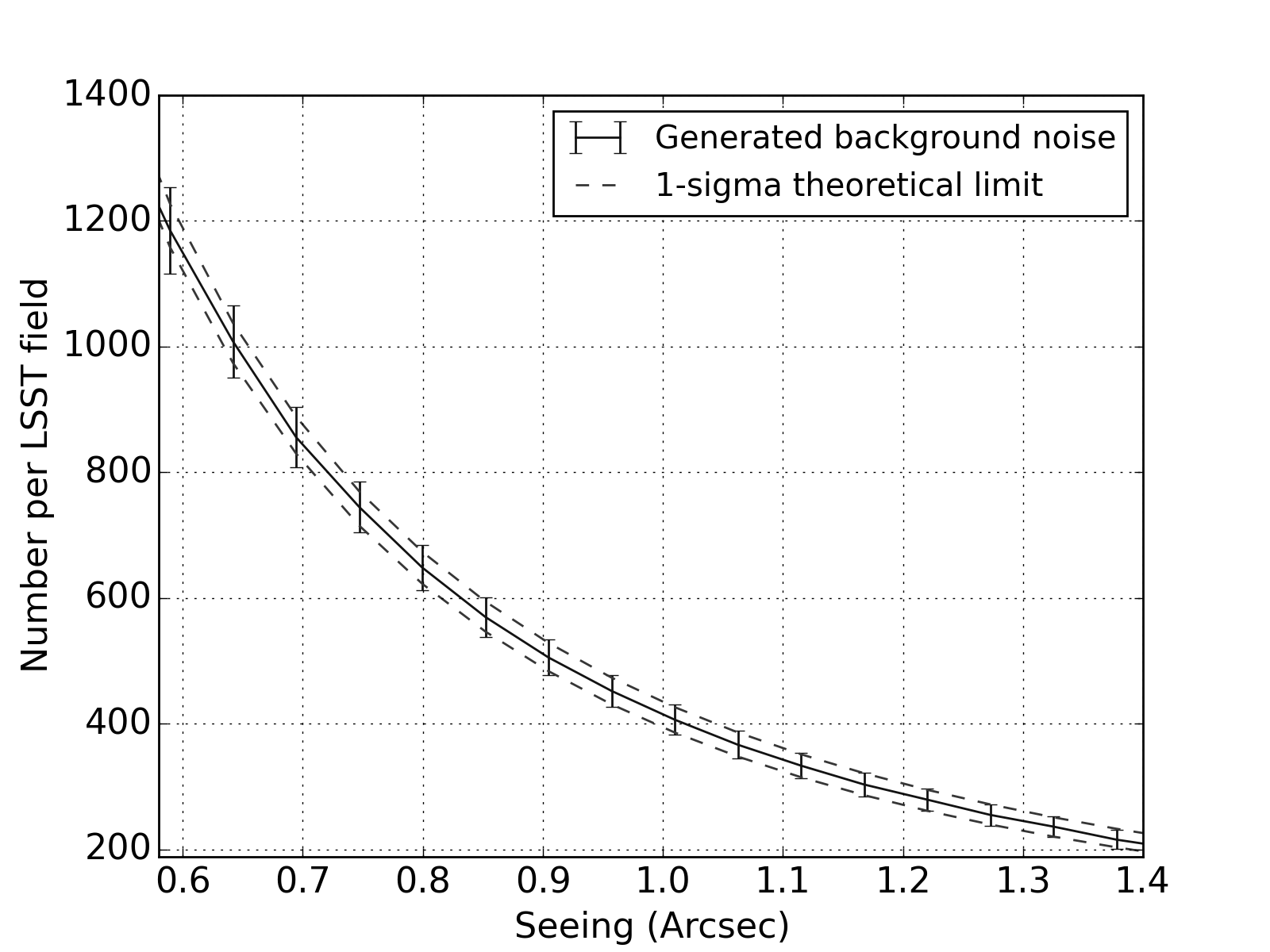}
     \caption{(top) The number of random noise in LSST field as a function of signal-to-noise ratio. Similarity to normal distribution is demonstrated by the dashed line. (bottom) The theoretical and generated numbers of random noise in LSST fields.}
    \label{fig.NS_noise}
\end{figure}

The remaining two categories of non-solar system transients consist of spurious detections arising from either random noise or image differencing artifacts, both of which will enter the MOPS input stream. The first source of false detections, from random fluctuations in the sky background and from detector noise, is driven by gaussian statistics at the individual pixel level. The number $N_{>\eta}$ of these $random$ sources above a given signal-to-noise threshold $\eta$ in the CCD image where Gaussian noise is convolved with a Gaussian PSF follows the formula by \citet{Kaiser04}
\begin{equation}
N_{>\eta}=\frac{S}{2^{5/2}\pi^{3/2}\sigma_g^2}\eta e^{-\eta^2/2},
\label{eq.Kaiser}\end{equation}
where $S$ is the total number of pixels in the focal plane array, $\sigma_g\simeq\Theta/2.35$, and $\Theta$ is the FWHM seeing measured in pixels. The number of random false positives depends strongly on the seeing (Figure~\ref{fig.NS_noise}), with the better the seeing the larger the number of random false positives. The average \enig seeing of 0.80 arcsecond leads to 650 random false positives with $\mathrm{SNR}>5$ in one LSST image. 

\begin{figure}[tbh]
  \centering
    \includegraphics[width=0.49\textwidth]{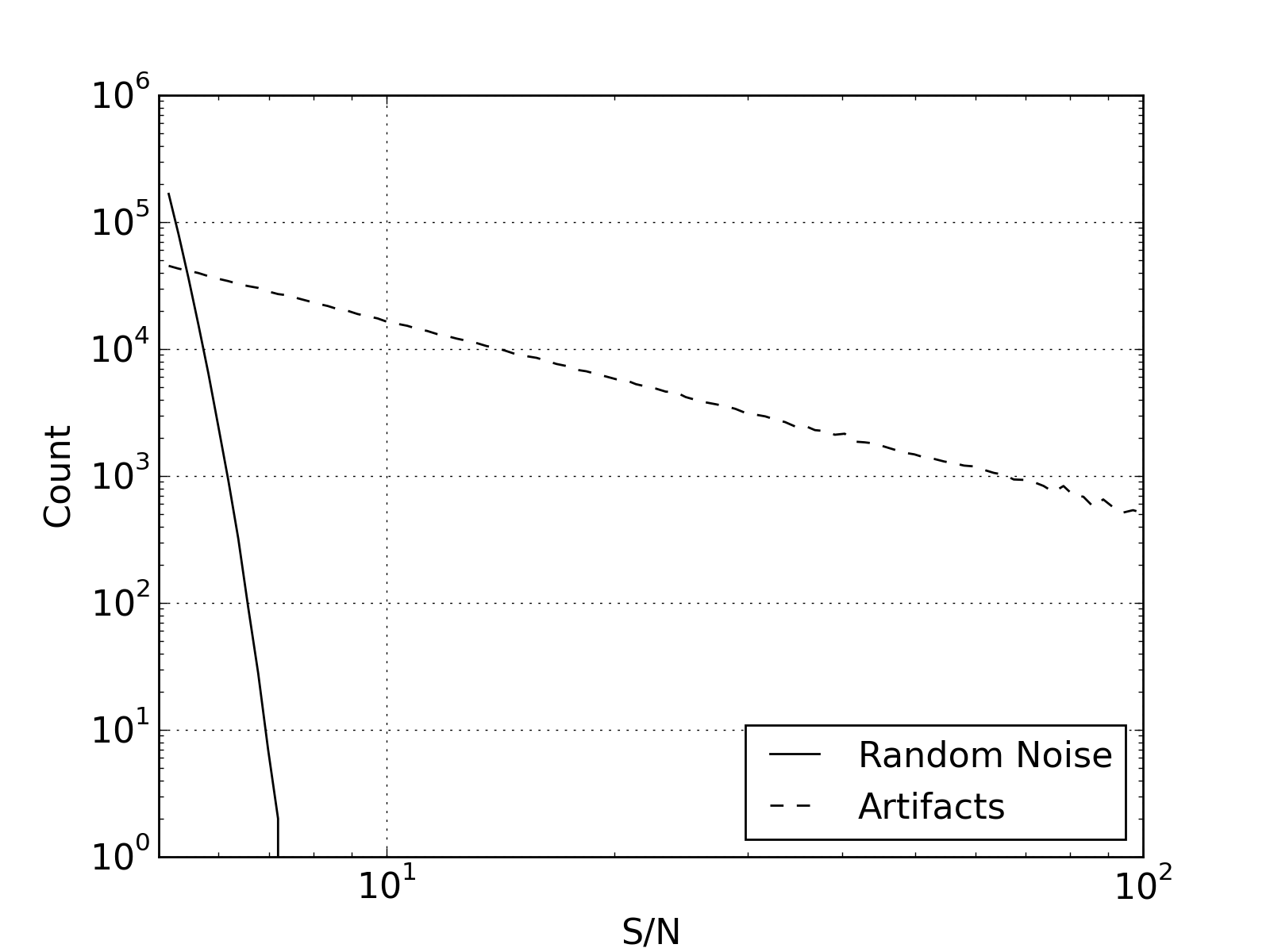}
    \includegraphics[width=0.49\textwidth]{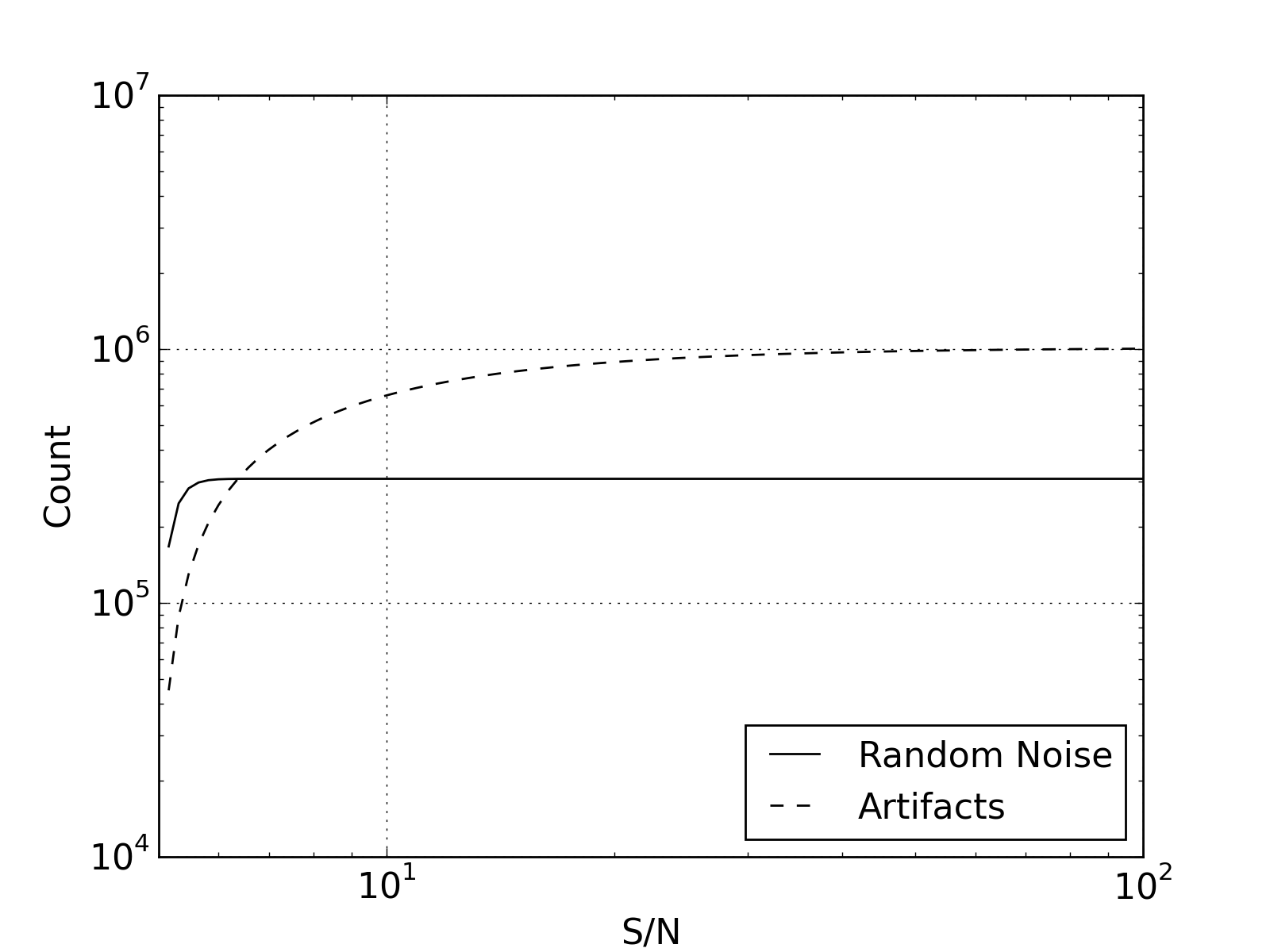}
     \caption{Histogram (left) and cumulative distribution (right) of random noise and artifacts on one night of LSST survey.}
    \label{fig.hist_noise}
\end{figure}

We generated random false positives following Equation~\ref{eq.Kaiser} in random x-y positions in the field. The number of random false positives for a given field was selected from a normal distribution with a mean and variance of $N$ from equation~\ref{eq.Kaiser}. Then, magnitudes were assigned to the generated random noise as follows: We generated a random number $p$ from a uniform distribution [0,1]. This number corresponds to the normalized cumulative distribution $N(>\eta)/N_{TOTAL}$. Then $\eta=\sqrt{\eta_{0}^{2}-2\log(1-p)}$ which can be directly transformed to a magnitude as $V=V_{LIM}-2.5\log(\eta/\eta_{0})$ where $V_{LIM}$ is the $m_5$ limiting magnitude at $\eta_{0}=5$. The number density of random false positives has a strong dependence on SNR; therefore, most of the random noise sources will be near the the limiting magnitude (Figure~\ref{fig.hist_noise}).

The second source of false detections comes from difference image artifacts, which arise from differencing a field image with a fiducial image of the static sky that has been derived from a stack of several (or a great many) images of the same field over some time period. This differencing removes stationary objects so that only transient sky phenomena, including moving objects, appear as detections in the difference image. However, registration errors across the field can leave dipole-shaped artifacts in the difference image at the location of a static source. Artifacts may also originate from a poor convolution kernel, variable seeing across the field, stray light in the optical system or reflections in the lenses. Artifacts are often concentrated around bright sources due to saturation or diffraction spikes, and masking around these sources can be an efficient means of substantially reducing the rate of artifacts. Although an improved optical configuration and machine learning can remove many of these false artifacts, some fraction will always remain in the data stream. 

For this work we assumed the estimated density of differencing artifacts derived by \citet{Slater}, who used actual imagery obtained by the Dark Energy Camera (DECAM) on Cerro Tololo \citep{2015AJ....150..150F} and processed them with a nascent version of the LSST image processing pipeline. \citet{Slater} report that the primary result of their study is that ``the LSST pipeline is capable of producing a clean sample of difference image detections, at roughly the 200--400 per square degree level.'' This is their final result, but our work used a preliminary estimate as the point of departure for our linking simulations. This earlier estimate allowed for roughly 90--380 artifacts per square degree, and we took the geometric mean of this range as the starting point, which leads to $185/\mathrm{deg}^2$ or 1777 artifacts per LSST field. \citet{Slater} did find far higher concentrations of artifacts near bright stationary sources, which they eliminated by masking the area around them, thus allowing the reported low artifact density. Following their result, we modeled bright source masking by reducing the effective fill factor by 1\%. 

\begin{figure}[tbh]
  \centering
          \includegraphics[width=0.6\textwidth]{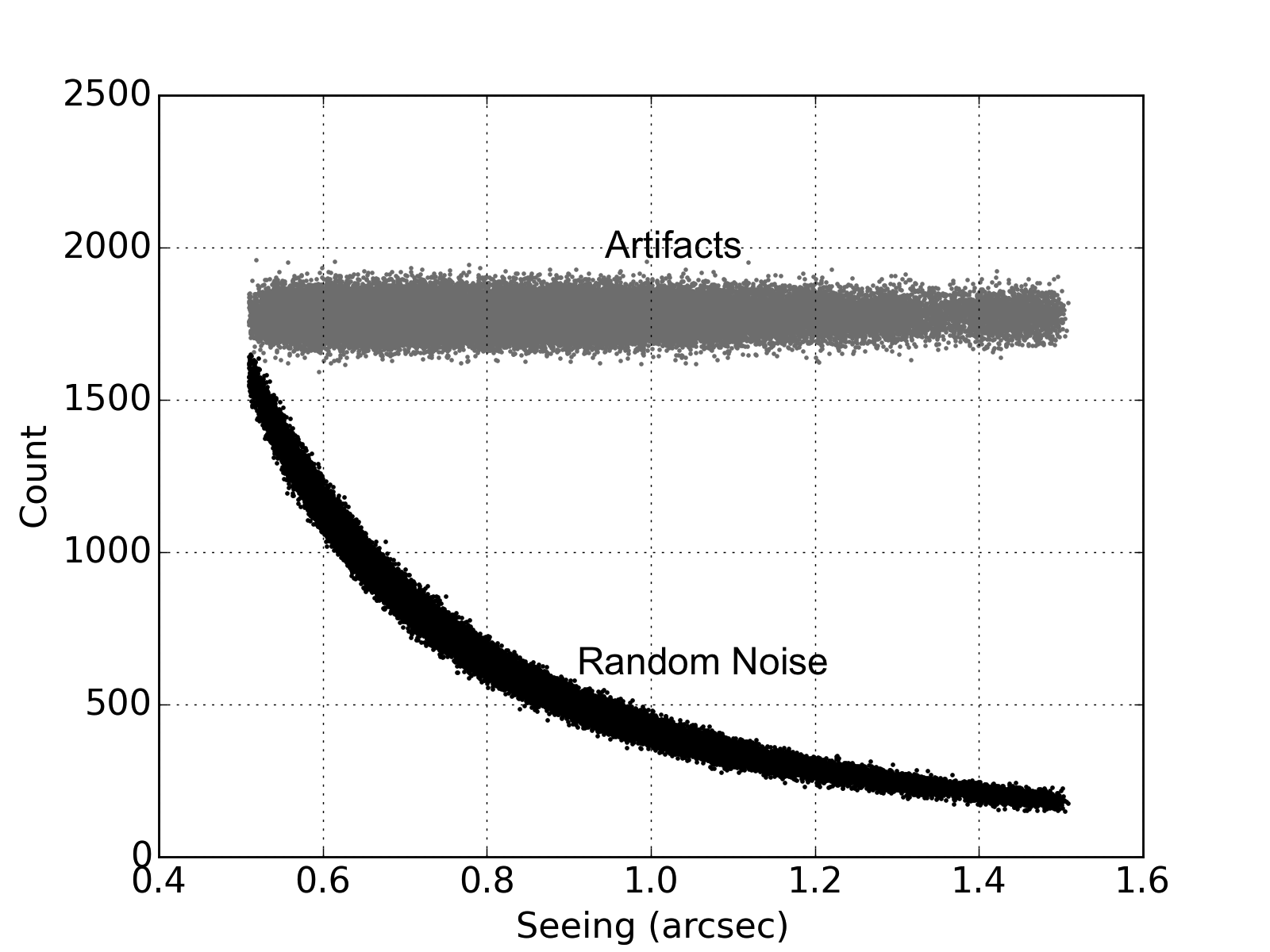}
     \caption{Random noise and artifact counts per individual field as a function of seeing during one month of the LSST survey.}
    \label{fig.scatter_noise}
\end{figure}

To seed the detection list with artifacts, we selected the number of artifacts in each field according to a gaussian distribution with mean and variance 1777 and distributed them randomly across the field. Thus our artifact rate was roughly $3\times$ the rate from random noise in typical seeing (Figure~\ref{fig.scatter_noise}), and about half of the upper bound derived by \citet{Slater} from processing actual DECam data. 

Our model for difference artifacts is independent of observing conditions such as seeing and field density. However, we note that the most dense regions of the galactic plane are relegated to the Galactic Plane proposal observations in \enig, which happens to be mostly covered by a single-visit-per-night cadence, and is anyway only a few percent of observing time. If we remove all Galactic Plane proposal fields from \enig there is a negligible effect on NEO completeness. Thus our linking and completeness results do not require or assume operation in star fields with extreme density.

Based on the \citet{Slater} report, we model that the SNR distribution of differencing artifacts follows $\propto\mathrm{\mathrm{SNR}}^{-2.5}$. The algorithm computes the SNR $\eta$ from $\eta=\eta_{0}{(1-p)}^{-2/3}$ where $p$ is a randomly generated number from a uniform distribution [0,1]. (See Figure~\ref{fig.hist_noise}.) The magnitude of a simulated artifact is then derived according to $V=V_{LIM}-2.5\log(\eta/\eta_{0})$ where $V_{LIM}$ is the $m_5$ limiting magnitude at $\eta_{0}=5$. Artifacts have much shallower dependence on $\eta$, and therefore tend to be far  brighter than random noise sources. Roughly half of modeled artifacts have $\mathrm{SNR}>10$, while virtually none of the random false detections had $\mathrm{SNR}>7$. 

The brightness distribution of artifacts suggests that at least some potential false tracklets that include artifacts can be immediately eliminated by enforcing consistency in the photometry. However, according to Figure~\ref{fig.hist_noise}, about 90\% of artifacts have $\mathrm{SNR}<20$, and if a bright artifact with $\mathrm{SNR}=20$ is paired with a faint asteroid detection having $\mathrm{SNR}=5$ the magnitude difference will be $\Delta m = 2.5\log_{10}\frac{20}{5}\simeq 1.5\,\mathrm{mag}$. As it happens, MOPS limits the photometric variation among tracklet components to $\Delta m < 1.5\,\mathrm{mag}$ by default, which suggests that few false tracklets in our simulation have been eliminated in this way. This criteria could be made more strict, which would reduce the false tracklet rate at the risk of removing real objects that are actually more interesting by virtue of a large light-curve amplitude. Thus, as a rule, the photometric consistency requirement should be as relaxed as much as feasible in order to avoid eliminating real tracklets. We suspect that this requirement can be dropped altogether without significantly impacting linking performance.

We note that our work neglects the possibility that artifacts are spatially correlated in RA-DEC, which could introduce difficulties in the linking process whereby artifacts could reappear near the same RA-DEC location and mimic the motion of asteroids. RA-DEC correlation among artifacts could arise from two causes, either camera defects or stationary sources. For LSST, the rotational dithering of the camera serves to break the correlation from any defects in the instrument, most of which would already be masked in processing, and the masking of bright stationary sources serves to remove them as a source of artifacts. \citet{jones2017} found that the rate of correlated detections in the DECam data stream was low enough to be negligible for our purposes, only $\sim2/\mathrm{deg}^2$. This no-correlation assumption is at variance with the Pan-STARRS1 experience, but appears to be well justified for LSST.

\section{Moving Object processing System}

A central question for this work is whether the linking of tracklets into tracks and orbits will prove successful with real LSST data. LSST MOPS will receive full-density lists of detections of moving and transient targets, including NEOs, MBAs and false detections. From these inputs MOPS must create tracklets, tracks and orbits, despite the fact that the data stream is contaminated by potentially large numbers of false detections, which leads to high rates of false tracklets. Our simulation synthesized detections in the LSST fields from a full-density NEO model ($\sim850,000$ orbits), an MBA model ($\sim11$ million orbits) and false detections (both random noise and differencing artifacts). The final detection lists were submitted to the MOPS \maket routine, and tracklets were created. Finally, tracklets were submitted to the linking stage, the most challenging step.

\subsection{Tracklets}

\begin{figure}[tb]

  \centering
    \epsscale{0.7}
    \plotone{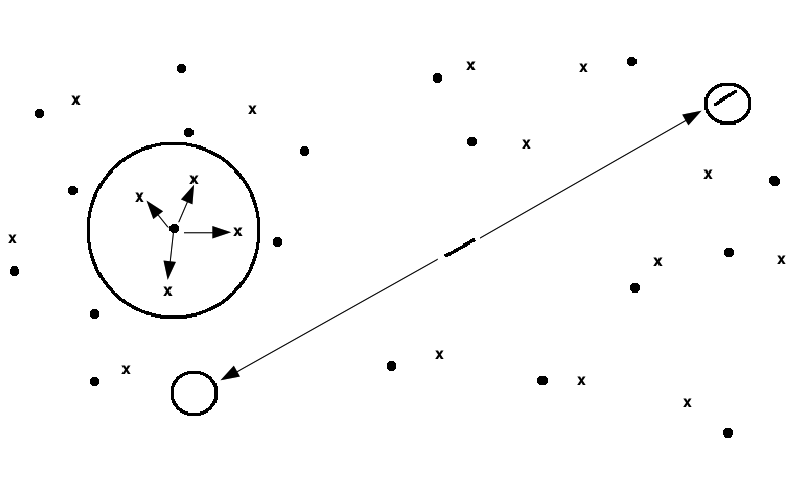}
     \caption{Schematic diagram for tracklet generation. Dots represent detections from the first image, x signs from the second one. The large circle represents the upper velocity limit for creating tracklets without rate information (up to $1.2\degree$/day). Arrows in that circle are all possible tracklets, connecting the first detections with all detections from the second image in the reach. Every  detection in the image has such a circle and corresponding set of tracklets. If the detection is faster than 1.2$\degree$/day it will be trailed (on the right), and information on the trail length and orientation can be used to search a smaller area for its counterpart in the second image (in two separate regions because the direction of motion is unknown). The matching detection must also be a trail with similar length and orientation.}
    \label{fig.tracklet_cartoon}
\end{figure}

The list of detections for a given field that has been multiple times in a night is submitted to the \maket part of MOPS. A tracklet is created for a detection in the first image if there is a second detection in its vicinity in the second image. The radius of the search circle is defined by the lower and upper velocity thresholds of \maket, which were set to 0.05$\degree$/day and 2.0$\degree$/day, respectively, in this study. If there are more possible connections in the circle, in addition to the ``CLEAN" tracklet, consisting of detections of one object, then a ``MIXED" tracklet consisting of detections of two objects or a ``BAD" tracklet that includes a false detection is created as well. Increasing the upper velocity limit increases the search area and thus the number of false tracklets. In some simulations, for velocities of 1.2--2.0$\degree$/day, we used the information on the trail length to limit the search area for companion detections. At 1.2$\degree$/day, a detection will have a non-PSF shape and its length will be 1.8 times the PSF for the average 0.86 arcsec seeing, and so its length and orientation can be determined. Thus, instead of a large circular search area around trailed detections, smaller regions consistent with the anticipated velocity and direction of the trails are searched, and any matching detections must have a compatible trail length and direction. See Figure~\ref{fig.tracklet_cartoon} for a graphical depiction.

\begin{figure}[tb]

  \centering
    \epsscale{0.9}
    \plotone{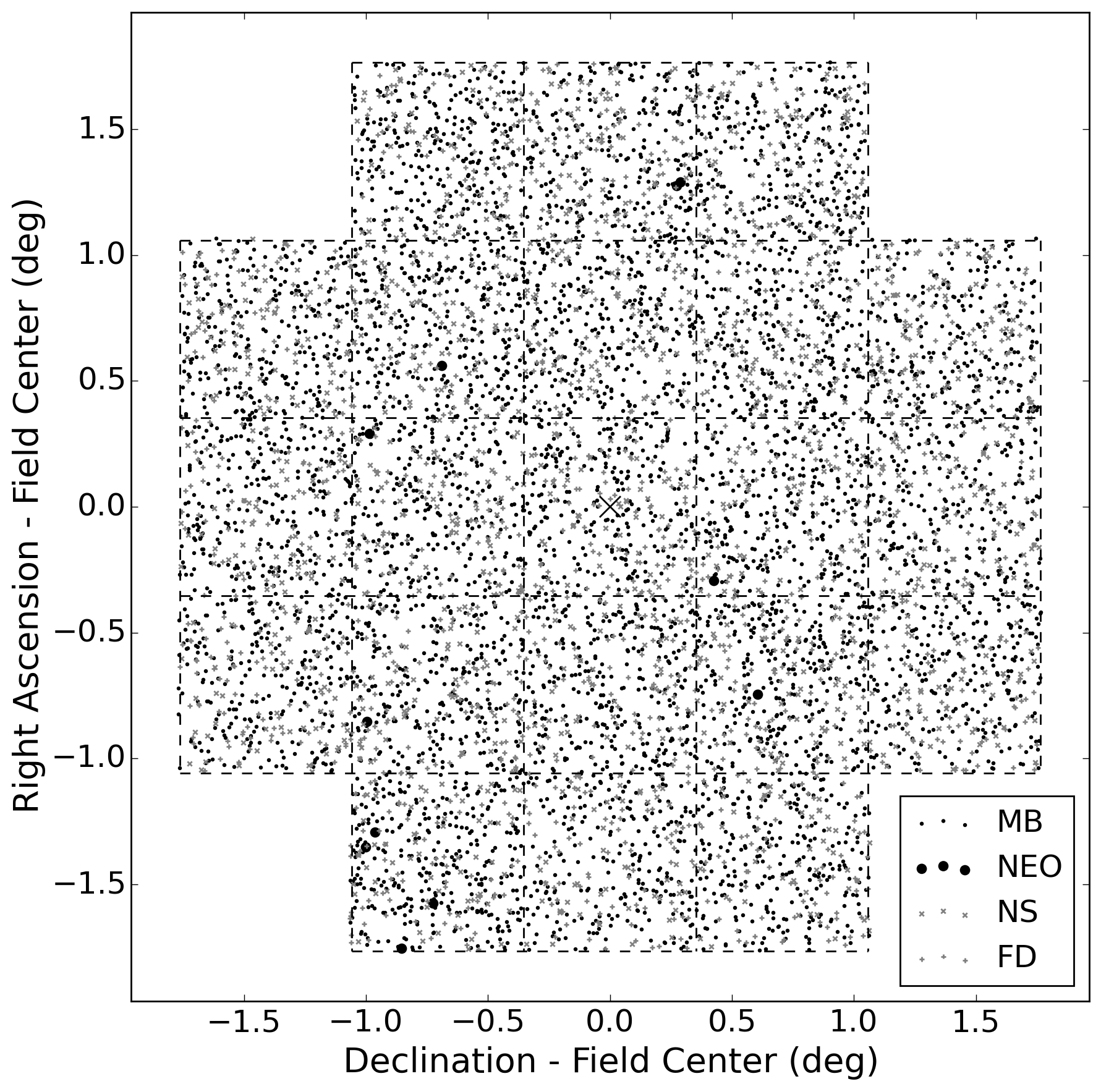}
     \caption{An example of a high-density LSST field from the \enig survey. The depicted field is number 1891 from night 3052, taken in the r filter with $m_5=24.79$, seeing 0.63 arcsec, airmass 1.04, and field center at opposition-centered ecliptic coordinates (Lat., Long.) $=(2.91\degree, 1.26\degree)$. Thus the field is near opposition in excellent conditions. The various types of detections referenced in the legend are ``MB''---main-belt asteroids, ``NEO''---near-Earth objects, ``NS''---false detections from random noise, and ``FD''---false detections arising from image differencing artifacts.}
    \label{fig.density}
\end{figure}

The number of tracklets depends on the density of detections, which can be large (Figure~\ref{fig.density}). To understand the feasibility of the simulation we gradually increased the number of detections in OC 28. The following steps are also summarized as Cases A-E in Table~\ref{Tab.confusion_FF}.

\begin{enumerate} 
\item Initially, we only used NEO orbits from Bottke's model (Case A, Table~\ref{Tab.confusion_FF}). Switching to Granvik's NEO model increased the number of detections by 35\% and tracklets by 55\% (Case B). Because Granvik's NEO model is more current and has many more objects we used that population in the simulations. At this stage, with only NEO orbits, nearly all tracklets were CLEAN, with only 4 MIXED tracklets (99.97\% tracklet purity). 

\item Adding the MBA population to Granvik's NEOs (Case C) increased the number of detections in one month to 15 million and the number of tracklets to 6 million. Most of the tracklets were for MBAs; however, about 17\% of tracklets were MIXED, i.e., derived from different objects. The large number of MIXED tracklets was substantially reduced by taking advantage of trail-related velocity information when in the velocity range 1.2--$2.0\degree/\mathrm{day}$ (Case D). In this dual velocity mode of \maket, $1.2\degree/\mathrm{day}$ is the upper threshold for creating a tracklet by searching in a circle. If the detection is trailed and the trail length implies a velocity $>1.2\degree/\mathrm{day}$, then its matching pair in the second image must be in a predicted location, based on the time between exposures, and the position and velocity of the first detection (Figure~\ref{fig.tracklet_cartoon}). Thus, the number of randomly linked detections in a large circle decreased dramatically. This increased the number of good NEO tracklets by 20\% and decreased the number of MIXED tracklets by a factor of 5. 

\item The next step added false detections from random noise to the full-density NEO and MBA detection list (Case E). This doubled the number of detections to 30 million, and so the synthetic to false detection ratio was about 1:1. However, the number of tracklets only increased from 6 million in case C to 7.5 million in case E. In this scenario tracklets were created up to the $2\degree$/day limit without the use of velocity information. In addition to 1 million MIXED tracklets, the simulation generated about 700,000 BAD tracklets (i.e., those with both synthetic and false detections) and 600,000 NONSYNTH tracklets consisting solely of false detections. 

\item The final, full-density simulation was achieved by also injecting differencing artifacts, which more than doubled again the total number of detections, to 66 million (Case F). Now, over 77\% of detections were false, and so the ratio between synthetic and false detections was about 1:3.5. NEOs represent only 0.07\% of the detection list. The full-density simulation was challenging for the tracklet stage. Therefore, we used trail-derived velocity information for tracklets created in the velocity range of 1.2--$2.0\degree$/day. Still, the total number of tracklets was very large, $\sim11.9$ million. Out of this sample, about 57\% of tracklets were somehow erroneous, either including at least one false detection or detections of different objects. This simulation revealed that artifacts related to false positives create the majority of the linking challenge. Though we did not directly test it, the use of trail-related velocity information presumably leads to a dramatic reduction in the false tracklet rate for the full-density simulation.
\end{enumerate}

\begin{sidewaystable}[ht!]
\small
\setlength\tabcolsep{1.5pt}
\caption{Number of detections and tracklets for OC 28 in various simulations. MIXED tracklets include detections from at least two distinct objects, BAD tracklets include detections from both false detections and moving objects, and NONSYNTH tracklets consist entirely of false detections.}
\begin{center}
\begin{tabular}{l|lcc|cccc|cccccc}
\tableline \tableline
     &     &     &            &\multicolumn{4}{c} {Detections} & \multicolumn{6}{c} {Tracklets}\\
Case & NEO & MBA & False Det  &Total &\%NEO & \%MBA & \%False   & Total &\%NEO & \%MBA & \%MIXED & \%BAD & \%NONSYNTH \\
\hline
A  &Bottke   &No  &None          &36k  &100 &0    &0    &11k  &100  &0    &0    &0   &0\\
B  &Granvik  &No  &None          &49k  &100 &0    &0    &17k  &100  &0    &0    &0   &0\\
C  &Granvik  &Yes &None          &15M  &0.3 &99.7 &0    &6.2M &0.23 &82.8 &16.9 &0   &0\\
D
\footnote{\label{1sttablefoot}Tracklet generation used rate information from 1.2--2.0$\degree$/day. Otherwise rate information was ignored over entire range 0.05--2.0$\degree$/day.}
   &Granvik  &Yes &None          &15M  &0.3 &99.7 &0    &5.4M &0.31 &94.8 &4.9  &0   &0\\
E  &Granvik  &Yes &Random only   &30M  &0.2 &50.6 &49.2 &7.5M &0.19 &68.2 &14   &9.7 &7.9\\
F
\footref{1sttablefoot}
   & Granvik &Yes &Random + artifacts &66M  &0.1 &22.6 &77.3 &12M  &0.14 &42.7 &2.2  &6.1 &48.8\\
\hline
\end{tabular}
\end{center}
\label{Tab.confusion_FF}
\end{sidewaystable}

\subsubsection{The Linking Process}
 
\begin{figure}[t]
  \centering
    \includegraphics[width=0.35\textwidth]{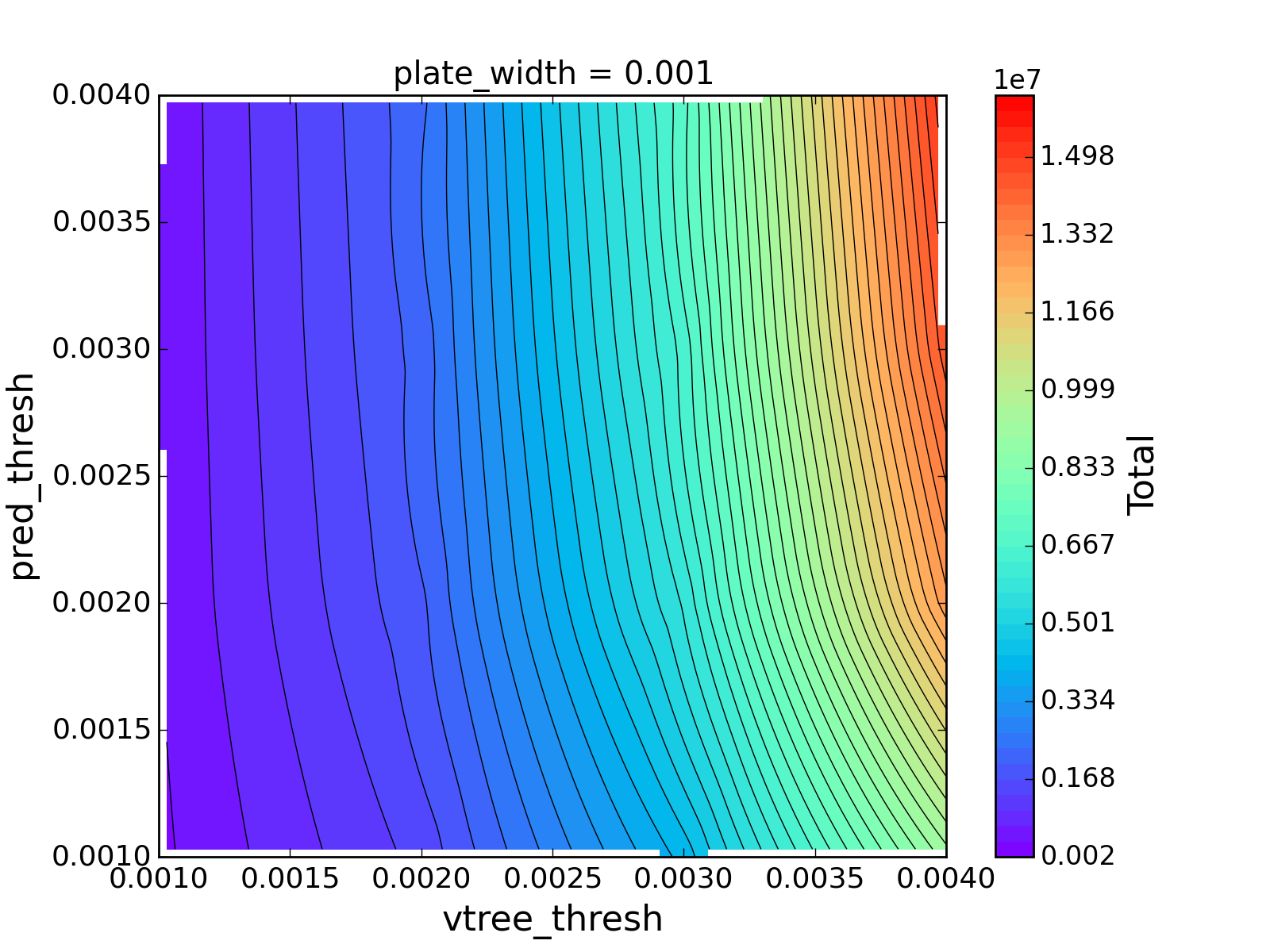}
    \includegraphics[width=0.35\textwidth]{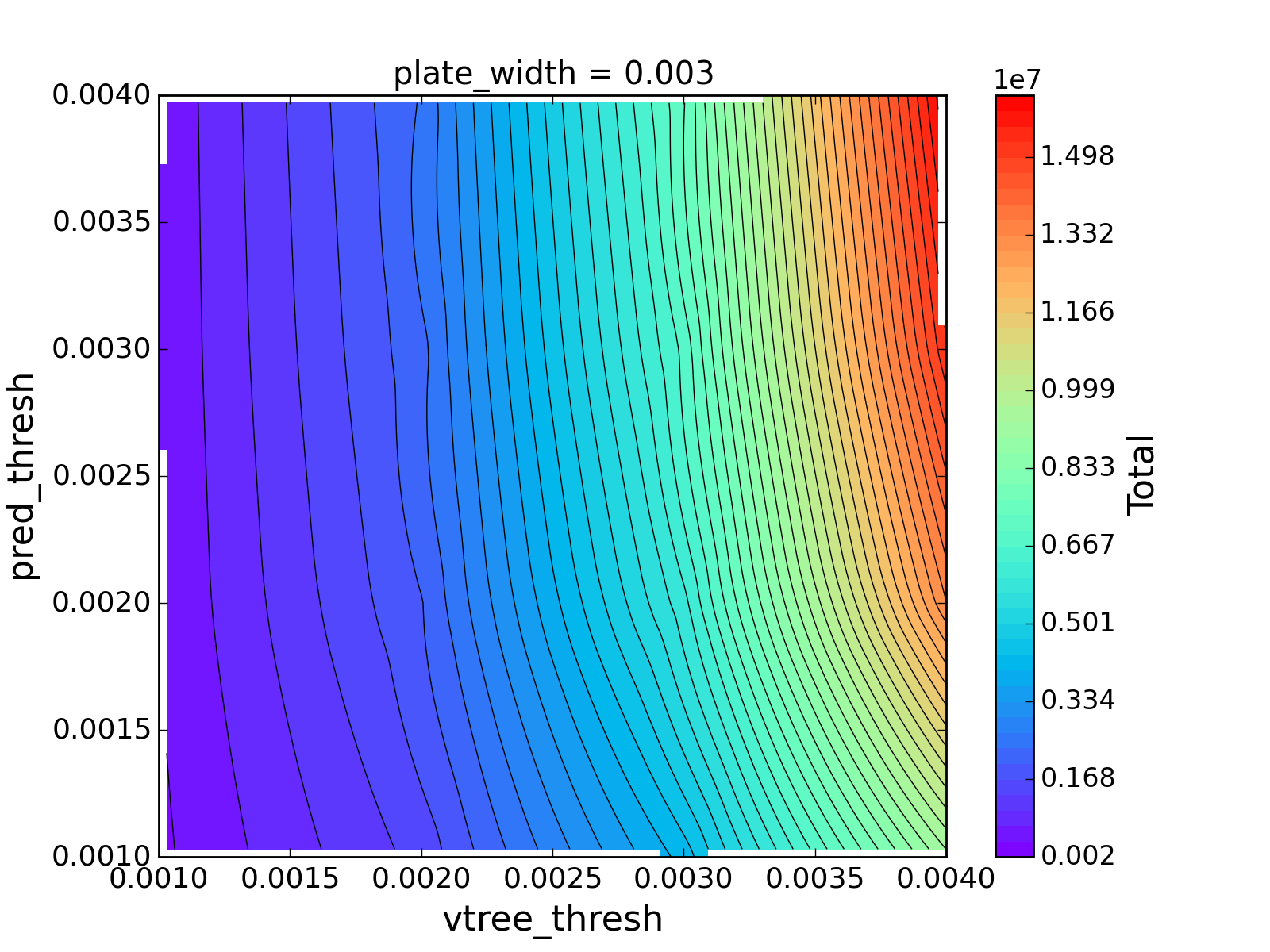}
    \includegraphics[width=0.35\textwidth]{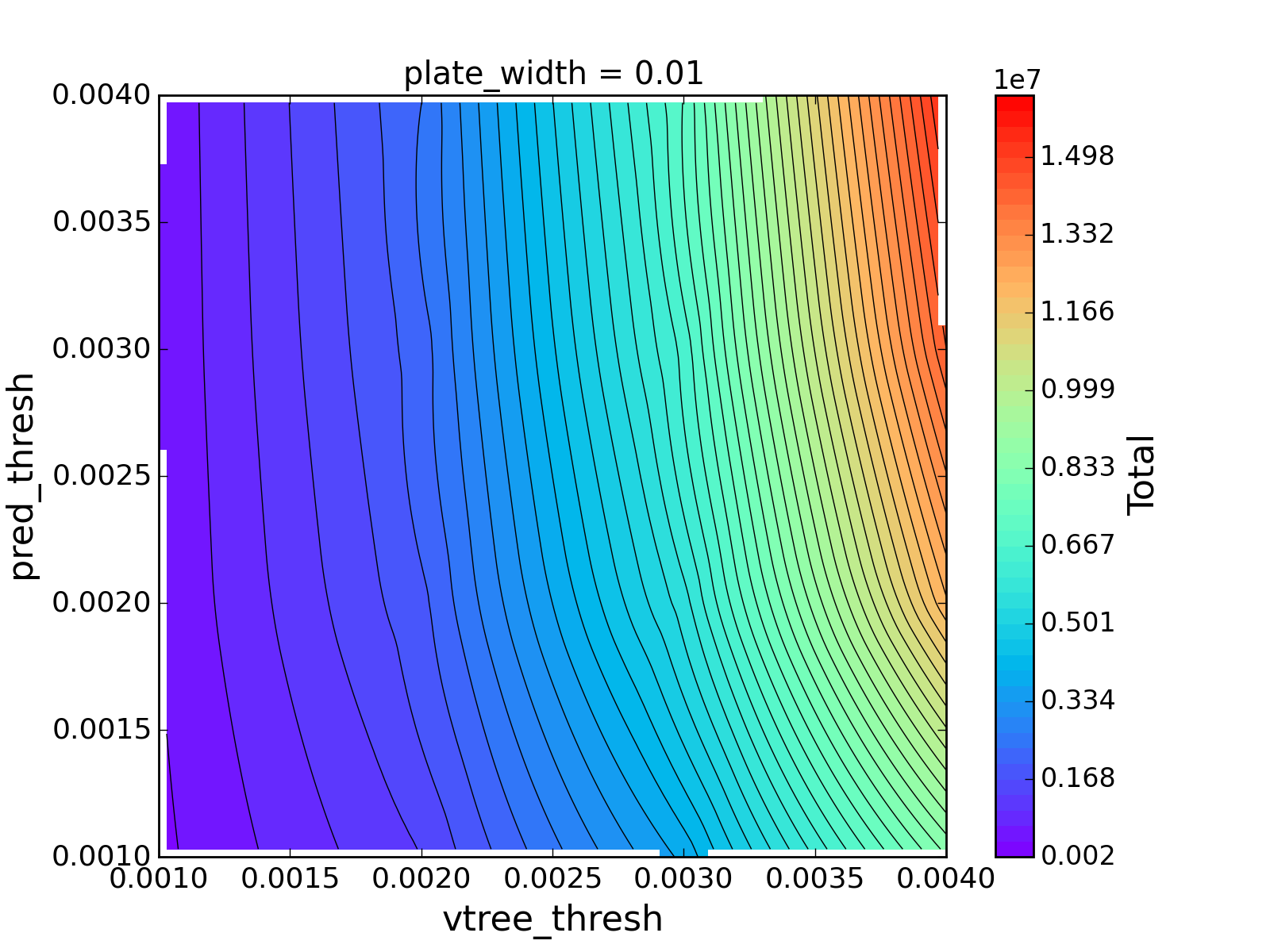}
    \includegraphics[width=0.35\textwidth]{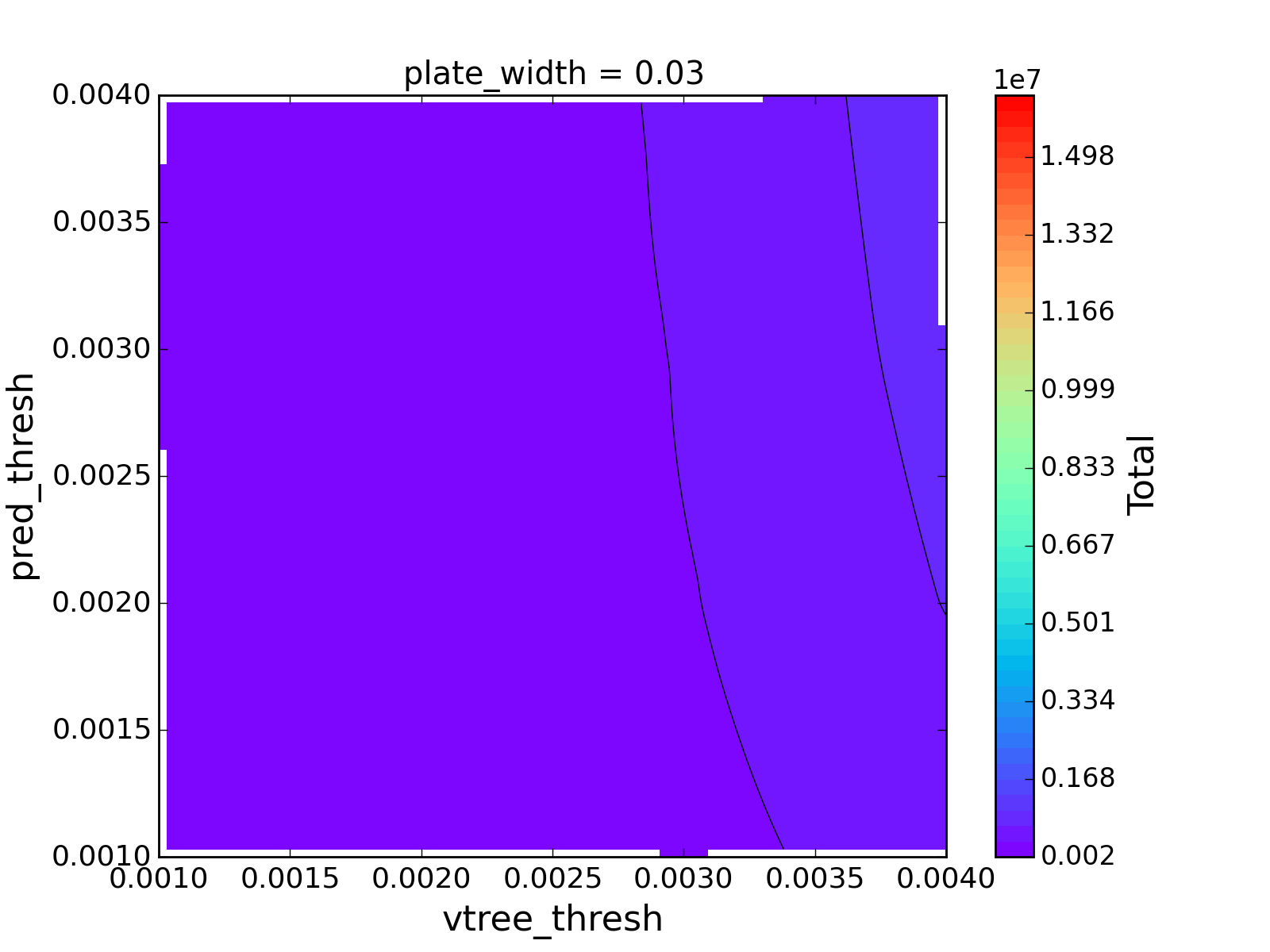}
     \caption{Total number of candidate tracks derived for a single, dense field as a function of the \vtree, \pred and \platew kd-tree linking parameters.}
    \label{fig.KD_total}
\end{figure}

\begin{figure}[tbh]
  \centering
    \includegraphics[width=0.35\textwidth]{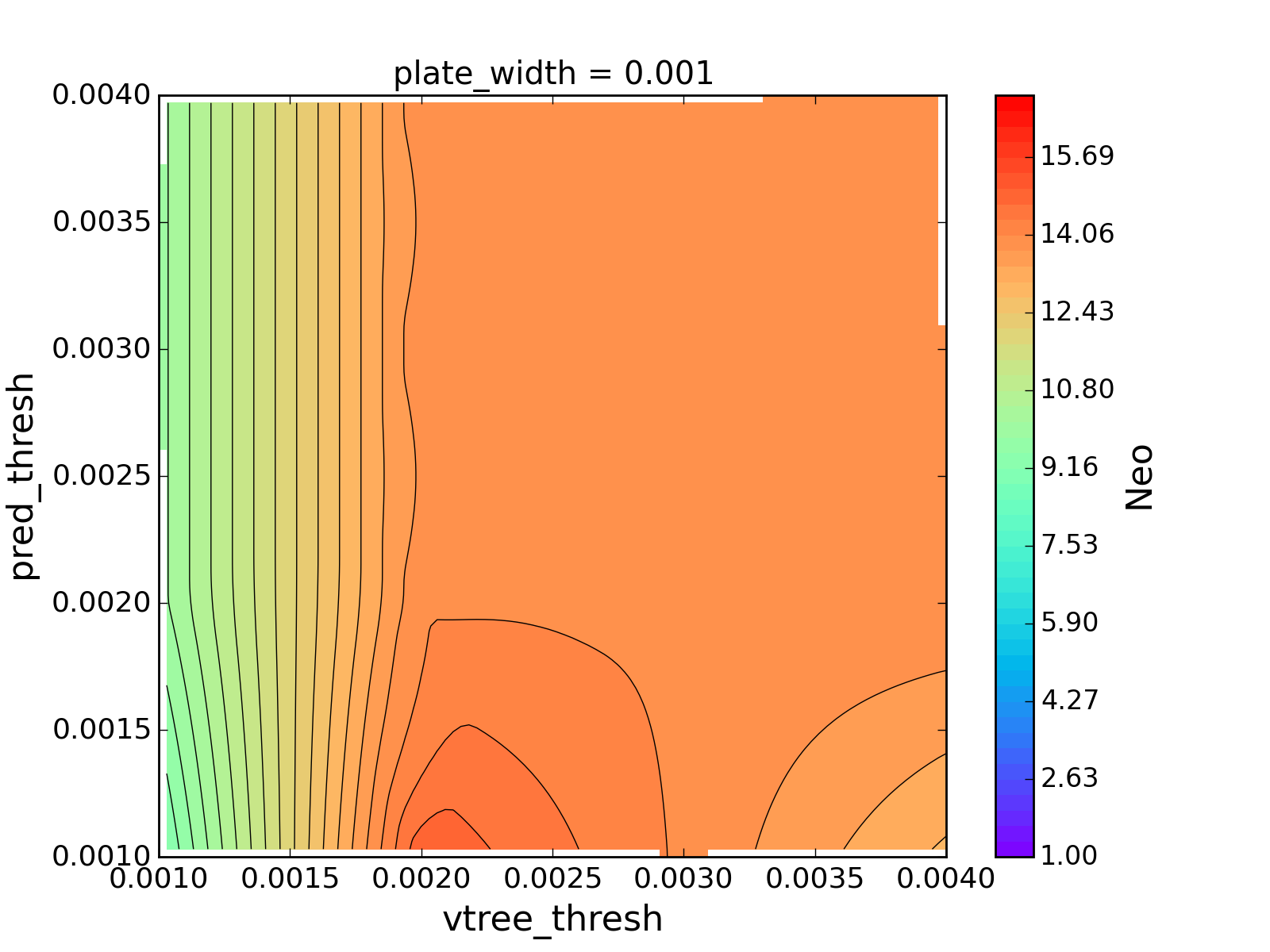}
    \includegraphics[width=0.35\textwidth]{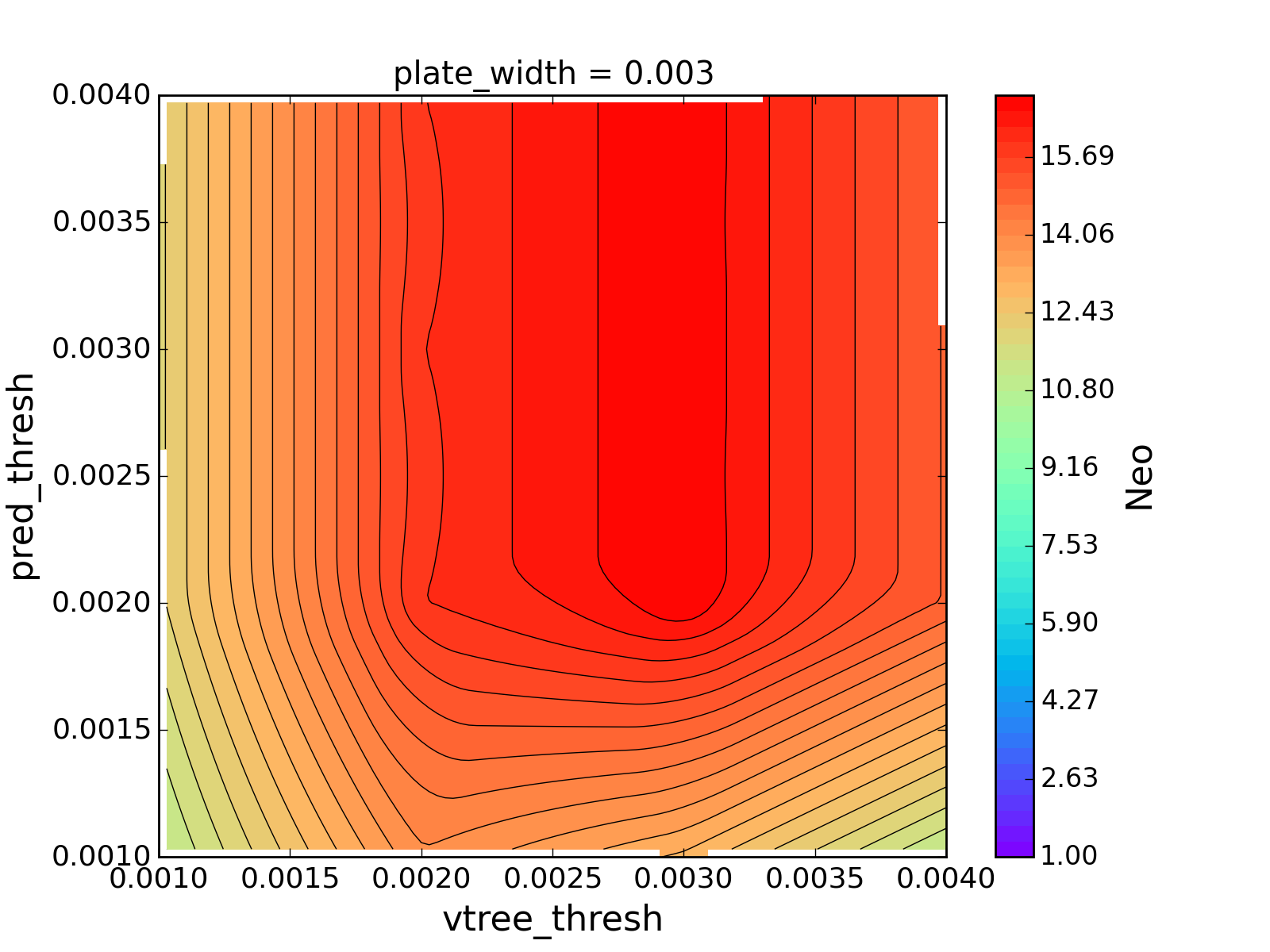}
    \includegraphics[width=0.35\textwidth]{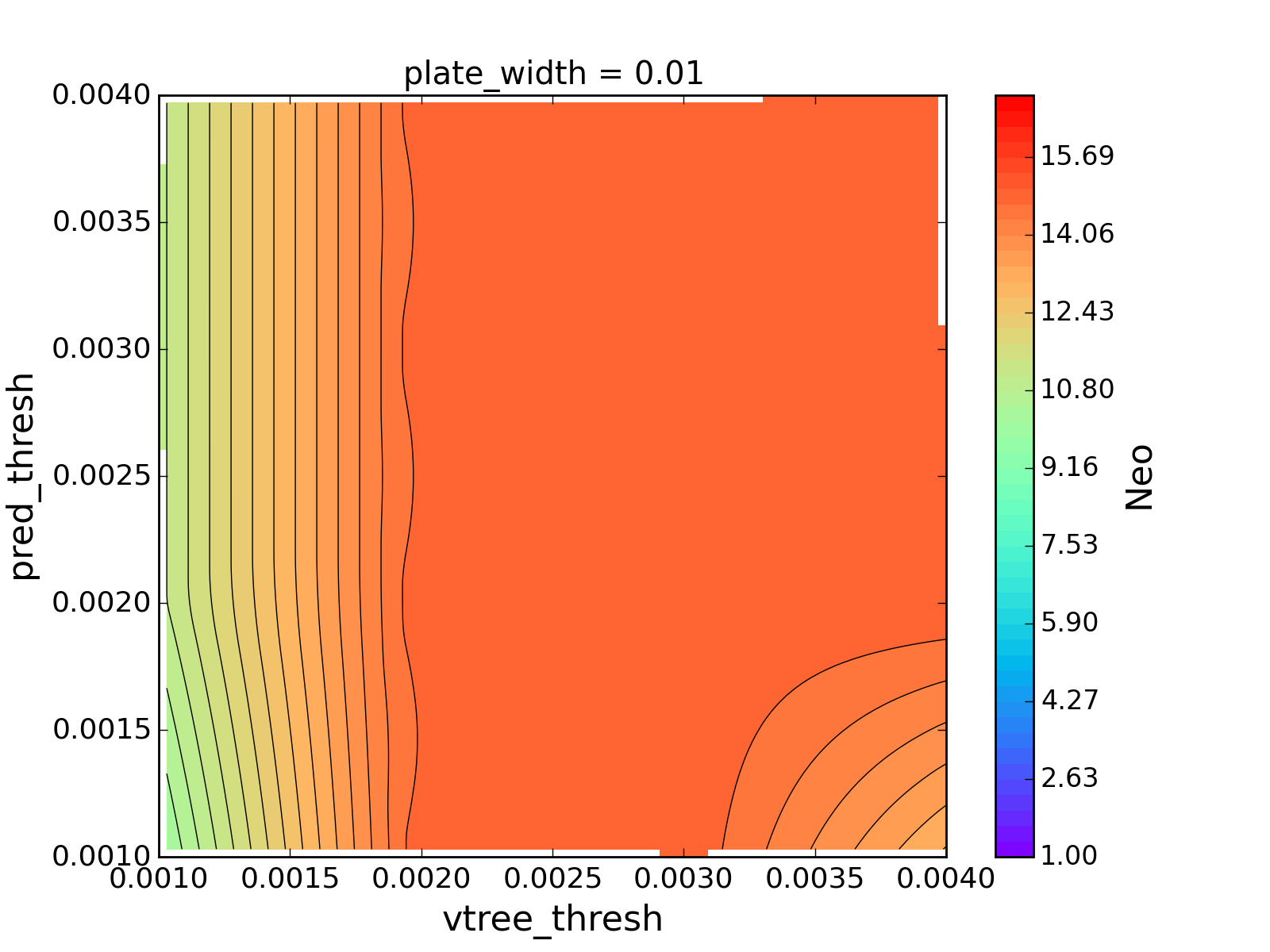}
    \includegraphics[width=0.35\textwidth]{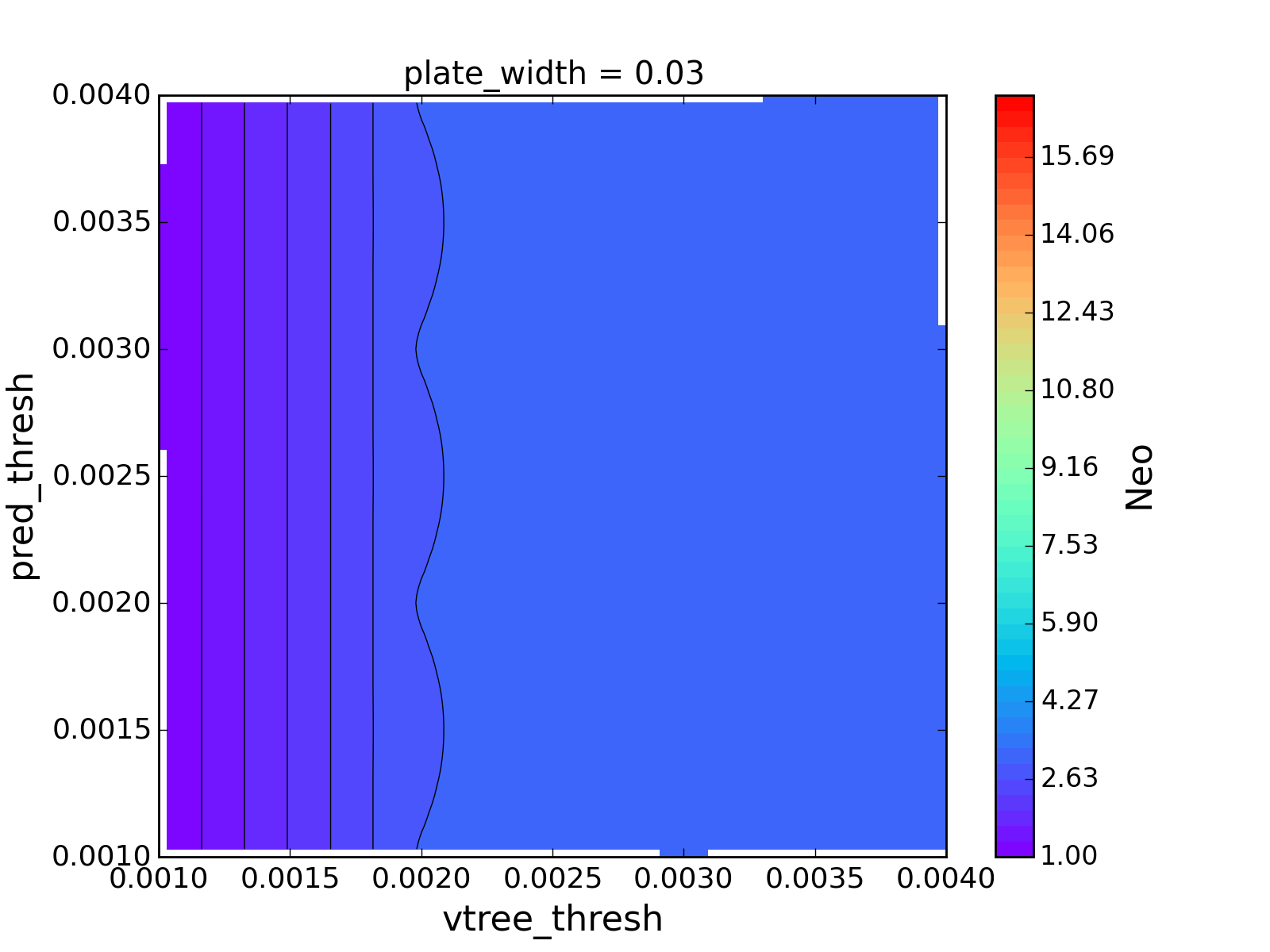}
     \caption{Total number of CLEAN NEO  tracks derived for a single, dense field as a function of the \vtree, \pred and \platew kd-tree linking parameters.}
    \label{fig.KD_neo}
\end{figure}

\begin{figure}[tbh]
  \centering
    \includegraphics[width=0.35\textwidth]{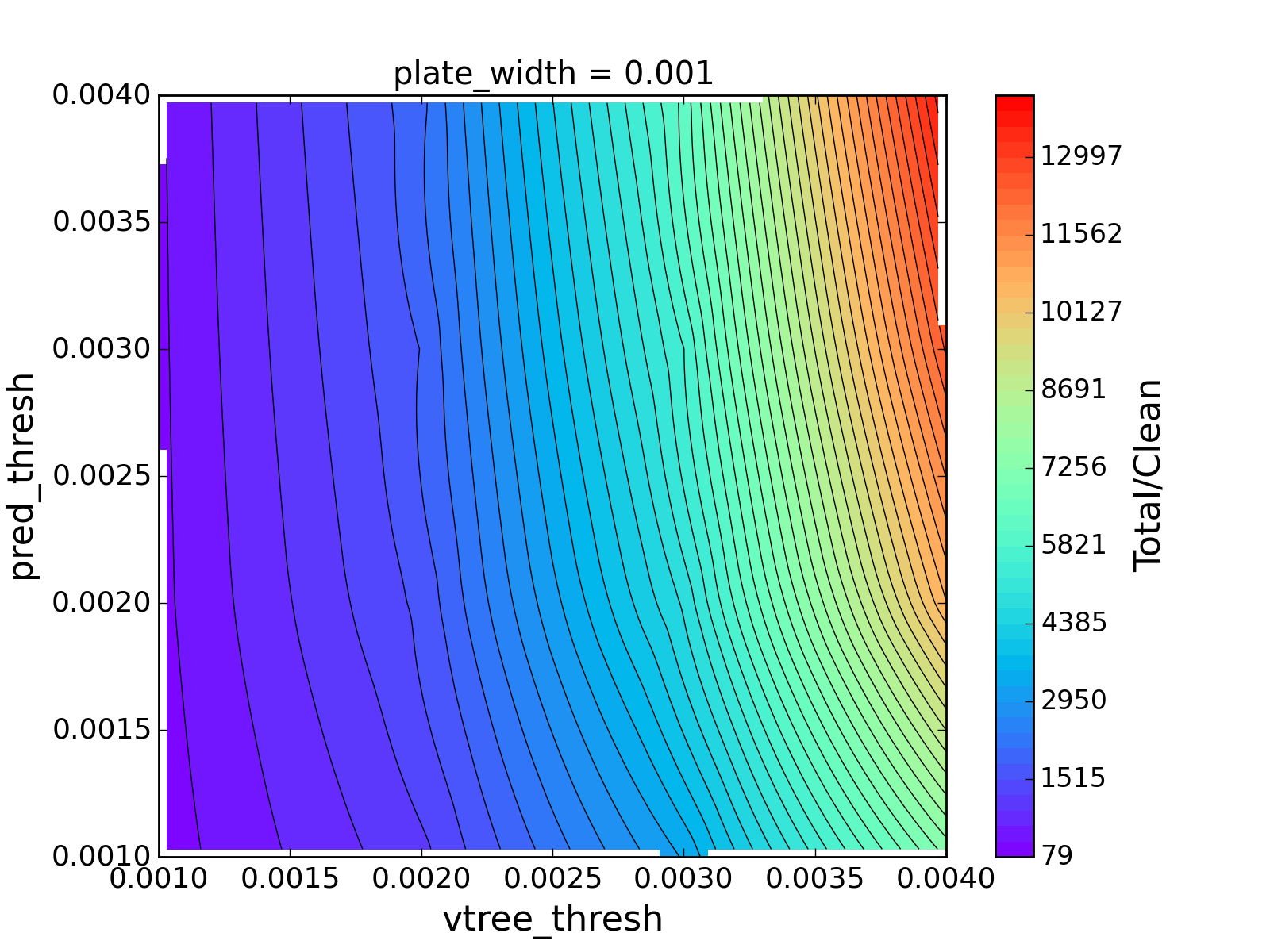}
    \includegraphics[width=0.35\textwidth]{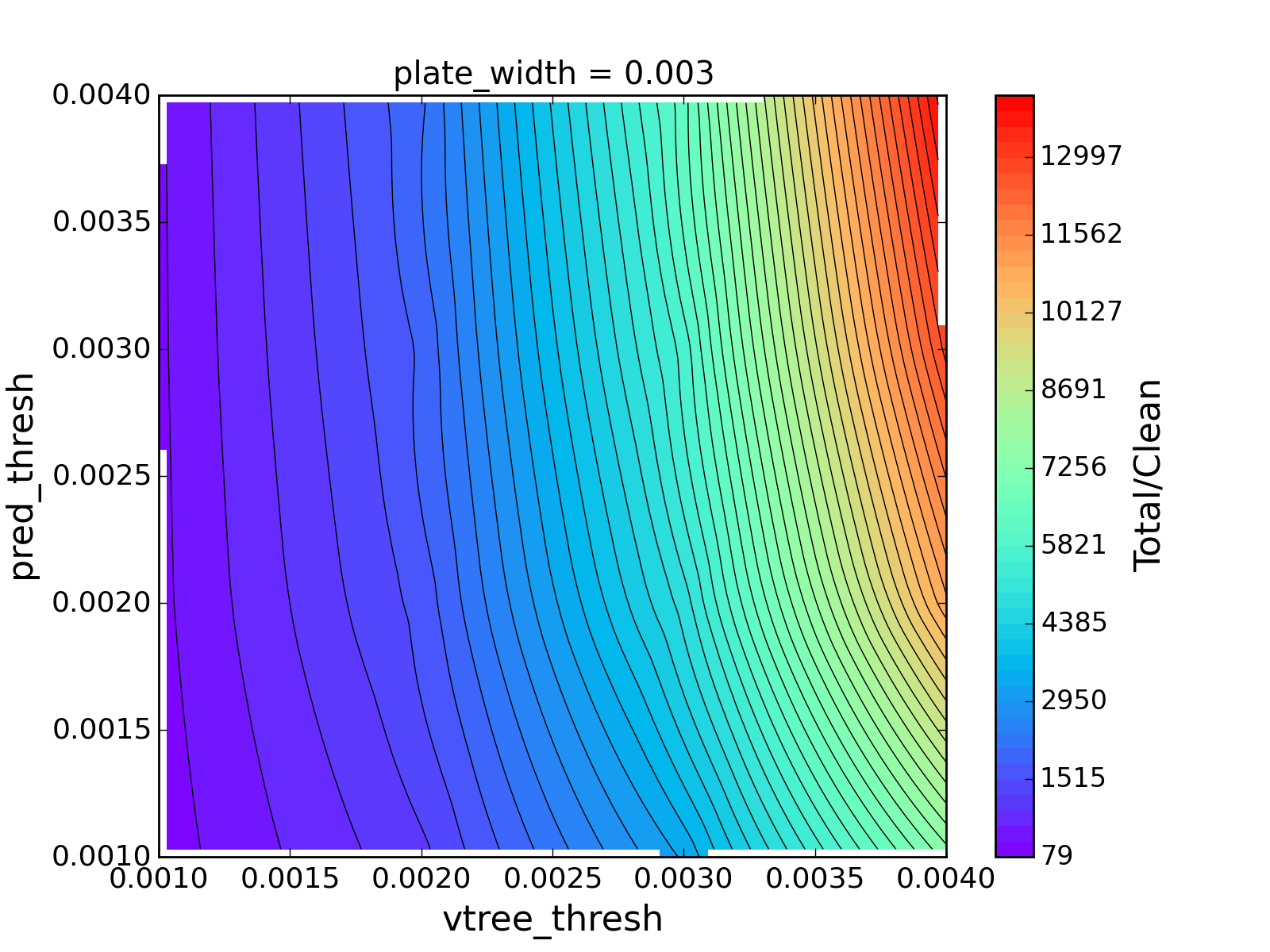}
    \includegraphics[width=0.35\textwidth]{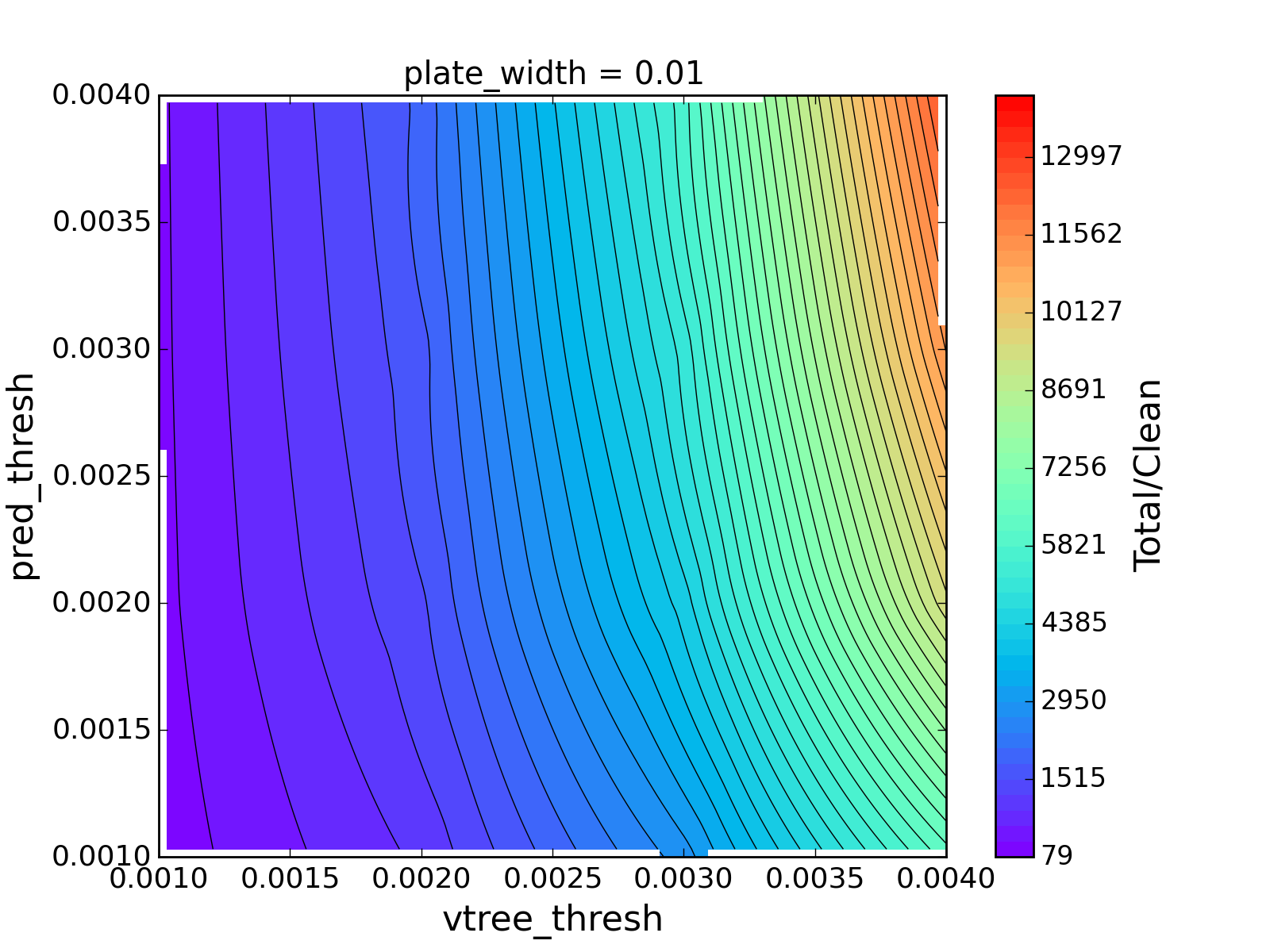}
    \includegraphics[width=0.35\textwidth]{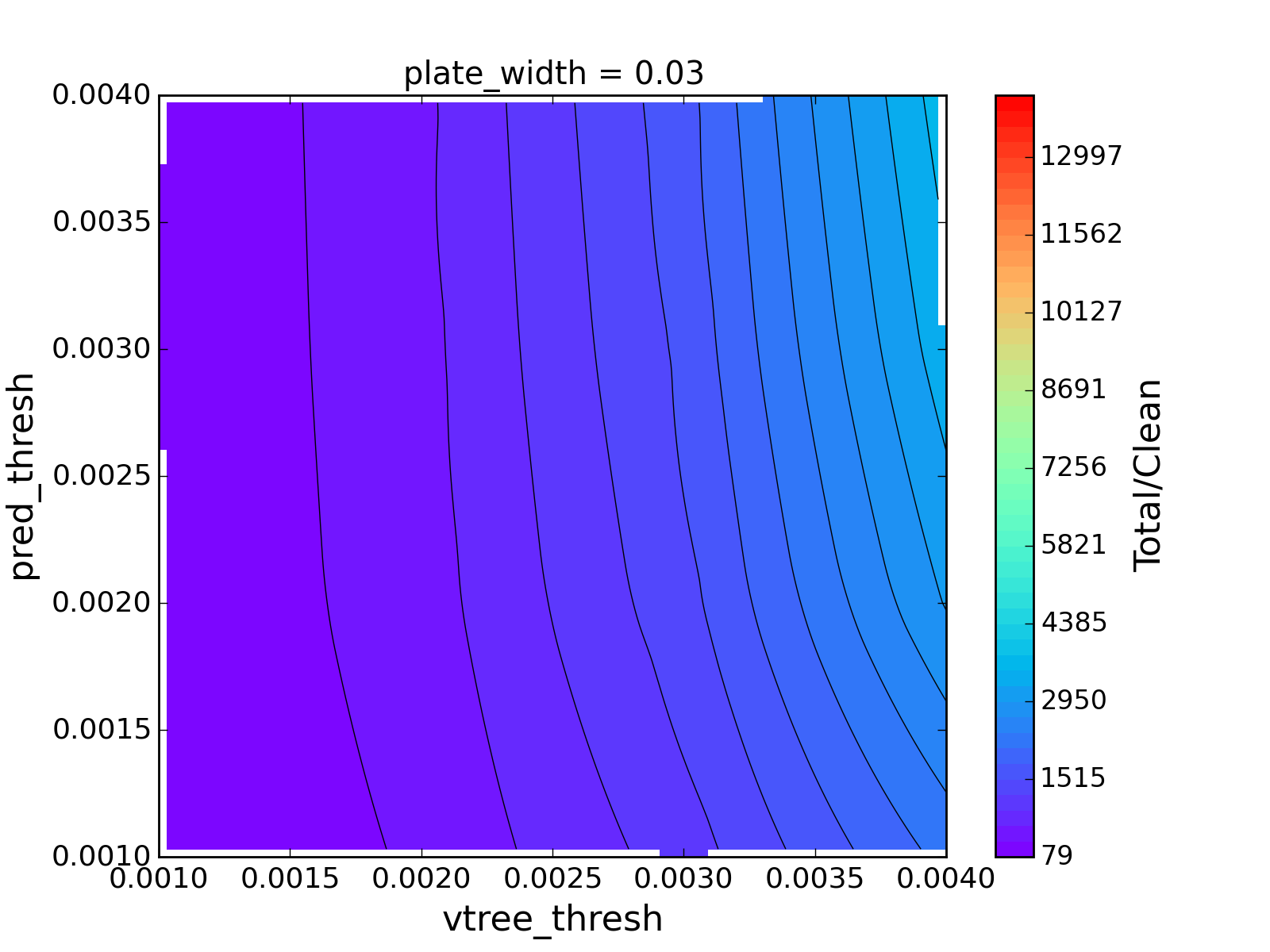}
     \caption{Ratio total number of tracks to number of CLEAN tracks derived for a single, dense field as a function of the \vtree, \pred and \platew kd-tree linking parameters.}
    \label{fig.KD_ratio1}
\end{figure}

\begin{figure}[tbh]
  \centering
    \includegraphics[width=0.35\textwidth]{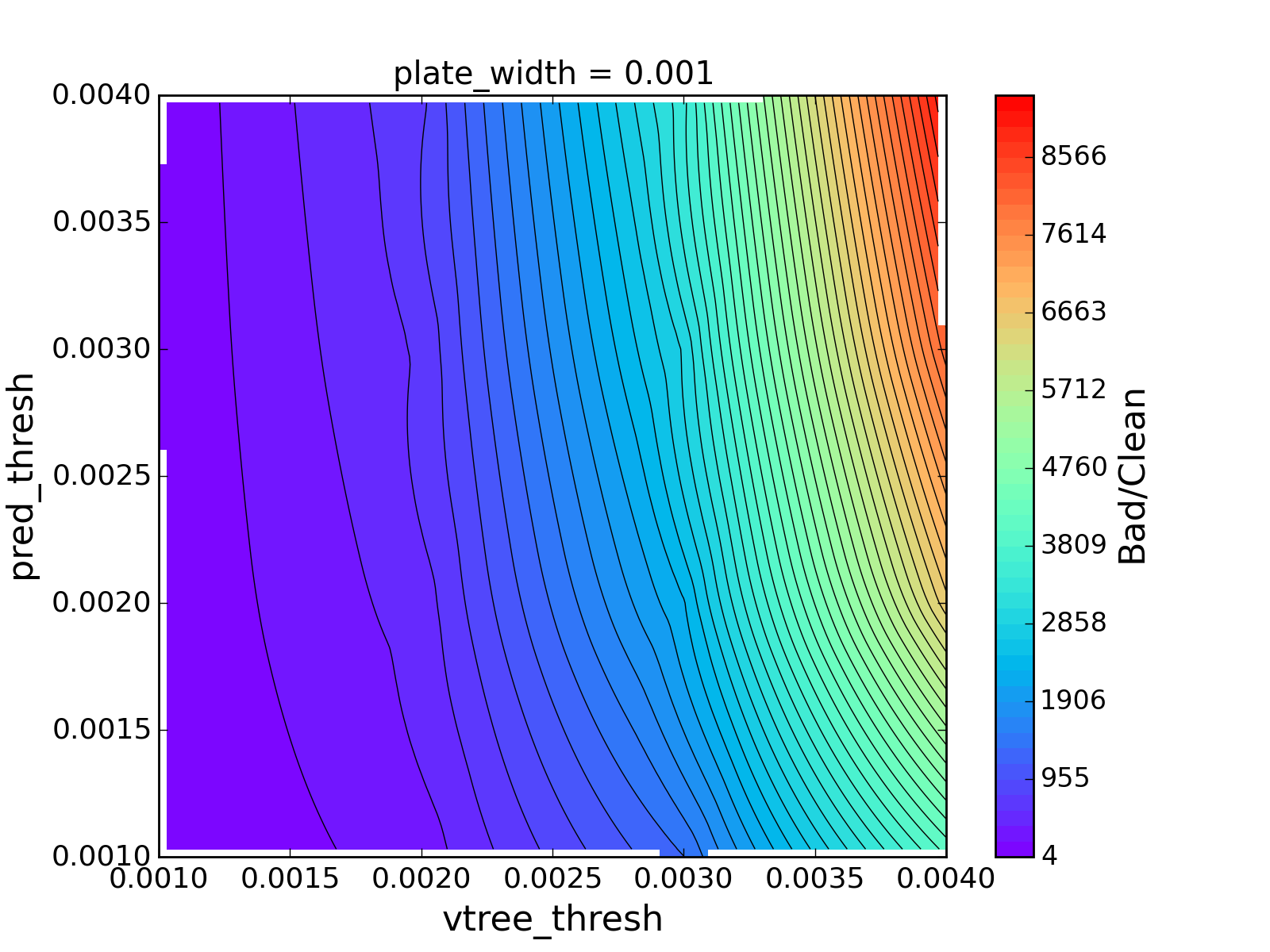}
    \includegraphics[width=0.35\textwidth]{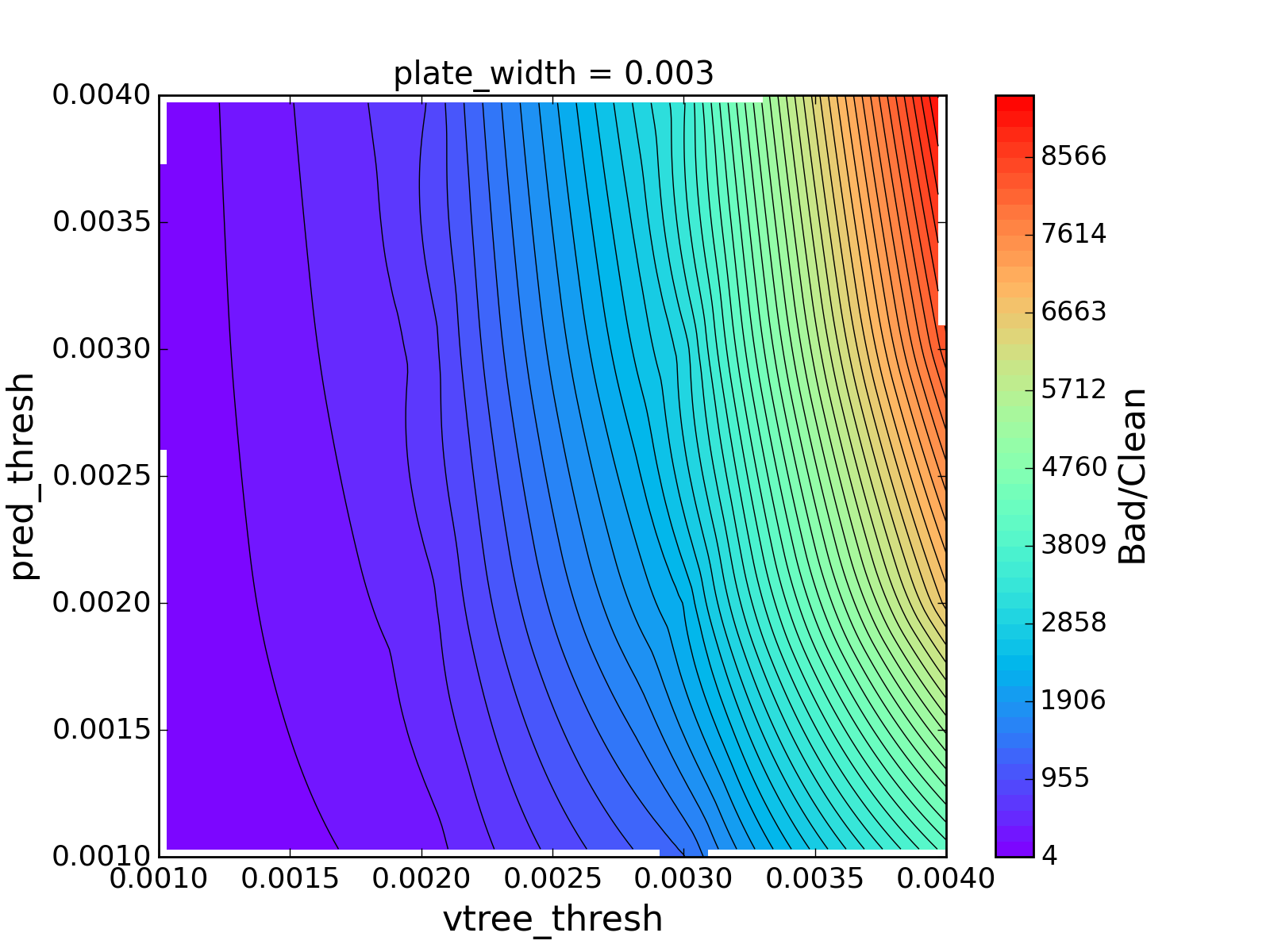}
    \includegraphics[width=0.35\textwidth]{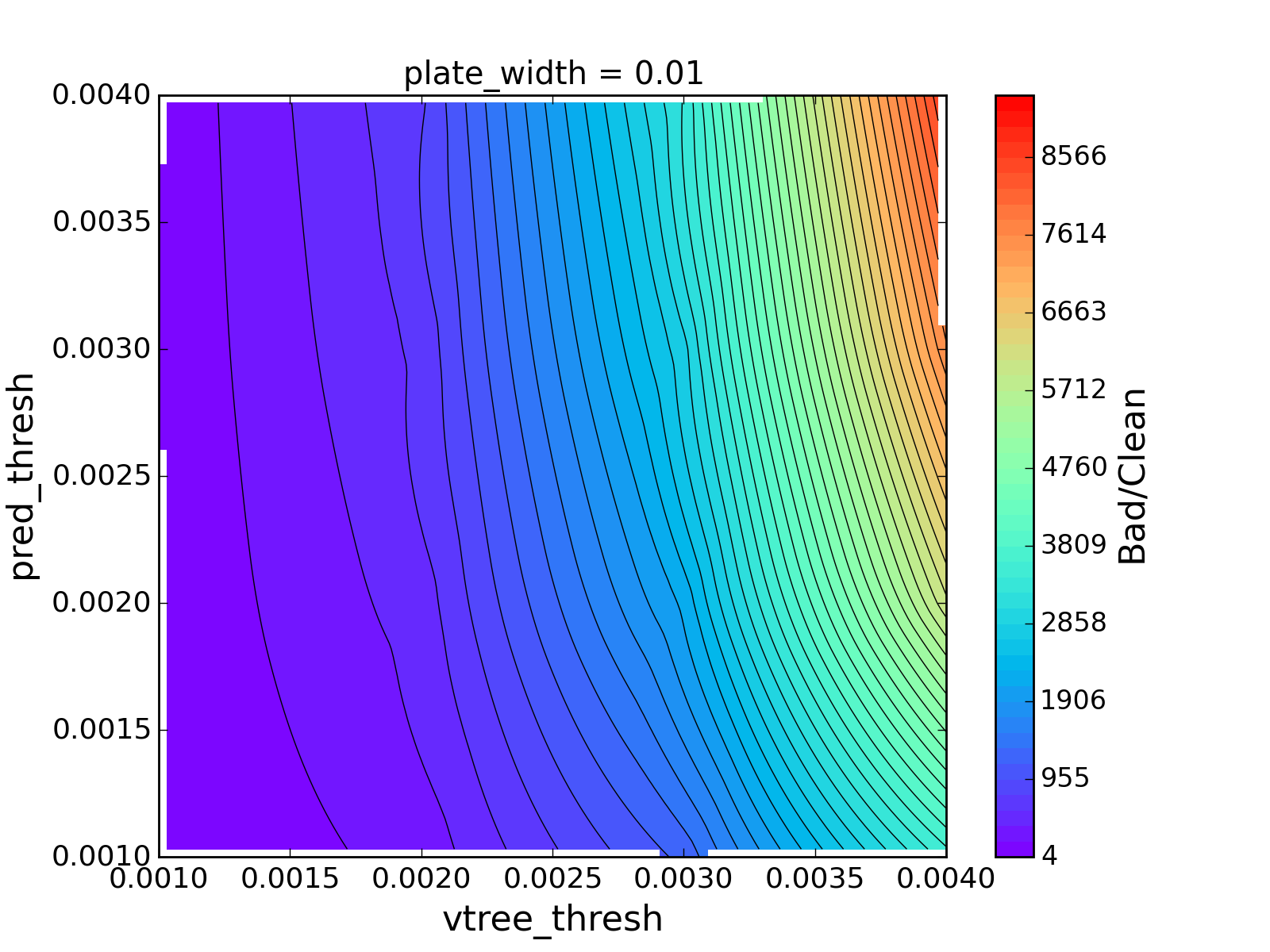}
    \includegraphics[width=0.35\textwidth]{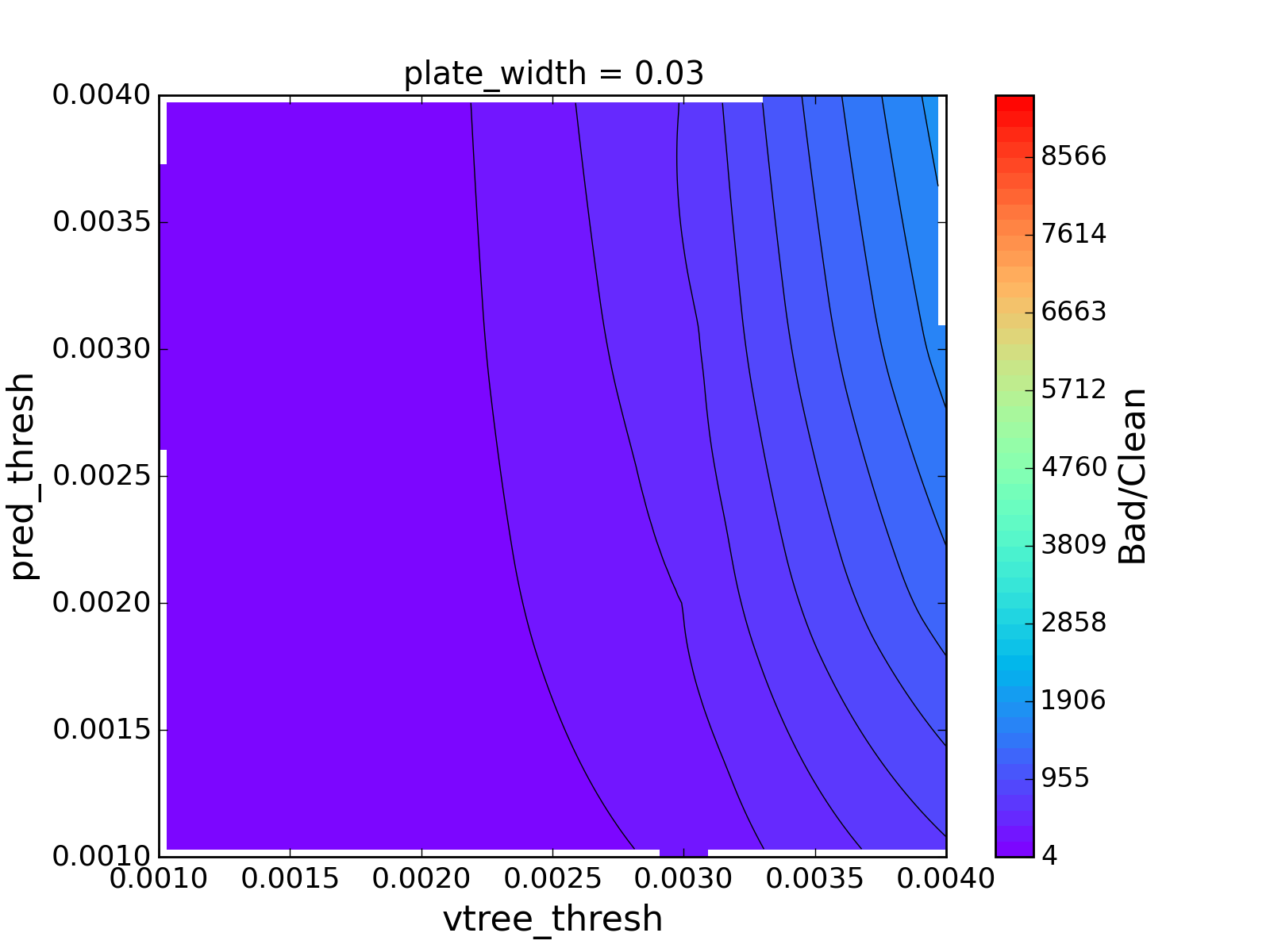}
     \caption{Ratio number of BAD tracks to number of CLEAN tracks derived for a single, dense field as a function of the \vtree, \pred and \platew kd-tree linking parameters.}
    \label{fig.KD_ratio2}
\end{figure}

\begin{figure}[tbh]
  \centering
    \includegraphics[width=0.35\textwidth]{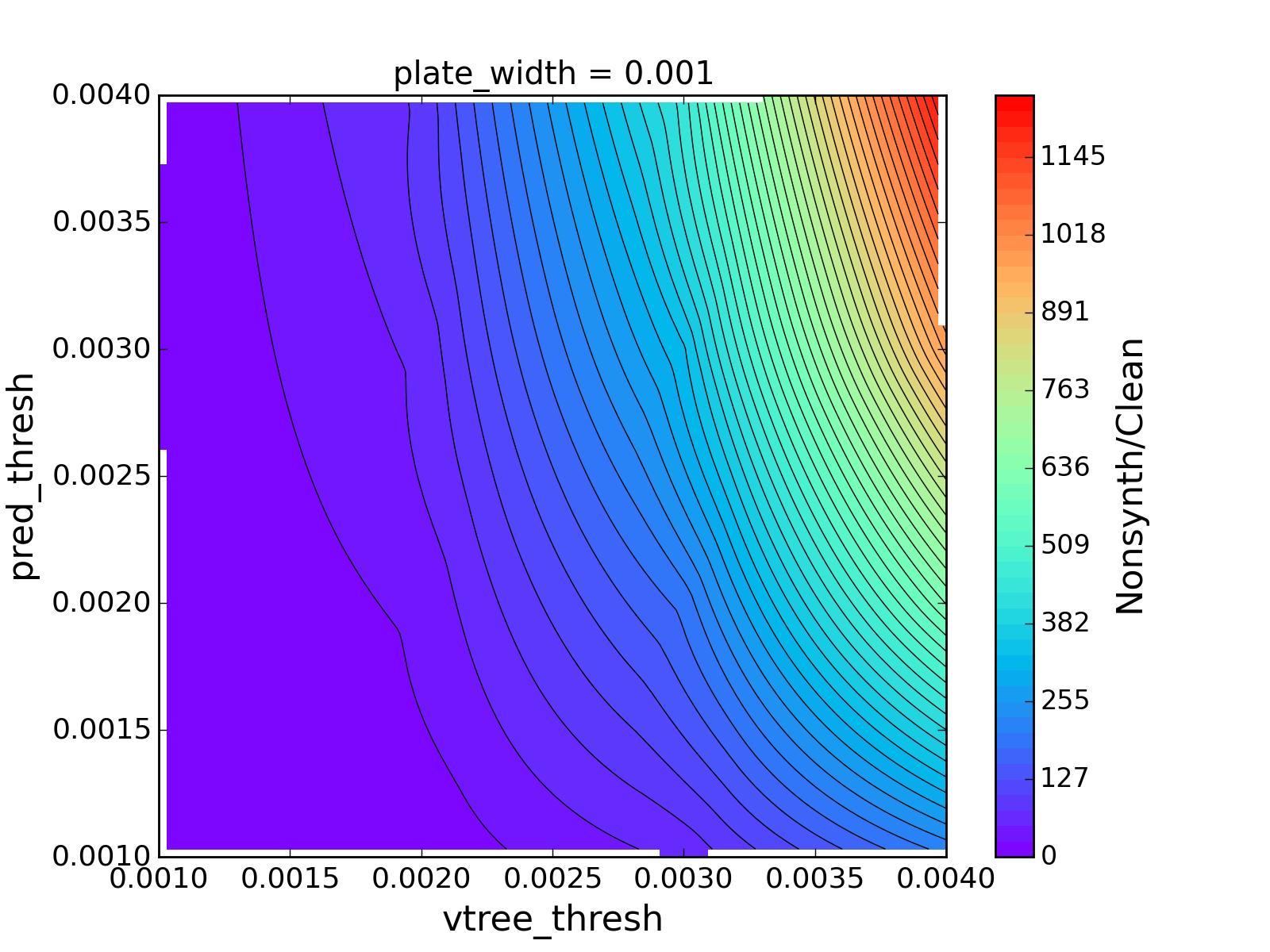}
    \includegraphics[width=0.35\textwidth]{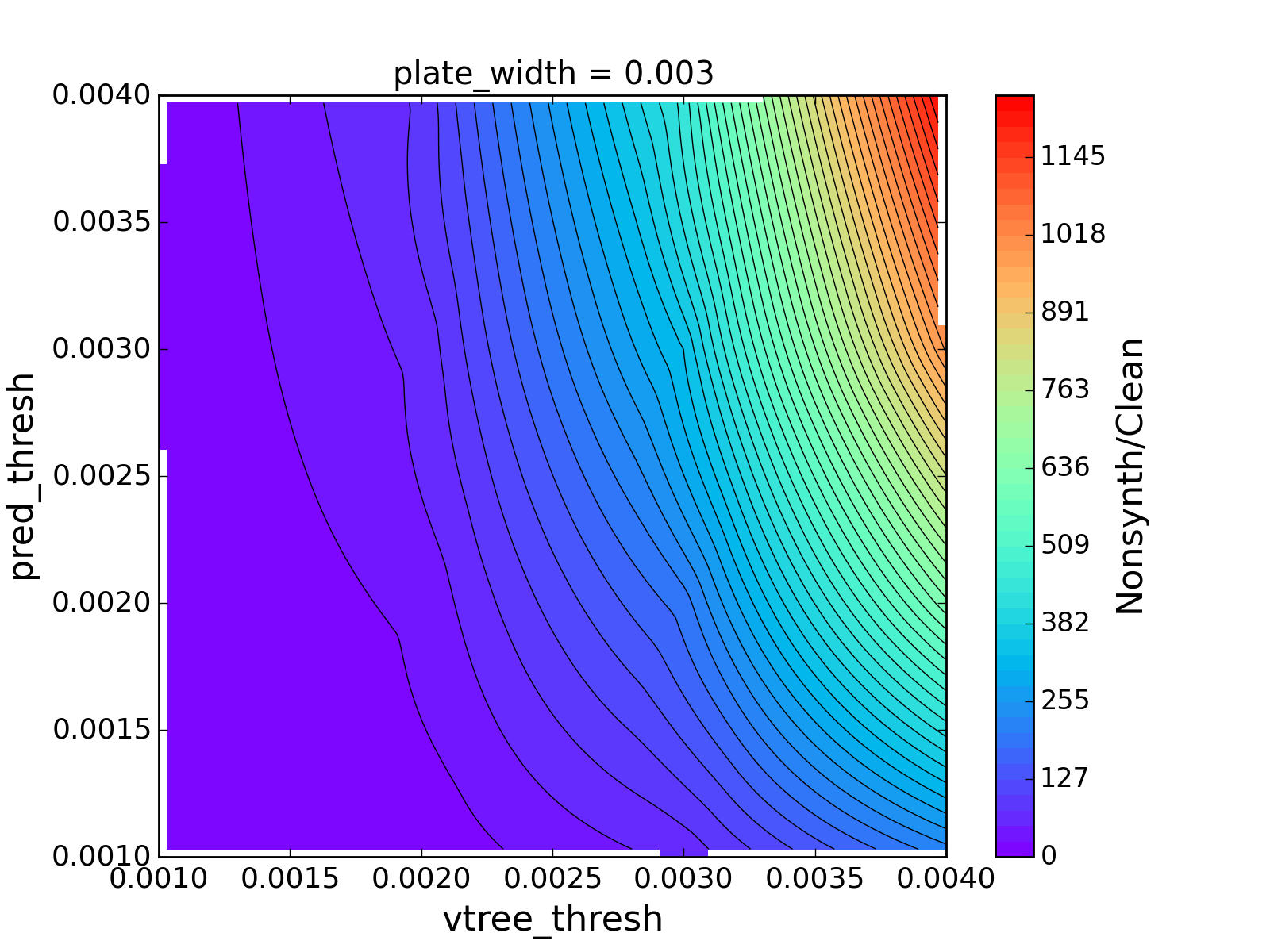}
    \includegraphics[width=0.35\textwidth]{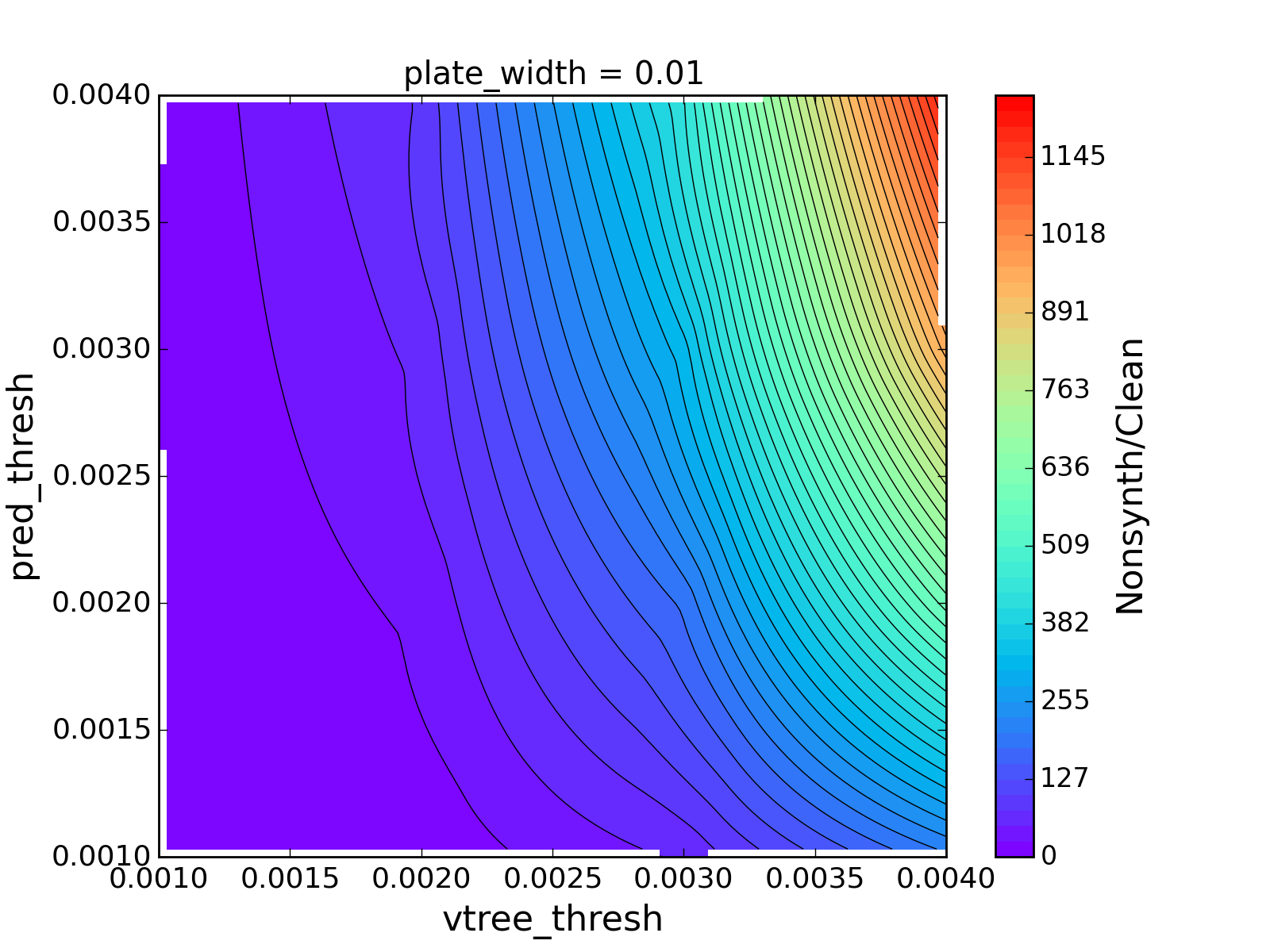}
    \includegraphics[width=0.35\textwidth]{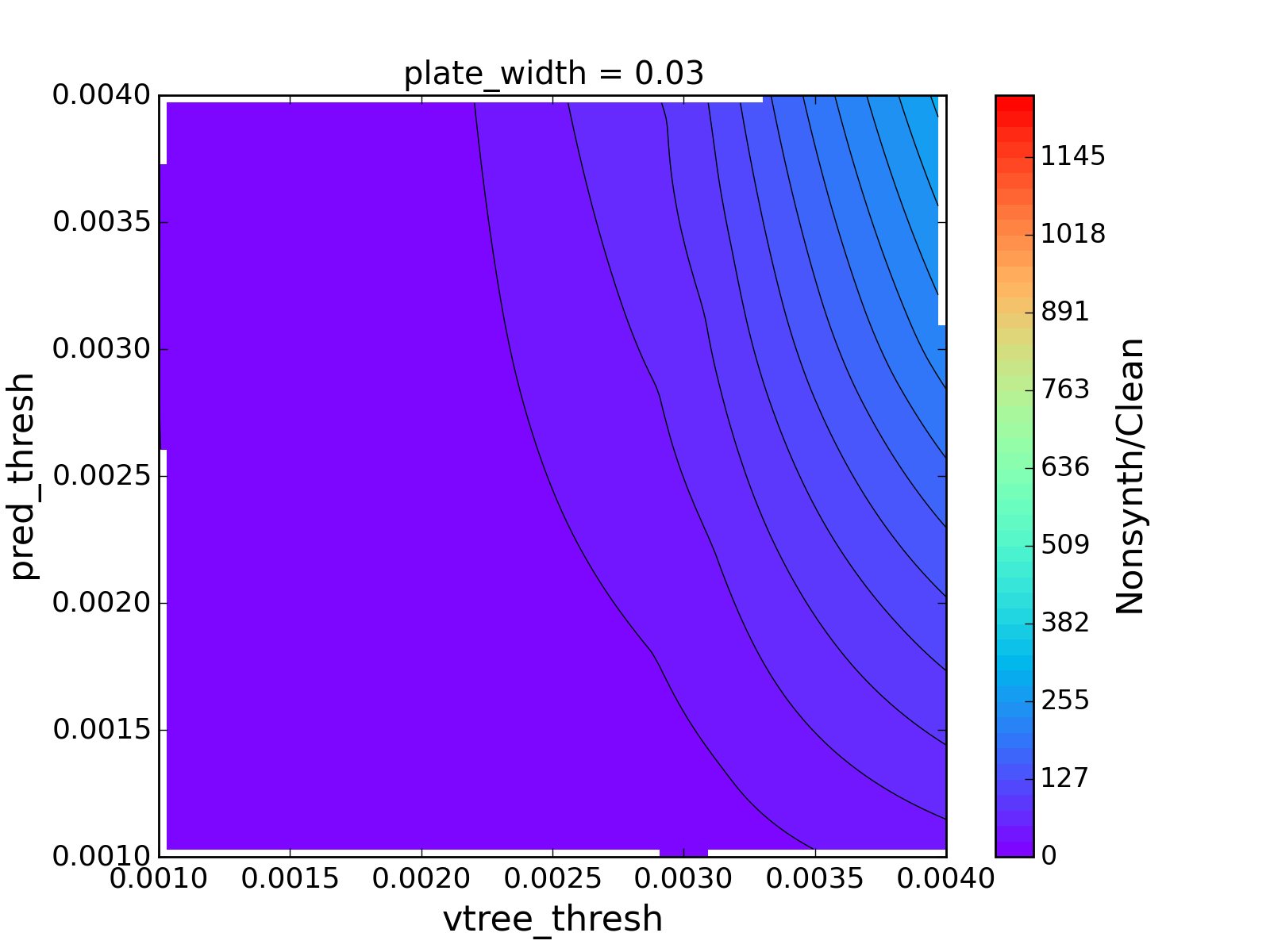}
     \caption{Ratio number of NONSYNTH tracks to number of CLEAN tracks derived for a single, dense field as a function of the \vtree, \pred and \platew kd-tree linking parameters.}
    \label{fig.KD_ratio3}
\end{figure}

\begin{figure}[tbh]
  \centering
    \includegraphics[width=0.35\textwidth]{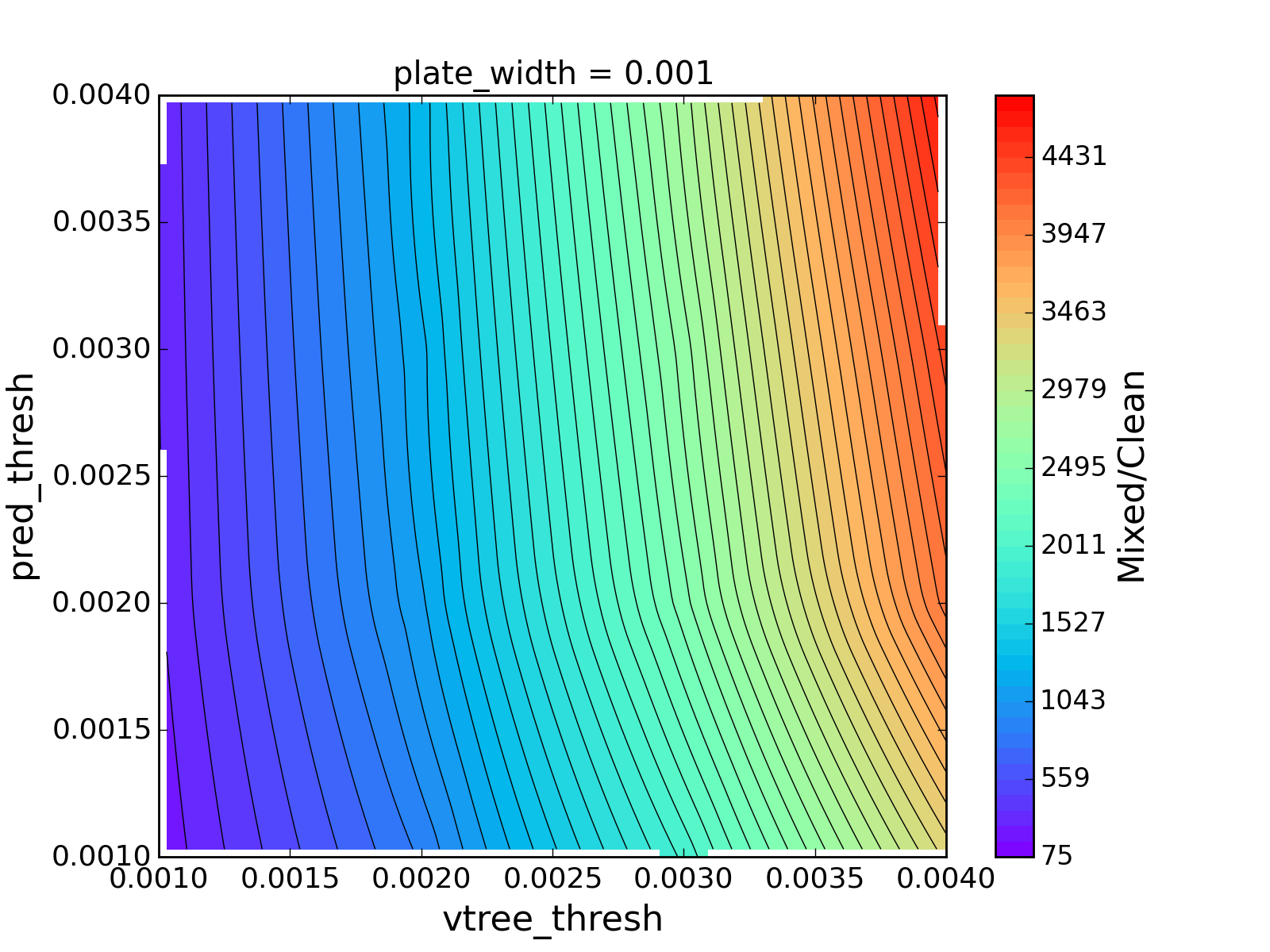}
    \includegraphics[width=0.35\textwidth]{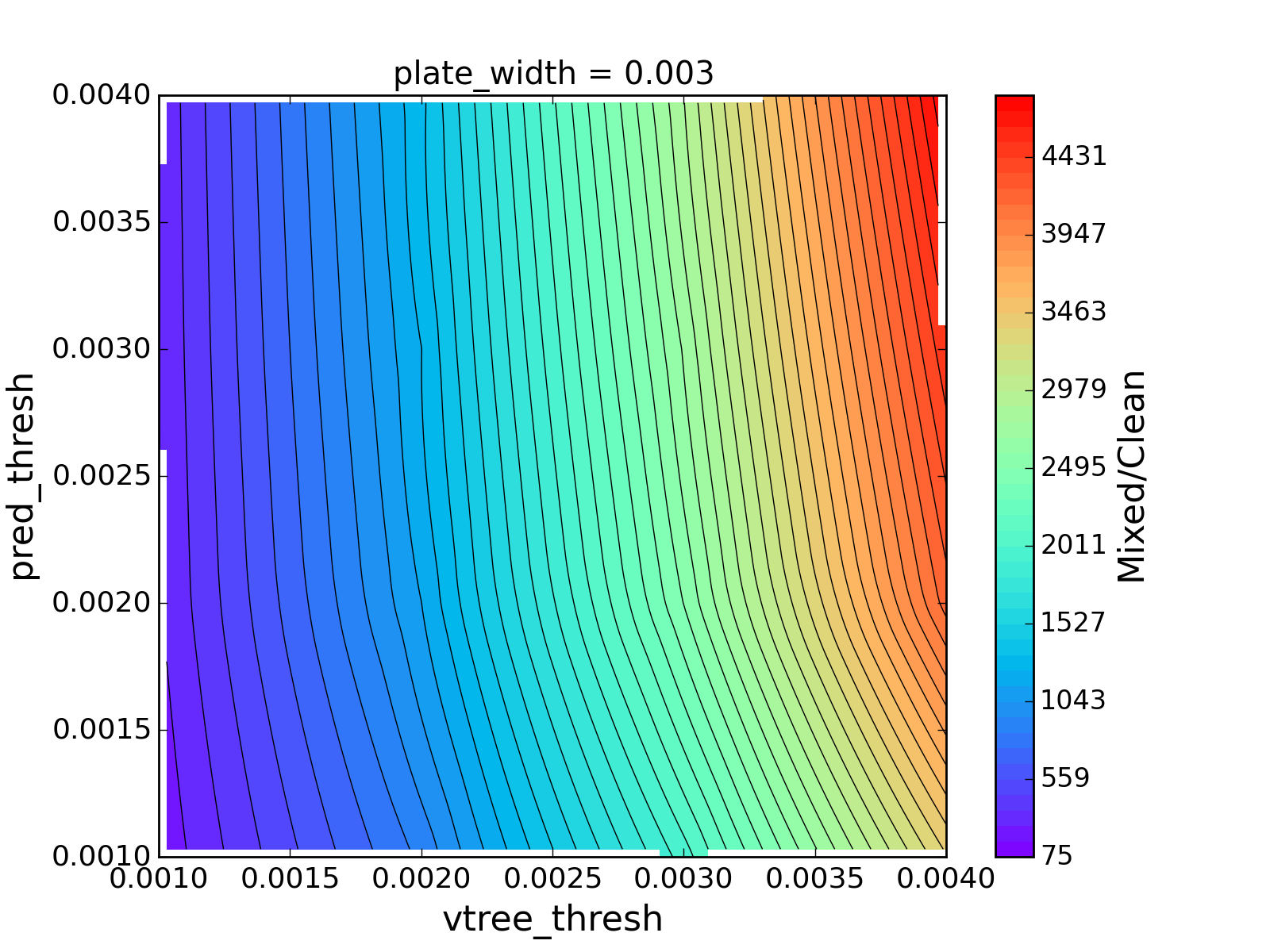}
    \includegraphics[width=0.35\textwidth]{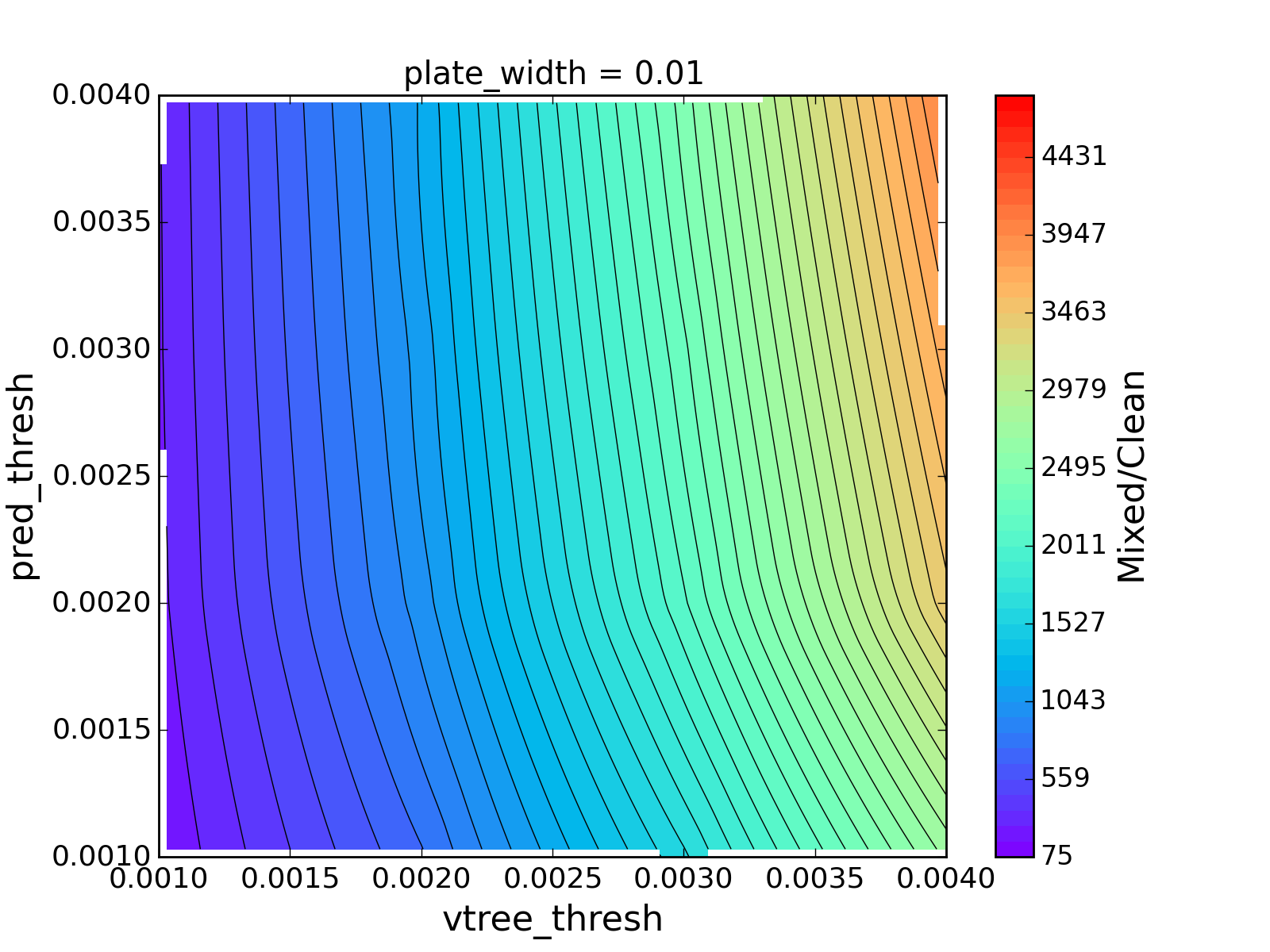}
    \includegraphics[width=0.35\textwidth]{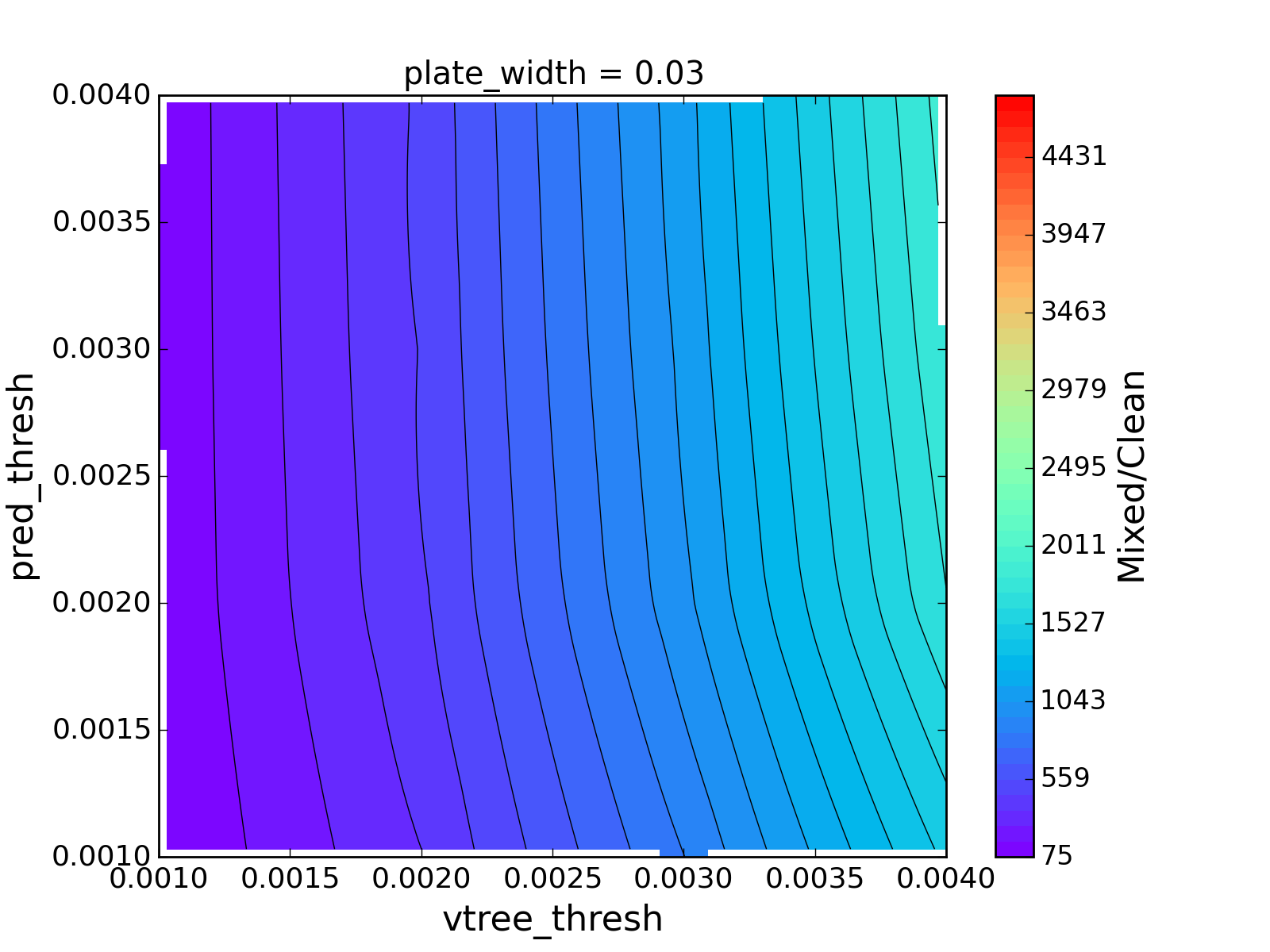}
     \caption{Ratio number of MIXED tracks to number of CLEAN tracks derived for a single, dense field as a function of the \vtree, \pred and \platew kd-tree linking parameters.}
    \label{fig.KD_ratio4}
\end{figure}

\begin{figure}[tbh]
  \centering
    \includegraphics[width=0.35\textwidth]{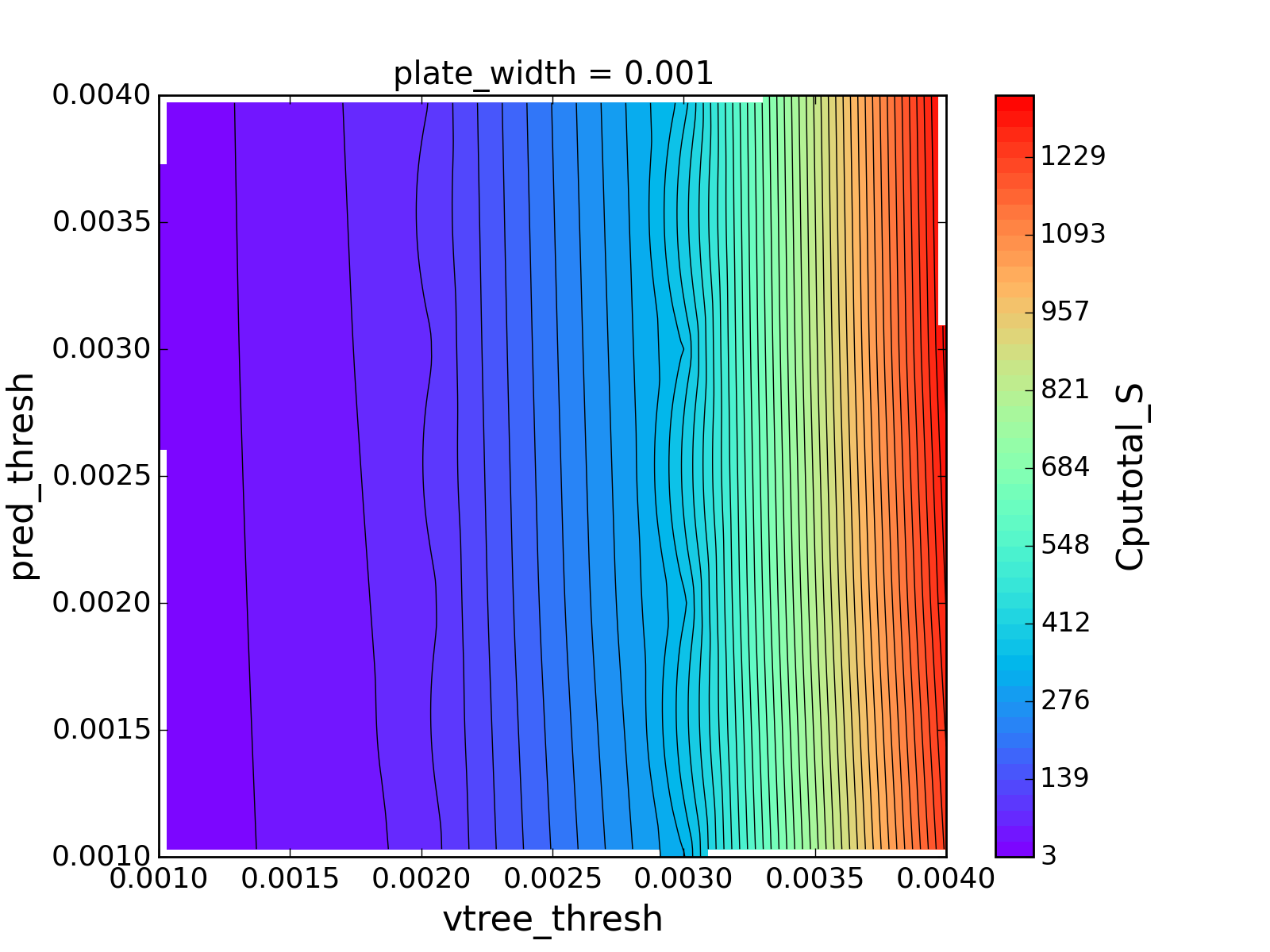}
    \includegraphics[width=0.35\textwidth]{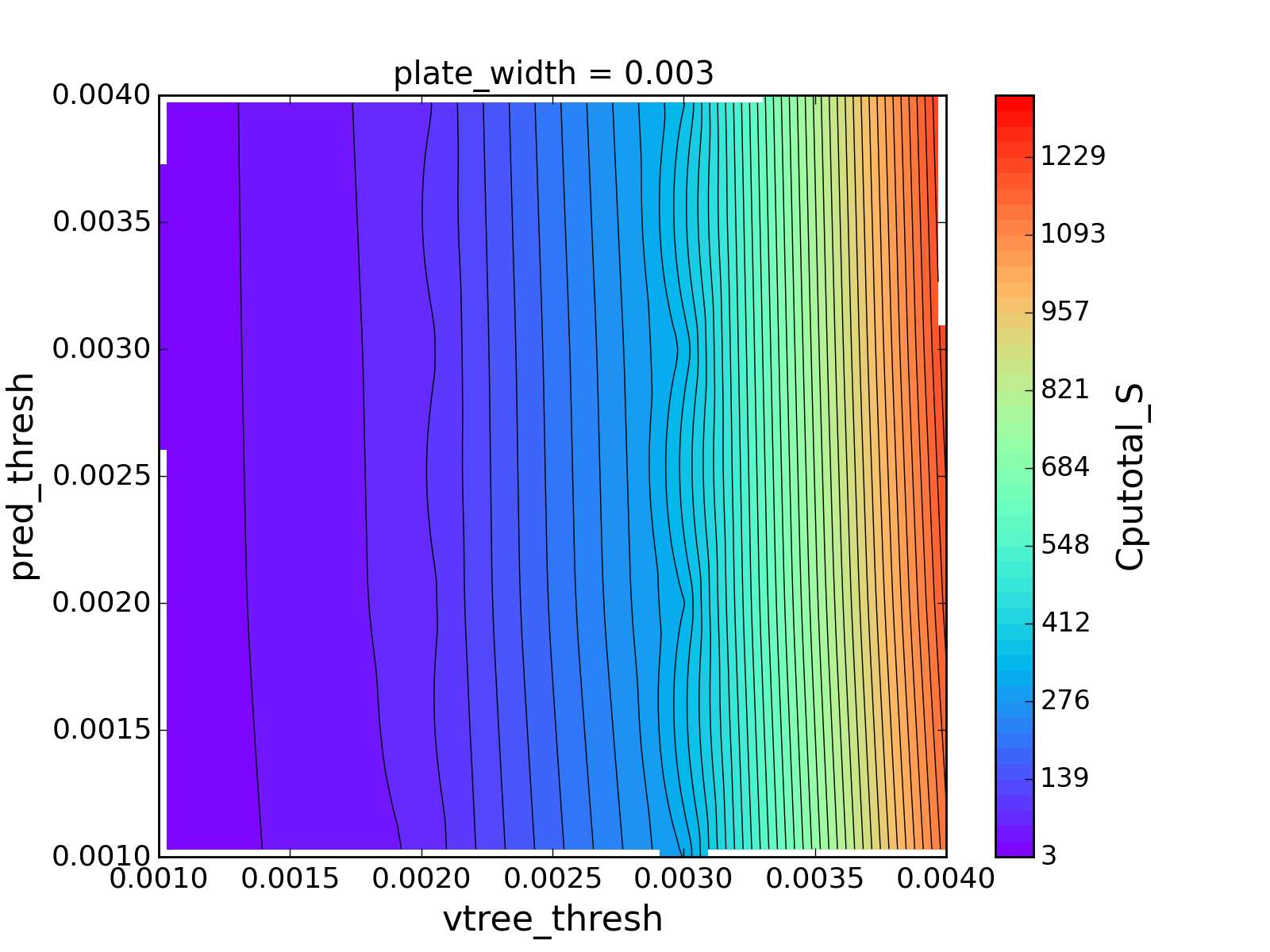}
    \includegraphics[width=0.35\textwidth]{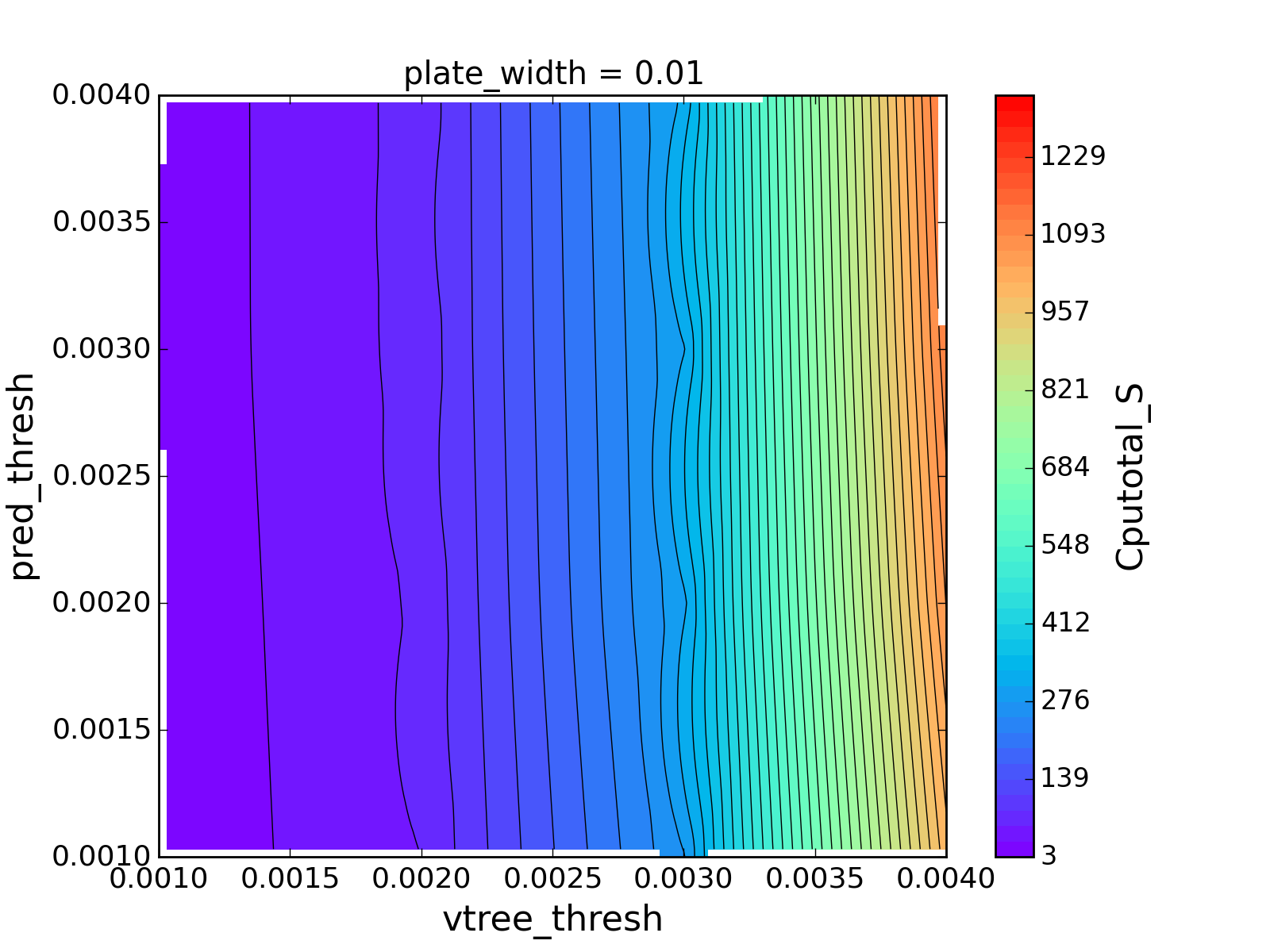}
    \includegraphics[width=0.35\textwidth]{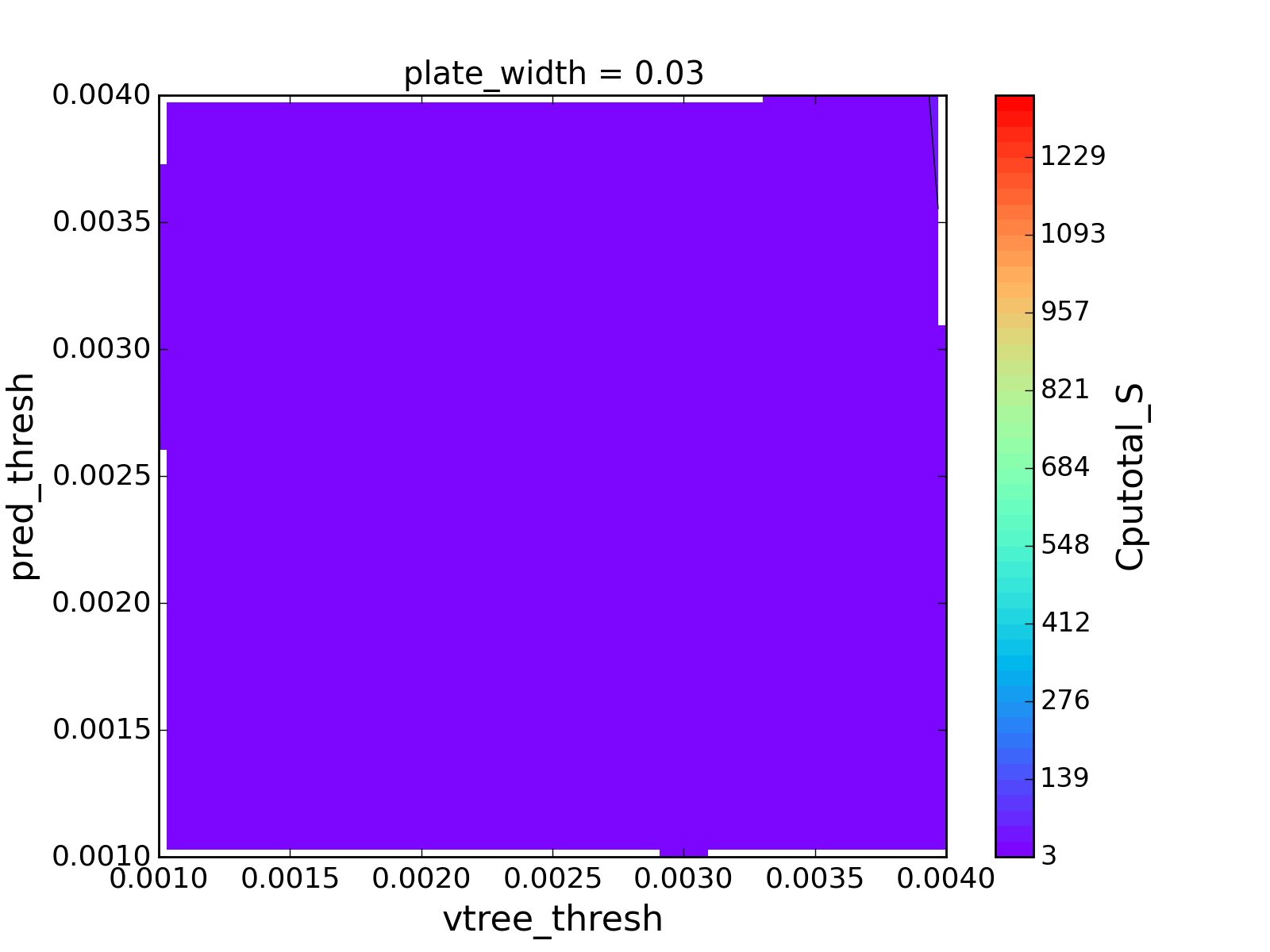}
     \caption{CPU time for running \linkt on a single, dense field as a function of the \vtree, \pred and \platew kd-tree linking parameters.}
    \label{fig.KD_cpu}
\end{figure}

\begin{figure}[tbh]
  \centering
    \includegraphics[width=0.35\textwidth]{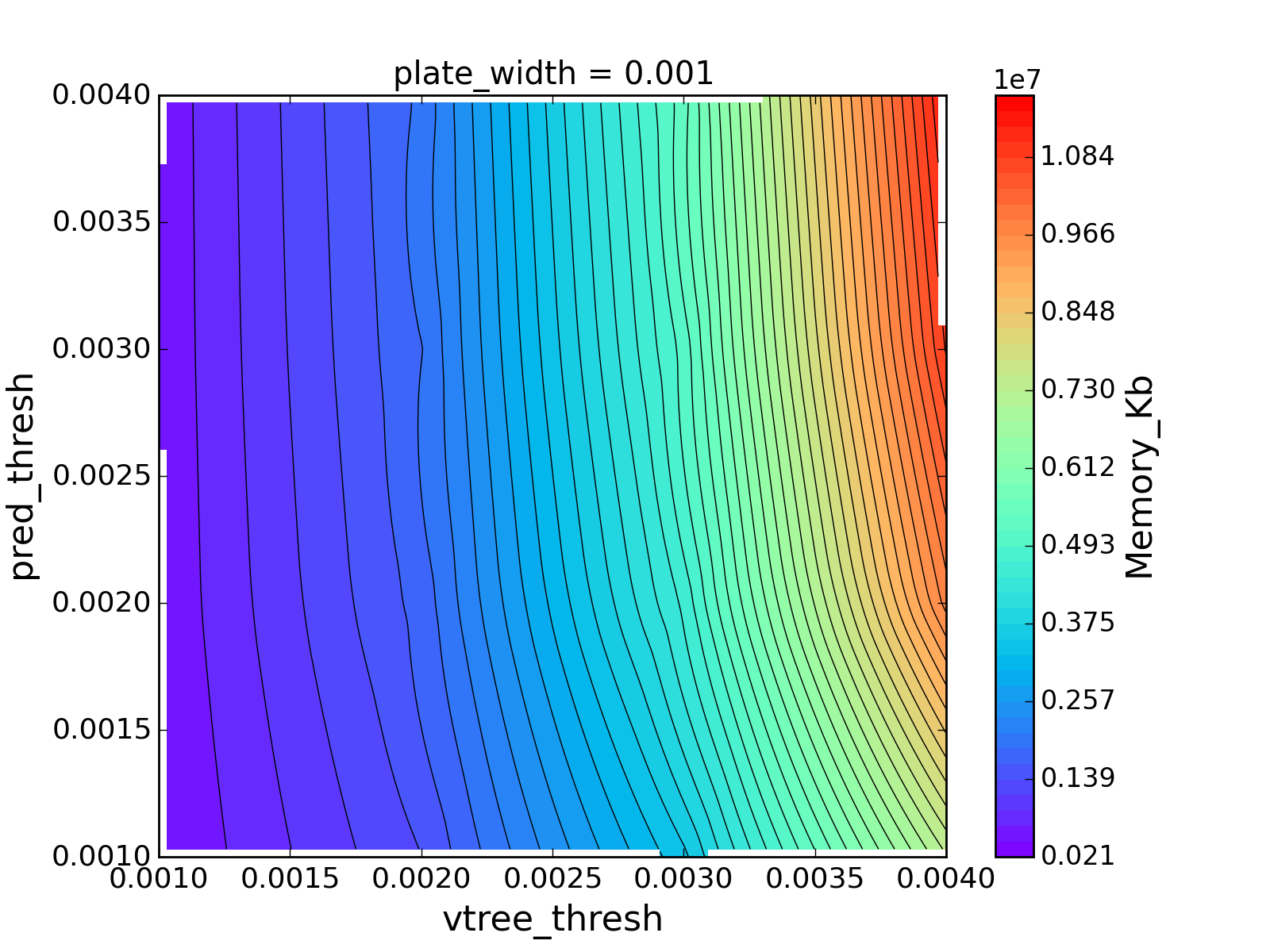}
    \includegraphics[width=0.35\textwidth]{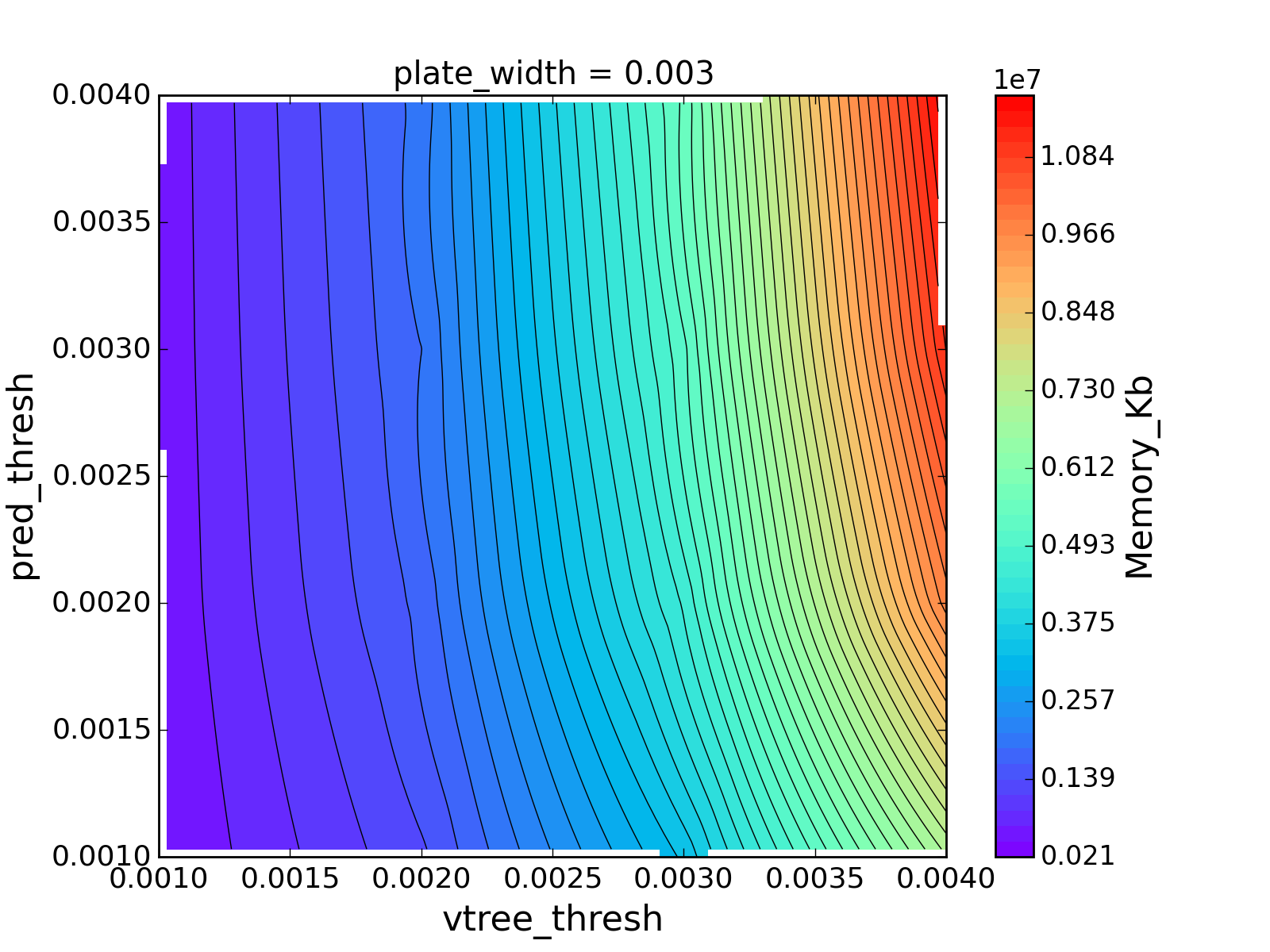}
    \includegraphics[width=0.35\textwidth]{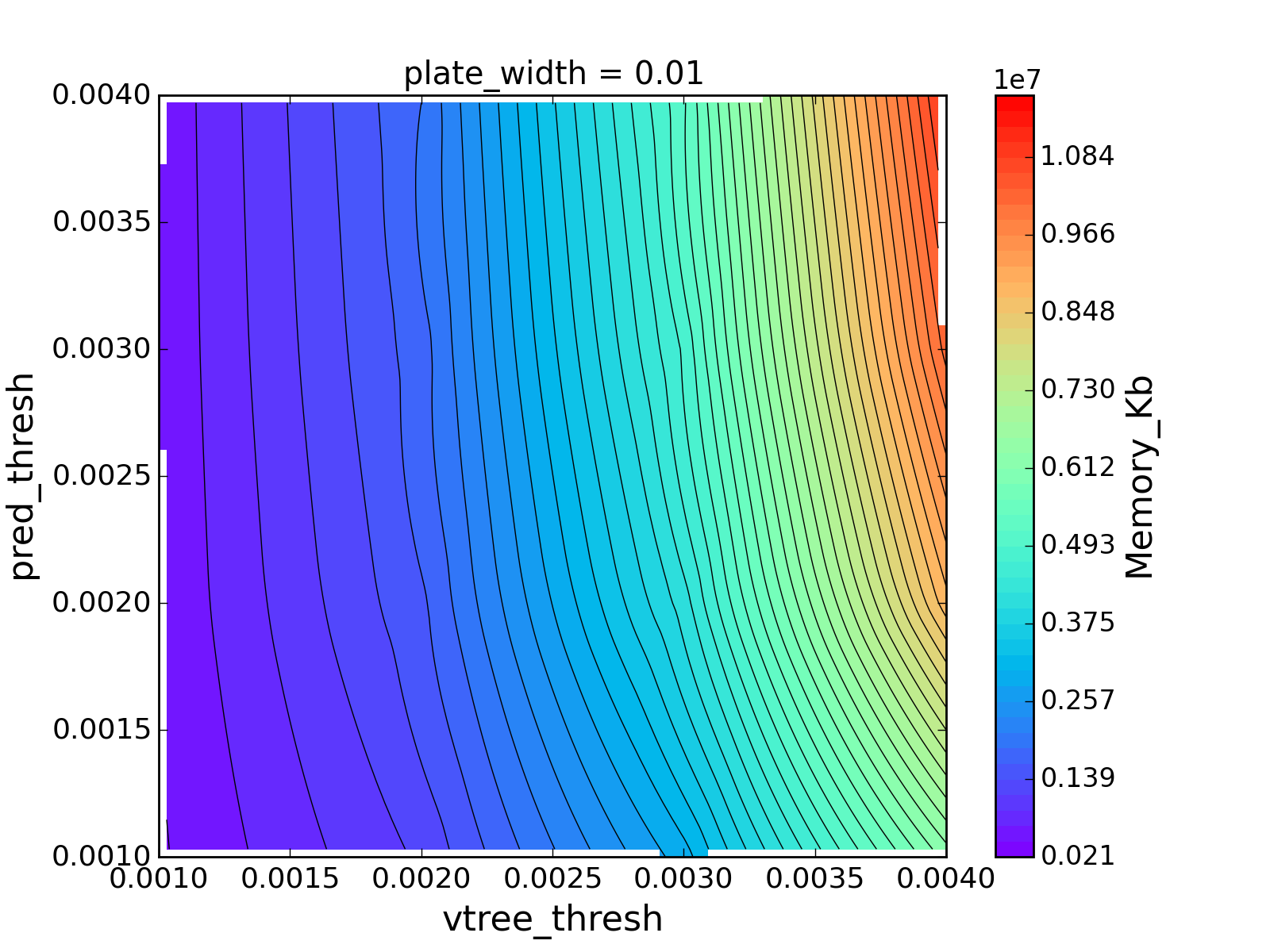}
    \includegraphics[width=0.35\textwidth]{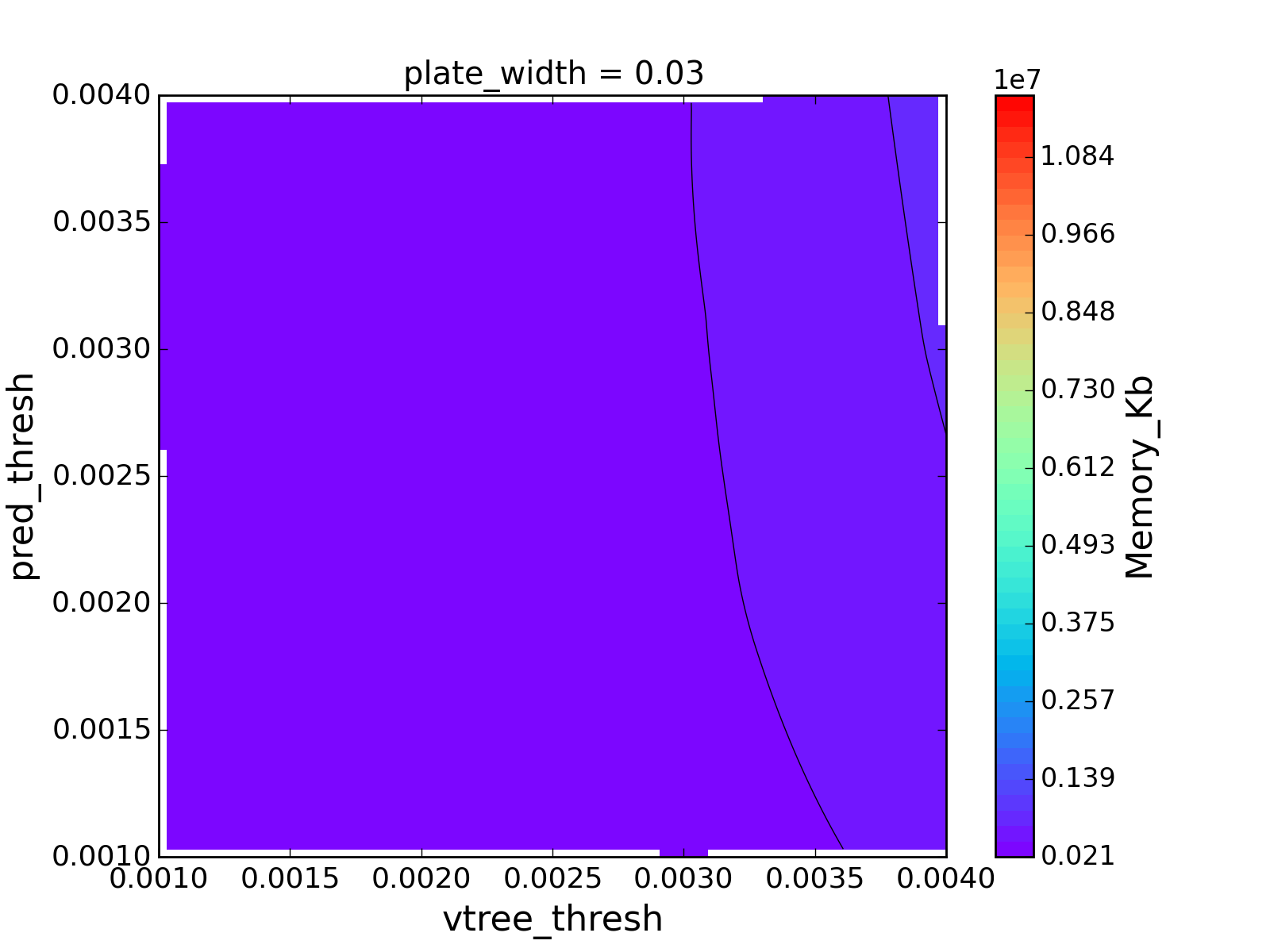}
     \caption{Memory usage of \linkt for a single, dense field as a function of the \vtree, \pred and \platew kd-tree linking parameters.}
    \label{fig.KD_memory}
\end{figure}

Automated linking of tracklets is a crucial element of LSST's NEO discovery pipeline. Without an automated linking stage, the NEO discovery rate would suffer and would rely heavily on follow-up observers, which will be impractical given the faint limit and volume of the LSST detections. The MOPS linking algorithm connects tracklets from three distinct nights into candidate tracks that are subsequently tested through orbit determination. The process consists of the following four distinct steps:

\begin{enumerate}
\item {\em Assemble tracklet list.} The first step collects, for a given field, all of the tracklets from the last $N$ nights for which the earlier position and velocity project into the destination field. The forward mapping of tracklets is based on linear motion, and acceleration that leads to nonlinear motion is not accounted for. Thus some NEO tracklets may be neglected, especially those very near the Earth with a rapid change in geometry and observable acceleration. 

\hspace{3mm} The combinatorics of linking strongly favor small $N$, but the objective of NEO completeness favors large $N$, which allows more missed detection opportunities. For LSST, $N$ usually ranges from 12--20, though 30 has been contemplated as a stretch goal. This work used $N=12$ days for linking tests, consistent with our objective of understanding whether linkage could be at all successful in the presence of large numbers of false detections. NEO linkage of nearby objects is not likely to succeed for large $N$ unless MOPS is extended so that some plane-of-sky acceleration is allowed when assembling the field-by-field tracklet lists. This would likely lead to a modest increase in the NEO discovery rate at the expense of many more false tracklets and increased linking overhead. 

\item {\em Assemble candidate track list.} The second step in linkage generates a list of candidate tracks based on the input tracklets. Generally, there are hundreds of available fields per night, each being processed in parallel. The \linkt algorithm is based on a kd-tree search \citep{2007ASPC..376..395K} that reduces the number of potential tracks to be tested from $n^2$ to $n\log n$, where $n$ is the number of tracklets available for linking on the given field. This saves a significant amount of computational resources, but the problem remains challenging. 

\hspace{3mm} \linkt has multiple tunable parameters, such as the minimum number of nights, the minimum number of detections, the minimum and maximum velocities, and some kd-tree linking parameters (\vtree, \pred, \platew). The ``vtrees" finds 2 compatible tracklets from which to estimate the endpoints of the track. The initial search pruning is done with respect to a maximum error denoted as \vtree. The track is confirmed when additional ``support tracklets" are found. \pred is a threshold for the goodness of fit for the support tracklets to the model estimated from the 2 initial tracklets. \platew in days flattens the tracklets epoch to the same time, if they fall within this margin. Different parameter values led to vastly different CPU and memory requirements, and markedly different numbers of candidate tracks. However, optimization of this stage is complex. The ideal parameter settings depended on the number of detections and varied from field to field. For instance, experiments with only synthetic NEO orbits led to 99\% linking efficiency. Adding noise and MBAs and running tests for selected target fields and tracks and varying \linkt parameters led to inconclusive results because the correct parameters depend on the field, and optimizing on a full lunation was infeasible. We explored the optimization of the kd-tree parameters on a single, dense field in the middle of OC28. The total number of candidate tracks increased as a function of \vtree and \pred, and there was only a weak dependence on \platew, at least for $\platew<0.01$ (Figure~\ref{fig.KD_total}). However, the most correct NEO tracks were derived for $\platew=0.003$ and$\vtree=0.003$ (Figure~\ref{fig.KD_neo}). Pushing the kd-parameters to obtain as many NEOs as possible led to an extreme increase in the number of false candidate tracks (Figures~\ref{fig.KD_ratio1}--\ref{fig.KD_ratio4}). Also, the memory and CPU load increased dramatically (Figures~\ref{fig.KD_cpu}--\ref{fig.KD_memory}). 

\hspace{3mm} This work was conducted with a single 8-core Linux workstation with 96 GB of memory (upgraded from 32 GB during the course of the work), and a crucial part of the challenge of linking was avoiding out-of-memory crashes. The final values utilized for the main linking simulation in this work were therefore a combination of feasibility and available computational resources: $(\vtree, \pred, \platew)= (0.001, 0.001, 0.003)$. This corresponds to the lower left corner of the upper right plot in Figures~\ref{fig.KD_total}--\ref{fig.KD_memory}. Better performance could have been obtained for, say, $(\vtree, \pred, \platew)= (0.003, 0.003, 0.003)$, but this would require use of a large cluster with more memory per core, something that will be readily available to LSST.

\item {\em Derive preliminary orbit.} The third step took the candidate tracks derived by \linkt and submitted them for Initial Orbit Determination (IOD). MOPS uses Gauss' method to generate potential initial orbits from the astrometry, and for each track the best fitting IOD is selected. Most false tracks were eliminated at this stage with no valid IOD.

\item {\em Perform differential corrections.} The fourth stage was Orbit Determination (OD), which used JPL OD routines to obtain converged orbits. This includes sophisticated fall-back logic to try to obtain 4- or 5-parameter fits if the 6-parameter orbit fit diverged. MOPS filtered out some false tracks at this stage based on rudimentary screening on post-fit residual statistics. As discussed below, MOPS's built-in orbit quality filtering is not strict and is agnostic regarding the expected errors in the astrometry, and thus relatively few false orbits were rejected at this stage. All orbits that passed the MOPS default quality screening were added to the MOPS derived object table, which was the basis for understanding the overall linking performance.
\end{enumerate}


\clearpage

\clearpage

\subsubsection{Linking Performance}

Linking tests were conducted on observing cycle 28 of the \enig baseline survey, with Granvik's NEO model, MBAs and the full false detection lists (Case F, Table~\ref{Tab.confusion_FF}). The NEO linking efficiency is defined as the number of unique NEOs present in the post-linking, derived-object catalog divided by the number of unique NEOs with possible 12-day tracks in the detection list. The linking efficiency was 93.6\% for $H<22$ NEOs and 84.0\% for all NEOs (i.e., $H<25$). These numbers were lower than the case without the false detections, where we achieved $>99\%$ linking efficiency, similar to previous work \citep{2007ASPC..376..395K,2013PASP..125..357D}. The lower efficiency for all NEOs arises from the fact that the vast majority of NEOs were of the smallest diameters, e.g., $23<H<25$. Also, smaller objects tend to have faster rates and greater acceleration because they are seen at closer geocentric distances, and they tend to have shorter observability windows. Note that the derived linking efficiency was for a single set of selected kd-tree parameters with a single 8-core workstation. With more powerful computational facilities and a more optimized kd-tree search (possibly on a per-field basis), there is excellent reason to believe that the linking efficiency can be significantly improved.

Many derived NEO orbits stemmed from objects in the MBA input catalog. Table~\ref{Tab.orbit_accuracy} shows the makeup of the 5348 NEO orbits (defined by $q<1.3\au$) derived from OC 28 alone. Among these orbits, 2222 originated from CLEAN linkages of actual NEOs, 1896 were CLEAN orbits associated with MBAs and 1230 were erroneous (``Not CLEAN") linkages. Nearly all of the erroneous linkages combined detections of different MBAs to form an NEO orbit; few were contaminated by false detections. At first blush this implies a purity of 77.0\% in the NEO catalog, but we describe below why this apparently low accuracy is mostly a manifestation of an ineffective orbit quality screening applied by MOPS. Correct interpretation of the orbits and improved screening increases the accuracy to 96\%. In contrast to the NEO orbits, Table~\ref{Tab.orbit_accuracy} reveals that the MBA catalog has 99.8\% purity already at this stage, without more refined filtering on orbit quality. Only 6 NEOs appear in the non-NEO orbit catalog, and most of these are borderline cases where $q\simeq1.3\au$.

\begin{table}[tbh]
\small
\caption{Accuracy of derived orbits from OC 28. The ``Incorrect Class.'' column indicates the number of objects for which the source object and the derived object had a different classification based on perihelion distance $q$. ``Not CLEAN'' indicates erroneous linkage of observations from either false detections or multiple objects.}
\begin{center}
\begin{tabular}{c|cccc}
\tableline \tableline
 Derived Classification& All & Incorrect Class. & Not CLEAN & Accuracy\\
 \hline
 NEO ($q\le1.3\au$) & 5348 & 1896  Non-NEO & 1230 &77.0\%\\
 Non-NEO ($q>1.3\au$) & 765,833 & 6 NEO & 1635 & 99.8\%\\
\hline
\end{tabular}
\end{center}
\label{Tab.orbit_accuracy}
\end{table}

\subsubsection{Orbit Quality Filtering}

The large fraction of erroneous linkages that appear in the NEO orbit catalog stem from a weak orbit quality filter implemented by MOPS, which requires the post-fit RMS of astrometric residuals to be less than 0.4 arcsec, a criterion that is too readily met for astrometry with a median error less than 0.05 arcsec. Moreover, because the RMS is not normalized by the reported astrometric uncertainty, it fails to take into account the varying quality of astrometry within and between tracklets in a candidate track. The upshot of this approach is that most such erroneous linkages show residuals clearly inconsistent with the astrometric uncertainty, and yet they pass the MOPS quality control test. Rather than modifying MOPS and re-running the simulation, we post-processed the post-fit astrometric residuals, with their associated uncertainties, to derive the sum of squares of the normalized residuals for each orbit in the NEO catalog. This provided the so-called $\chi^2$ of residuals, from which it is straightforward from classical statistics to calculate the probability $p_\mathrm{val}$ that the fit is valid, which is to say, the likelihood of of getting a higher value of $\chi^2$ by chance. A higher post-fit $\chi^2$ naturally leads to a lower $p_\mathrm{val}$ because the increased residuals reflect a poorer fit that has a lower probability.

\begin{figure}[tbh]
  \centering
    \plotone{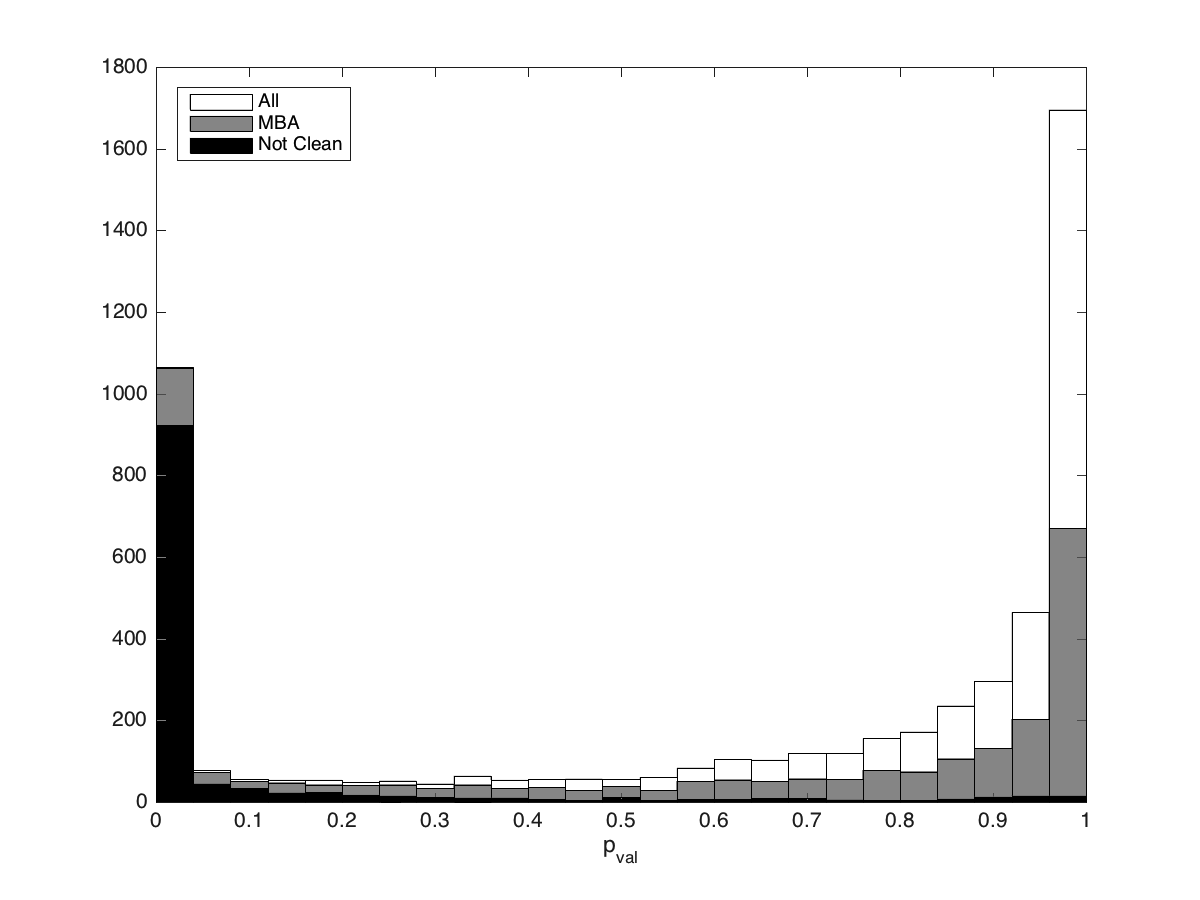}
     \caption{Histogram of postfit residual statistics of derived NEO orbits. In most cases, Not CLEAN NEO candidates can be easily distinguished.}
    \label{fig.neo_quality}
\end{figure}

Figure~\ref{fig.neo_quality} depicts the distribution of $p_{val}$ among the 5348 cataloged NEO orbits. The histogram reveals that few erroneous linkages appear for $p_{val}>0.25$ and that few NEOs appear for $p_{val}<0.25$, thus we selected 25\% as the $p_{val}$ cutoff for acceptable orbits. This criterion led to rejection of 7\% of clean and 87\% of not clean orbits. Most of the clean orbits that were filtered out were MBAs mis-classified as NEOs, 14\% of which were filtered out. Only 2\% of clean NEO orbits were removed by this filter. As tabulated in Table~\ref{Tab.MBconfusion}, more aggressive $p_\mathrm{val}$ filtering---at the 50\% or 90\% level---is less effective at removing erroneous linkages, even as the loss of clean NEOs becomes unacceptable. Thus a modest modification of MOPS is necessary to allow a more statistically rigorous orbit quality filtering, but the rudimentary approach described here leads to a 96\% purity (3816/3979, see Table~\ref{Tab.MBconfusion}) in the NEO catalog. In the context of accuracy, the clean MBAs that appear in the NEO orbit catalog are accounted as correctly linked, which is, in fact, the case.

\begin{table}[tbh]
\small
\caption{The number of cataloged NEO orbits of various classifications for varying values of the $p_{val}$ orbit quality filter. Here ``Non-NEO'' refers to MBAs that appear in the derived NEO catalog with $q<1.3\au$.}
\begin{center}
\begin{tabular}{l|rrrr}
\tableline \tableline
& \multicolumn{4}{c} {$p_{val}$ cutoff}\\
Classification & 0\% & 25\% & 50\% & 90\%    \\
\hline
All&5348&3979&3636&2314\\
CLEAN&4118&3816&3532&2279\\
Not CLEAN&1230&163&104&35\\
w/False Detection&35&3&1&1\\
CLEAN NEO&2222&2180&2062&1375\\
CLEAN MBA&1896&1636&1470&904\\
Not CLEAN NEO&2&0&0&0\\
Not CLEAN MBA&1228&163&104&35\\
\hline
\hline
\end{tabular}
\end{center}
\label{Tab.MBconfusion}
\end{table}

The rate of contamination of NEO orbits by false positives is extremely low, despite the large numbers of false positives injected into the detection stream. As shown in Table~\ref{Tab.confusion_noise}, after filtering at $p_{val}>25\%$, only 5 false detections appear in the NEO catalog. This can be compared to the total of over 29,000 detections that form the NEO catalog and the 51M false detections polluting the data stream. This result demonstrates that NEOs can be successfully linked with high efficiency and high accuracy when surveying with the baseline LSST cadence, even in the presence of significant numbers of false detections. 

\begin{table}[tbh]
\small
\caption{Number of detections of various classifications from OC 28. The total number in the input detection list and the number that were linked into the derived NEO catalog are shown.}
\begin{center}
\begin{tabular}{l|rrr}
\tableline \tableline
&Total&\multicolumn{2}{c}{---Derived NEO Catalog---}\\
& &All&$p_{val}<25\%$\\
\hline
Total &65,900,928&39,188&29,288\\
MBA&14,899,279&20,680&11,868\\
NEO &48,628&18,446&18,060\\
False&50,953,021&62&5\\
\% False&77.3\%&0.16\%&0.02\%\\
\hline
\hline
\end{tabular}
\end{center}
\label{Tab.confusion_noise}
\end{table}

\subsubsection{Confusion from MBAs}

\begin{figure}[tbh]
  \centering
    \epsscale{1.2}
    \plotone{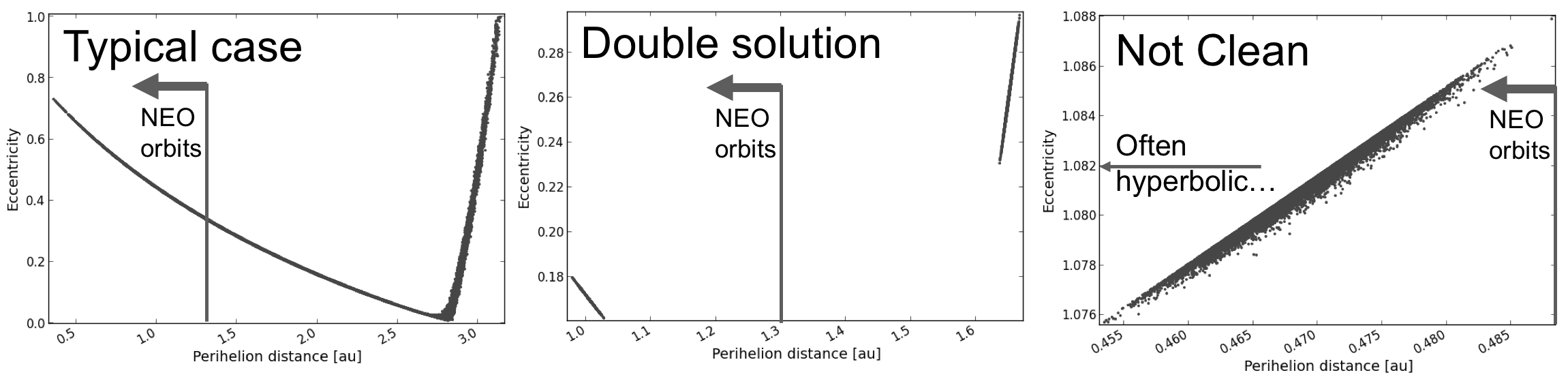}
     \caption{Examples of typical uncertainty regions for misclassified or erroneous linkages in the derived NEO orbit catalog. The plots depict Monte Carlo samples from systematic ranging that reflect the extent of possible solutions in perihelion distance $q$ and eccentricity $e$. The plots show the typical case of an MBA discovery (left) where the data are compatible with orbits spanning the NEO and MBA orbital regimes. In some of such cases two disjoint solutions are present, one NEO and one MBA (center). Erroneous linkages of two different MBAs often lead to NEO orbits with a small uncertainty, though many such cases are also hyperbolic.}
    \label{fig.MB_conf3}
\end{figure}

\begin{figure}[tbh]
  \epsscale{0.7}
  \centering
    \plotone{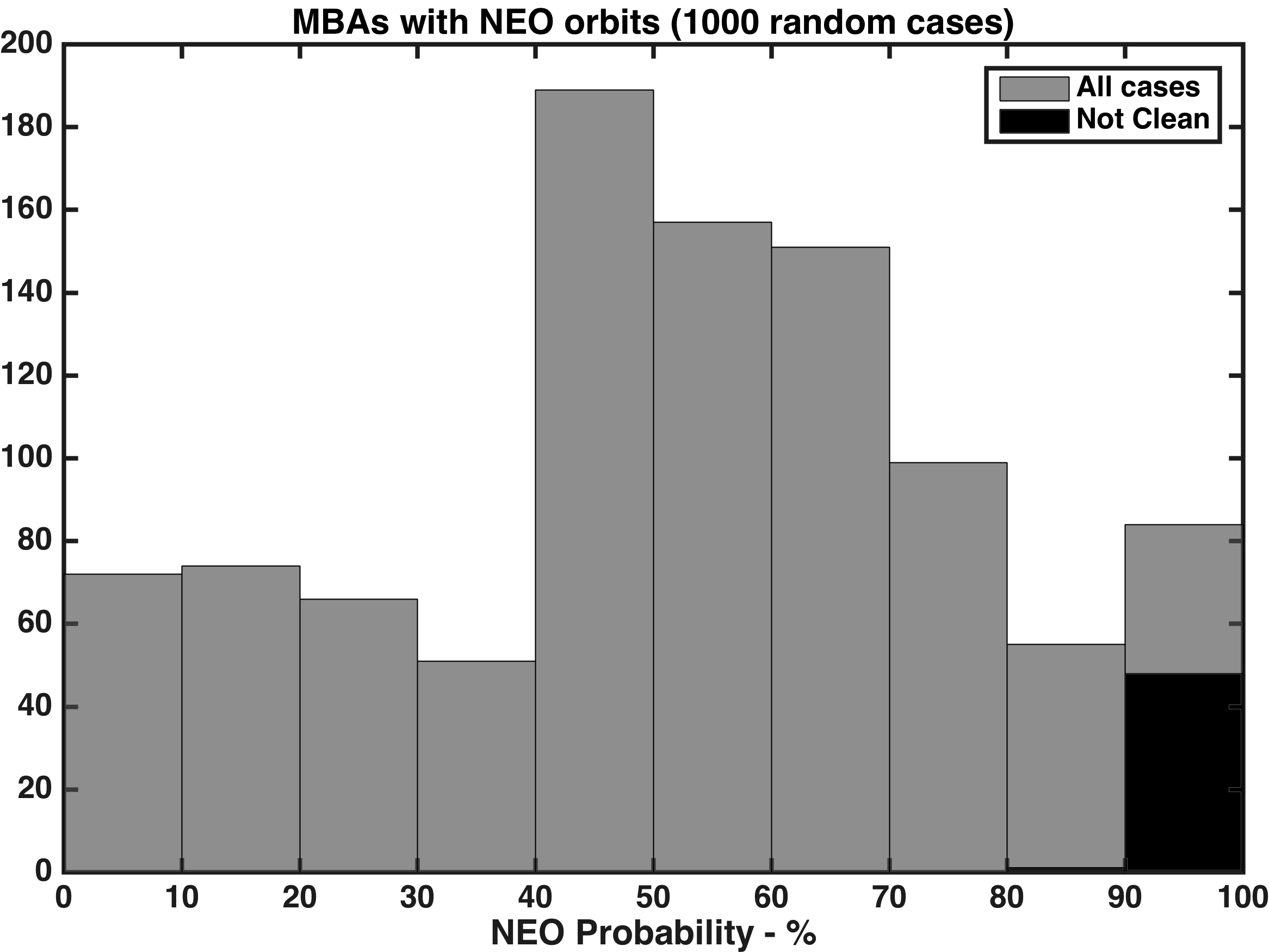}
     \caption{Histogram of computed probability that a track derived from MBA tracklets relates to an NEO orbit, as derived from systematic ranging analyses.}
    \label{fig.neo_prob}
\end{figure}

\begin{figure}[tbh]
  \epsscale{0.7}
  \centering
    \plotone{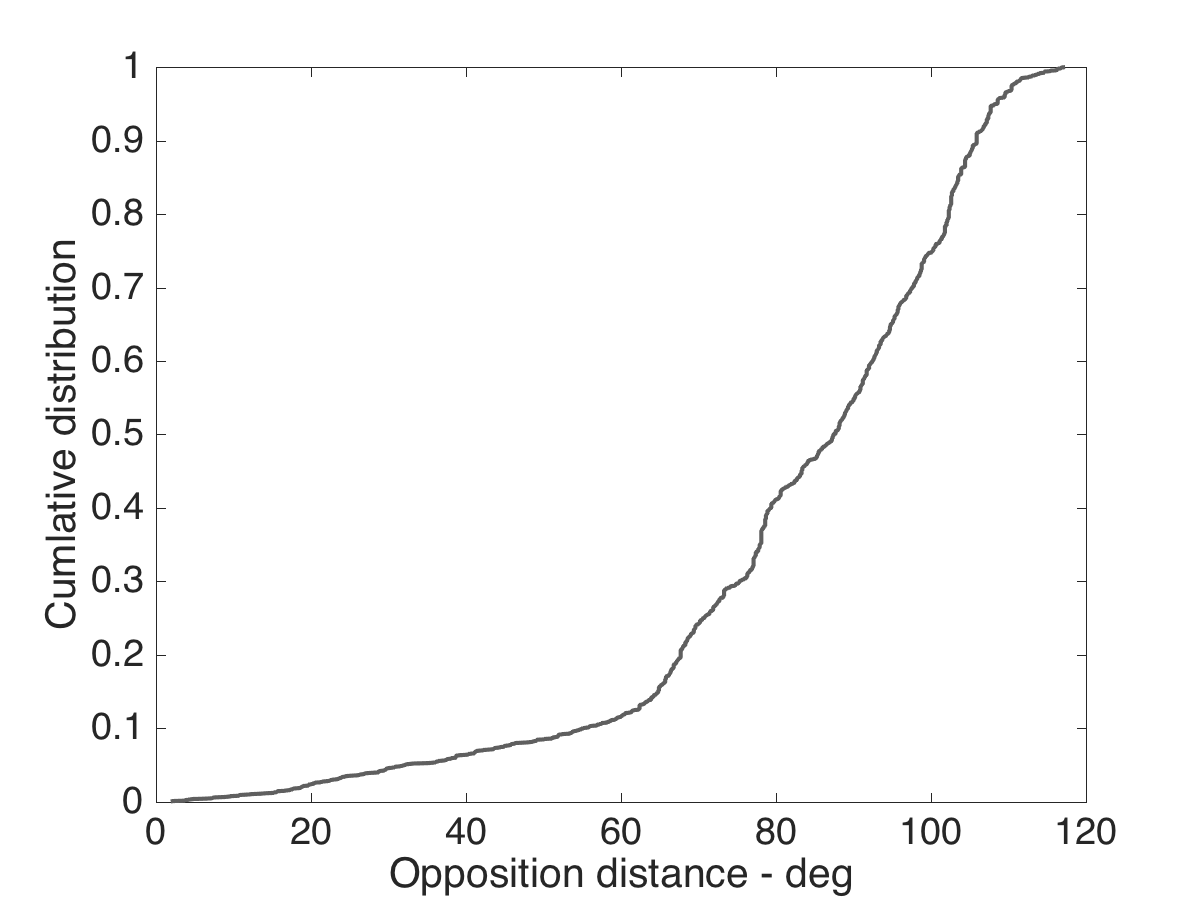}
     \caption{Cumulative distribution of opposition distance for MBAs that appear in the NEO orbit catalog with $p_{val}>25\%$. The distribution shows that this main-belt confusion is largely limited to detections made far from opposition, i.e., with low solar elongation.}
    \label{fig.opp-dist}
\end{figure}

To better understand the issue of the large fraction of NEO orbits stemming from correctly linked non-NEO objects, we used systematic ranging to explore the full orbit determination problem for these cases.  Systematic ranging is an orbit estimation technique designed to analyze poorly constrained orbits, typically with only one or a few nights of astrometry, for which the conventional least squares orbit determination can fail due to nonlinearity \citep{2015Icar..258...18F}. We tested hundreds of cases and found that nearly all showed a characteristic ``V''-shaped orbital uncertainty pattern in $e$ vs. $q$ that allowed both NEO and MBA orbits (left panel, Figure~\ref{fig.MB_conf3}). In some cases the ``V'' shape was broken at the vertex so that there were two distinct orbital solutions (center panel, Figure~\ref{fig.MB_conf3}). The systematic ranging technique affords a statistically rigorous estimate of the probability that the track represents an NEO orbit, and for these correctly-linked MBAs that appear with NEO orbits, few have high NEO probabilities, reflective of the fact that the data are compatible with the non-NEO (truth) orbits (Figure~\ref{fig.neo_prob}). It is also important to note that most of these MBAs that appear as NEOs are detected far from opposition. Figure~\ref{fig.opp-dist} shows that only $\sim10\%$ of these cases are found within $60\degree$ from opposition, and that about half are detected at $80\degree$ or farther from opposition. This result is merely reflecting the classical result that orbital ambiguities result from three-night orbits of objects far from opposition. It is an unavoidable feature of observing at low solar elongations, and is generally corrected after a fourth night of data is obtained. However, as described below, the current MOPS configuration does not efficiently attribute a fourth night of data to the already cataloged orbit, and so the ambiguity is often not resolved in our simulations. We note also that this confusion is an artifact of simulating only a single observing cycle. In actual operations, MBAs seen at low solar elongation would eventually move into the opposition region and appear even brighter there. These MBAs would be readily cataloged with their correct orbits because there is little ambiguity in the opposition region, at which point it becomes straightforward to link to the ambiguous orbits arising from near-sun detections. 

We also conducted systematic ranging analyses on some of the erroneous linkages leading to NEO orbits, almost all of which were erroneous MBA-MBA linkages, and these revealed a very different characteristic pattern in the $e$ vs. $q$ uncertainty space (right panel, Figure~\ref{fig.MB_conf3}). The uncertainty region was typically very small, leading to a high computed probability that the orbit is of an NEO (``Not Clean'' in Figure~\ref{fig.neo_prob}). In these cases, the uncertainty regions were also elongated and with one side having a sharp cutoff. In many such cases the heliocentric orbits were hyperbolic. This points to a likelihood that more effective screening tests can be developed to eliminate these false MBA-MBA linkages, despite the fact that some pass even strict orbit quality tests. For example, Table~\ref{Tab.MBconfusion} shows that even for $p_{val}>90\%$ a few dozen erroneous linkages remain in the NEO catalog. However, most of these erroneous MBA-MBA linkages are readily repaired when the individual MBAs are eventually re-observed at other epochs and correctly linked through other tracklets.

\subsubsection{Duplicate Orbits}

\begin{figure}[tbh]
  \centering
    \plottwo{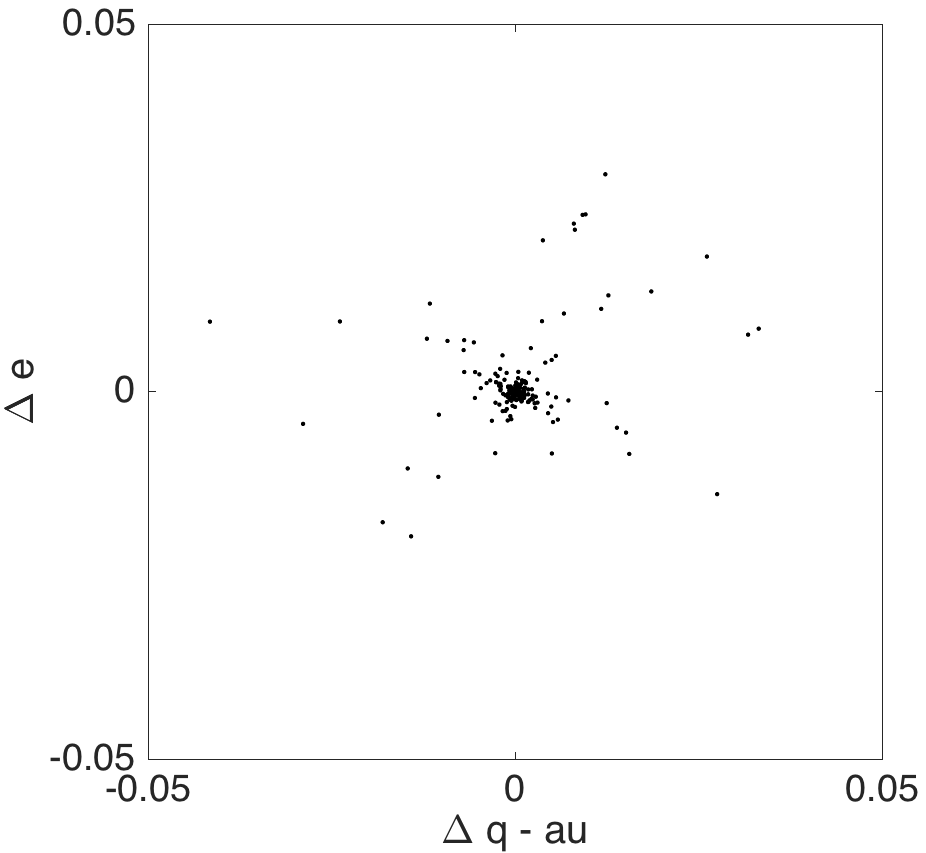}{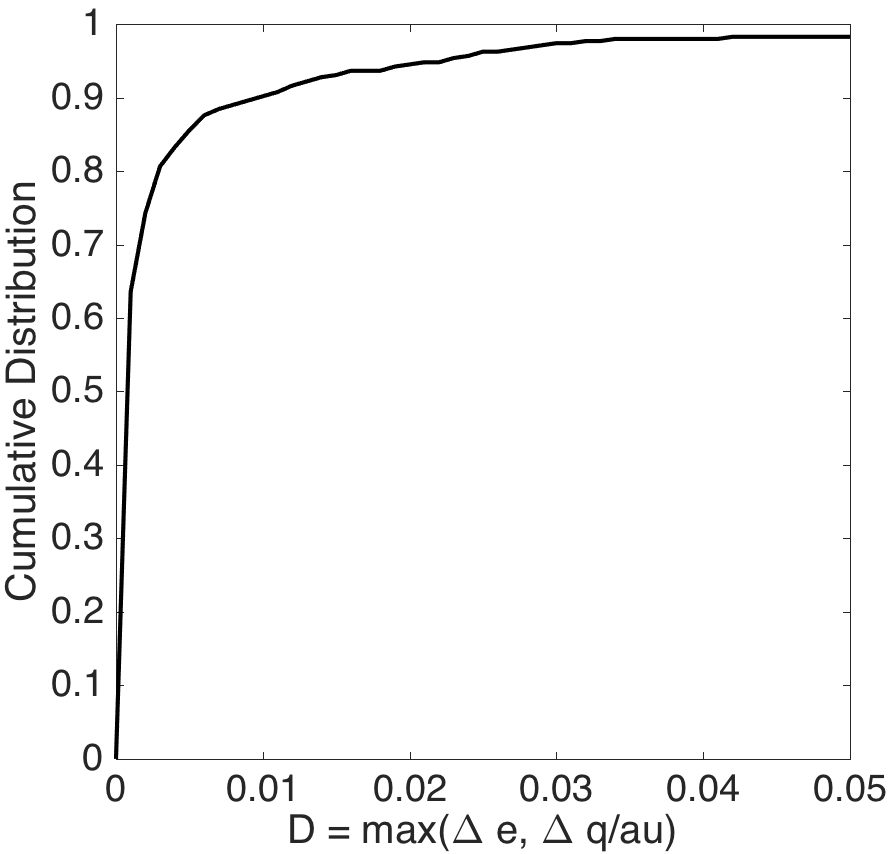}
    \caption{(left) Scatter plot of $\Delta q$ and $\Delta e$ between duplicate NEO orbits. (right) Cumulative distribution of duplicate separation in the $q$ and $e$ phase space. }
    \label{fig.duplicates}
\end{figure}

Table~\ref{Tab.orbit_accuracy} indicates that there were 4118 clean linkages in the NEO catalog, but not all of these are unique. Table~\ref{Tab.orbit_types} shows that 8.7\% of these are actually duplicate entries of the same object. In Figure~\ref{fig.duplicates} we see that the duplicate NEO entries are of almost identical orbits, with 95\% of duplicates matching in both eccentricity and perihelion distance (in au) to within 0.02. The non-NEO catalog has an even greater rate of duplication (17.3\%). 

\begin{table}[tbh]
\small
\caption{Duplication among derived orbits.}
\begin{center}
\begin{tabular}{c|cccc}
\tableline \tableline
 Class & Clean & Unique & Duplicates & Fraction\\
 \hline
 NEO & 4118 & 3758 & 360 & 8.7\%\\
 Non-NEO & 764,198 & 632,298 & 131,900 & 17.3\%\\
\hline
\hline
\end{tabular}
\end{center}
\label{Tab.orbit_types}
\end{table}

Virtually all of these duplicates are readily linked with standard orbit-to-orbit identifications techniques \citep{2000Icar..144...39M}, which are already part of MOPS. Most duplicates can be avoided altogether with a more efficient application of the MOPS attribution algorithm \citep{2001Icar..151..150M}. Within the linking process, a tracklet is first checked to see if it is can be attributed to an object already in the catalog. If so then it is linked to that object and removed from the tracklet list so that it is not passed along to kd-tree linking. The fact that so many objects in our simulation are linked into multiple independent tracks in a single observing cycle implies, first, that there are at least six tracklets in the lunation, indicating a very solid discovery, and second, that the attribution algorithm can easily be tuned to attribute these extra tracklets before they are even linked into tracks. Not only would such a re-tuning keep the orbit catalog cleaner, it would also cut down on the computational expense of kd-tree searches by removing tracklets from the search that are associated with already discovered objects. The problem of duplicate orbits is likely to be easily resolved through testing and tuning of existing MOPS functionality.


\section{Discussion}

We performed a high-fidelity simulation of linking NEO and MBA detections into orbits in a realistic density scenario with false detections and constraints of the LSST survey in one observing cycle. Tracklet generation created false tracklets at a rate of 57\% being false. This rate can be larger if one neglects the information on trail length and orientation when creating tracklets. We used this velocity information for the velocity range of 1.2--2.0 deg/day. 

Optimization of kd-tree parameters to provide maximum number of clean tracks is correlated with large number of false tracks and varies from field to field. It is also CPU and memory intensive, though it can be managed by distributed and multi-core or cloud computing.

On a single-lunation, full-density simulation, with NEOs, MBAs and false detections, we obtained a linking efficiency of 93.6\% for $H<22$ NEOs with 12-day tracks. Linking efficiency on the full population down to $H<25$ was lower. We believe that, with modest revision and tuning of the MOPS linking algorithms and  an appropriate allocation of computational resources that this number can be significantly increased, probably to  99\% or more. 

On the same simulation, the derived NEO catalog was comprised of 96\% correct linkages. The remaining 4\% of linkages were almost exclusively incorrect MBA-MBA links, most of which should be eliminated over a longer duration simulation. Less than 0.1\% of orbits in the derived NEO catalog included false detections.

Some enhancements to MOPS are needed in the linking stage to eliminate duplicate and false orbits. This includes improving the orbit quality filter and tuning of the attribution, precovery\footnote{Here ``precovery'' refers to a search of the MOPS database for tracklets observed previously that did not form a derived object because not enough tracklets were observed at the time. It is similar to attribution of new detections, but operates on past observations.} and orbit-orbit identification modules. Together with optimization of the kd-tree track search, this would increase the linking efficiency and thus increase the number of cataloged NEOs. The linking efficiency directly affects the discovery completeness as discussed in \citet{2017Veres_1}.

\acknowledgements
\begin{center}
{\em Acknowledgments} 
\end{center}

The Moving Object Processing System was crucial to the successful completion of this study. This tool is the product of a massive development effort involving many contributors, whom we do not list here but may be found in the author list of \citet{2013PASP..125..357D}. This report identifies a few deficiencies in MOPS, but our remarks should not be viewed in a pejorative sense. The software has so far never been fielded for its designed purpose, and we expect that minor improvements and tuning can resolve the issues that we have mentioned. We thank Larry Denneau (IfA, Univ. Hawaii) for his tremendous support in installing and running the MOPS software.

This study benefited from extensive interactions with Zeljko Ivezic, Lynne Jones and Mario Juric, all from the University of Washington. As members of the LSST project, they provided vital guidance in understanding the performance and operation of LSST. They also provided important insight into the expected interpretation and reliability of LSST data. And they reviewed with us their early results on DECam image processing, which allowed us to include credible image differencing artifacts in the simulated LSST detection stream.

Davide Farnocchia (JPL) supported the systematic ranging analyses of linking products described in this report.

Mikael Granvik (Univ. Helsinki) kindly provided an early version of the \citet{2016Natur.530..303G} NEO population model, which was used extensively in this study.

This  research  was  conducted  at  the  Jet  Propulsion  Laboratory, California Institute of Technology, under a contract with the National Aeronautics and Space Administration.

\vspace{.3cm}

\noindent\copyright\ Copyright 2017 California Institute of Technology. Government sponsorship acknowledged.

\end{document}